\documentclass[twocolumn,floatfix,showpacs,prd,aps,tightenlines,superscriptaddress]{revtex4}
\usepackage{amsmath}
\usepackage{graphicx}
\usepackage{psfrag}
\usepackage{color}
\usepackage{dcolumn}
\usepackage{bm}
\usepackage{longtable}
\usepackage[mathscr]{eucal}
\usepackage{mathrsfs}

\def\msun{\rm M_{\odot}}
\def\kms{\rm km \, s^{-1}}

\def\simlt{\mathrel{\rlap{\lower 3pt\hbox{$\sim$}}\raise 2.0pt\hbox{$<$}}}
\def\simgt{\mathrel{\rlap{\lower 3pt\hbox{$\sim$}} \raise 2.0pt\hbox{$>$}}}
\def\lsim{\mathrel{\rlap{\lower 3pt\hbox{$\sim$}}\raise 2.0pt\hbox{$<$}}}
\def\gsim{\mathrel{\rlap{\lower 3pt\hbox{$\sim$}} \raise 2.0pt\hbox{$>$}}}

\def\mbulge{M_{\rm Bulge}}
\def\msunpc3{\msun~{\rm {pc^{-3}}}}
\newcommand{\be}{\begin{equation}}
\newcommand{\ee}{\end{equation}}
\def\kms{{\rm\,km\,s^{-1}}}

\newcommand{\bea}{\begin{eqnarray}}
\newcommand{\eea}{\end{eqnarray}}
\newcommand{\beq}{\begin{equation}}
\newcommand{\eeq}{\end{equation}}
\newcommand{\KMS}{\rm km\ s^{-1}}

\begin{document}

\def\fun#1#2{\lower3.6pt\vbox{\baselineskip0pt\lineskip.9pt
  \ialign{$\mathsurround=0pt#1\hfil##\hfil$\crcr#2\crcr\sim\crcr}}}
\def\lap{\mathrel{\mathpalette\fun <}}
\def\gap{\mathrel{\mathpalette\fun >}}
\def\kms{{\rm km\ s}^{-1}}
\def\vk{V_{\rm recoil}}

\title{Gravitational Recoil From Accretion-Aligned Black-Hole Binaries}

\author{
Carlos O. Lousto,
Yosef Zlochower
}
\affiliation{Center for Computational Relativity and Gravitation,\\
and School of Mathematical Sciences, Rochester Institute of
Technology, 85 Lomb Memorial Drive, Rochester, New York 14623}
\author{Massimo Dotti}
\affiliation{Universit\`a di Milano Bicocca, Dipartimento di Fisica G. Occhialini, \\
Piazza della Scienza 3, I-20126, Milano, Italy}
\author{Marta Volonteri}
\affiliation{Astronomy Department, University of Michigan, Ann Arbor 48109, USA
and Institut d'Astrophysique de Paris, 98 bis Bd Arago, Paris, 75014, France
}

\begin{abstract}

We explore the newly discovered ``hangup-kick'' effect, which greatly
amplifies the recoil for configurations with partial spin-/ orbital-angular
 momentum alignment, by studying a
set of 48 new simulations of equal-mass, spinning black-hole binaries. 
We propose a
phenomenological model for the recoil that takes this new effect into
account and then use this model, in conjunction with statistical
distributions for the spin magnitude and orientations, based on
accretion simulations, to find the probabilities for observing recoils
of several thousand $\KMS$. In addition, we provide initial parameters,
eccentricities, radiated linear and angular momentum, precession rates
and remnant mass, spin, and recoils for all 48 configurations. Our
results 
indicate that surveys exploring peculiar (redshifted or blueshifted)
differential
line-of-sight velocities should observe at least one case 
above $2000\ \KMS$ out of
four thousand merged galaxies. On the other hand, the  
probability that a remnant BH recoils in any direction at a velocity
 exceeding the $\sim 2000\ \KMS$ escape velocity of large elliptical 
galaxies is $0.03\%$.
Probabilities of recoils exceeding the escape velocity
quickly rise to 5\% 
for galaxies with escape velocities of $1000\ \KMS$ and nearly
20\% for galaxies with escape velocities of
$500\ \KMS$. In addition the direction of these large recoils is
strongly peaked toward the angular momentum axis, with very low
probabilities of recoils exceeding $350\ \KMS$ for angles larger
than $45^\circ$ with respect to the orbital angular momentum axis.

\end{abstract}

 \maketitle

\section{Introduction}\label{sec:Introduction}

Speculations about the relevance of gravitational recoils in
astrophysical black-hole binary (BHB) mergers can be traced back at
least thirty years \cite{1984RvMP...56..255B,Redmount:1989}.  The
crucial scale of the problem is when those recoils reach velocities
comparable to the escape velocities of the relevant structures, i.e.\
globular clusters, which have escape velocities of 10s of $\KMS$,
 and dwarf, spiral, and giant elliptical galaxies, which have escape
velocities from 100s to $\sim1000$ $\KMS$ for normal galaxies.
 For large galaxies undergoing major
mergers the effective escape velocity can be up to a factor of a few
higher at the time of coalescence, as the central potential well
deepens rapidly at that time. Once the merger is complete and the
stellar systems begins to relax, the potential becomes shallower
\cite{Blecha:2011}.

Early attempts to compute recoil velocities from BHB mergers
used perturbative \cite{1983MNRAS.203.1049F,1984MNRAS.211..933F} and
post-Newtonian approximations (see \cite{Blanchet:2005rj} for a review up
to 2005) and found recoils up to a few hundred $\KMS$,
 but uncertainties in the 
computations were of the same order of magnitude as those velocities
(see \cite{LeTiec:2009yg} for a more current review). 
The first computation that used full numerical
simulations within the Lazarus approach produced similar results 
\cite{Campanelli:2004zw}.

The accurate computation of recoil velocities had to wait for the 2005
breakthroughs~\cite{Pretorius:2005gq, Campanelli:2005dd, Baker:2005vv}
in Numerical relativity (NR), since it proved to be a genuinely
strong-field, 
highly nonlinear General Relativistic phenomenon.
The first systematic study of recoil velocities considered unequal mass, 
nonspinning BHBs~\cite{Gonzalez:2006md}. That study found
that the maximum recoil velocity for non-spinning BHBs is 175 $\KMS$,
which occurs for a mass ratio near 1:3.

Unexpectedly, spinning BHBs, with individual spins anti-aligned
with each other and both parallel to the angular momentum direction
were found to produce recoils
of over a factor two larger than the unequal-mass maximum \cite{Herrmann:2007ac, Koppitz:2007ev}, 
and a revolution occurred when it was discovered~\cite{Campanelli:2007ew} 
that a configuration of the spins lying in the orbital plane led to recoils
of almost~\cite{Campanelli:2007cga, Lousto:2011kp} $4000\ \KMS$. This last
figure caught the attention of observational astronomers who
began to look for these highly-recoiling BHs by searching 
the spectral data of galaxies for differential redshifts of several
thousand $\KMS$. The idea there was that gas close to the BH would
remain bound to it, while gas further out in the accretion disk would
be left behind. The two gas components would then have different relative
redshifts.
 Initial searches produced the first supermassive recoiling  BH
candidates
\cite{Komossa:2008qd,Shields:2008va,Bogdanovic:2008uz,Civano:2010es} 
and now more thorough surveys
have increased the numbers of potential candidates to several dozen
\cite{Eracleous:2011ua,Tsalmantza:2011ju}. These observations may provide
the first confirmation of a general relativistic strong field, 
highly-dynamical, full-numerical prediction.

A recent study~\cite{Lousto:2011kp} pointed out that configurations
with partially aligned spins, which we call the ``hangup-kick''
configuration, can lead to even larger recoil velocities, of nearly
5000 $\KMS$. More importantly, these configurations are favored with
respect to the ``spin in the orbital plane'' configuration by the
effects of accretion on the BHs during very early orbital (Newtonian)
stages \cite{Bogdanovic:2007hp,Dotti:2009vz}.  We address this
question in more detail in this paper.

This paper is organized as follows. In Sec.~\ref{Sec:Numerical} we
review
the numerical methodology to perform the simulations.
In Sec.~\ref{sec:Runs}, we  describe the initial configurations of
a family of BHBs chosen to model ``hangup-kicks''. In
Sec.~\ref{Sec:Results}, we provide the main results of these
evolutions in tables of radiated energies, angular and linear momenta, as 
well as final remnant mass and spin. We then model recoils using
empirical fitting formulas.
In Sec.~\ref{Sec:Accretion} we describe smoothed particle hydrodynamics (SPH)
 simulations that model accretion
onto BHBs to obtain the spin magnitude and direction distribution of
the individual spins in merging BHBs
for our full numerical simulations. 
Using these spin direction and magnitude distributions, we give
predictions for the recoil distribution and the probabilities of
observing large recoils.
We discuss the consequences and future extensions of
these techniques in Sec.~\ref{Sec:Discussion}.

\section{Numerical Relativity Techniques}\label{Sec:Numerical}

We use the TwoPunctures thorn~\cite{Ansorg:2004ds} to generate initial
puncture data~\cite{Brandt97b} for the black-hole binary (BHB) simulations
described below. These data are characterized by mass parameters $m_p$, which are
not the horizon masses, of each BH, as well as the momentum and spin of each BH.
 We evolve these
BHB data-sets using the {\sc
LazEv}~\cite{Zlochower:2005bj} implementation of the moving puncture
approach~\cite{Campanelli:2005dd,Baker:2005vv} with the conformal
function $W=\sqrt{\chi}=\exp(-2\phi)$ suggested by Ref.~\cite{Marronetti:2007wz}.
For the runs presented here,
we use centered, eighth-order finite differencing in
space~\cite{Lousto:2007rj} and a fourth-order Runge Kutta time integrator. (Note that we do
not upwind the advection terms.)

Our code uses the {\sc Cactus}/{\sc EinsteinToolkit}~\cite{cactus_web,
einsteintoolkit} infrastructure.
We use the {\sc Carpet}~\cite{Schnetter-etal-03b} mesh refinement driver to
provide a ``moving boxes'' style of mesh refinement. In this approach
refined grids of fixed size are arranged about the coordinate centers
of both holes.  The {\sc Carpet} code then moves these fine grids about the
computational domain by following the trajectories of the two BHs.

We obtain accurate, convergent waveforms and horizon parameters by
evolving this system in conjunction with a modified 1+log lapse and a
modified Gamma-driver shift
condition~\cite{Alcubierre02a,Campanelli:2005dd,vanMeter:2006vi}, and an initial lapse
$\alpha(t=0) = 2/(1+\psi_{BL}^{4})$, where $\psi_{BL}$ is the
Brill-Lindquist conformal factor and is given by
$$
\psi_{BL} = 1 + \sum_{i=1}^n m_{i}^p / (2 |\vec r- \vec r_i|),
$$
where $\vec r_i$ is the coordinate location of puncture $i$.
The lapse and shift are evolved
with
\begin{subequations}
\label{eq:gauge}
  \begin{eqnarray}
(\partial_t - \beta^i \partial_i) \alpha &=& - 2 \alpha K,\\
 \partial_t \beta^a &=& (3/4) \tilde \Gamma^a - \eta \beta^a \,,
 \label{eq:Bdot}
 \end{eqnarray}
 \end{subequations}
where we use $\eta=2$ for all simulations presented below.

We use {\sc AHFinderDirect}~\cite{Thornburg2003:AH-finding} to locate
apparent horizons.  We measure the magnitude of the horizon spin using
the Isolated Horizon algorithm detailed in Ref.~\cite{Dreyer02a}.
Note that once we have the
horizon spin, we can calculate the horizon mass via the Christodoulou
formula
\begin{equation}
{m_H} = \sqrt{m_{\rm irr}^2 + S_H^2/(4 m_{\rm irr}^2)} \,,
\end{equation}
where $m_{\rm irr} = \sqrt{A/(16 \pi)}$ and $A$ is the surface area of
the horizon, and $S_H$ is the spin angular momentum of the BH (in
units of $M^2$).
In the tables below, we use the variation in the measured horizon
 irreducible mass and spin during the simulation 
as a measure of the error in these quantities.
We measure radiated energy, linear momentum, and angular momentum, in
terms of the radiative Weyl Scalar $\psi_4$, using the formulas provided in
Refs.~\cite{Campanelli:1998jv,Lousto:2007mh}. However, rather than using
the full $\psi_4$, we decompose it into $\ell$ and $m$ modes and solve
for the radiated linear momentum, dropping terms with $\ell \geq 5$.
The formulas in Refs.~\cite{Campanelli:1998jv,Lousto:2007mh} are valid at
$r=\infty$.
We extract the radiated energy-momentum at finite
radius and extrapolate to $r=\infty$ using both linear and quadratic
extrapolations. We use the difference of these two extrapolations as
a measure of the error.

\section{Simulations}\label{sec:Runs}

We evolved a set of 48 equal-mass, spinning, quasicircular,
``hangup-kick'' configurations, with 30 simulations having individual BH
spins of magnitude $\alpha=1/\sqrt{2}$ and $18$ simulations having BH
spin magnitudes of $\alpha=0.9$, where $\vec \alpha$ is the
dimensionless
spin of the BH ($\vec \alpha = \vec S_{H}/M_{H}^2$, where $\vec S_{H}$
is the spin angular momentum and $M_{H}$ is the mass of the BH).
 In the ``hangup-kick'' configuration, the $z$ components
of the individual spins are equal, while the projections of the
individual spins onto the orbital plane are equal in magnitude but
opposite in direction.  The $\alpha=1/\sqrt{2}$ configurations were
split into five sets of 6, where the runs in each individual set had
the same initial angle $\theta$ between the spin direction and orbital
angular momentum direction (here we chose $\theta=22.5^\circ$,
$45^\circ$, $60^\circ$, $120^\circ$, $135^\circ$). In each set with as
given $\theta$, we chose the initial orientation $\phi_i$ between the
in-plane spin and linear momentum to be $0^\circ$, $30^\circ$,
$90^\circ$, $130^\circ$, $210^\circ$, and $315^\circ$. For the
$\alpha=0.9$ runs, we used the same initial 6 $\phi_i$ configurations
for $\theta=60^\circ$, $\theta=30^\circ$, and $\theta=15^\circ$.  We
combine these results with the simulations of~\cite{Lousto:2010xk}
(which have $\theta=90^\circ$) in order to perform our analysis below.

\begin{figure}
  \includegraphics[width=\columnwidth]{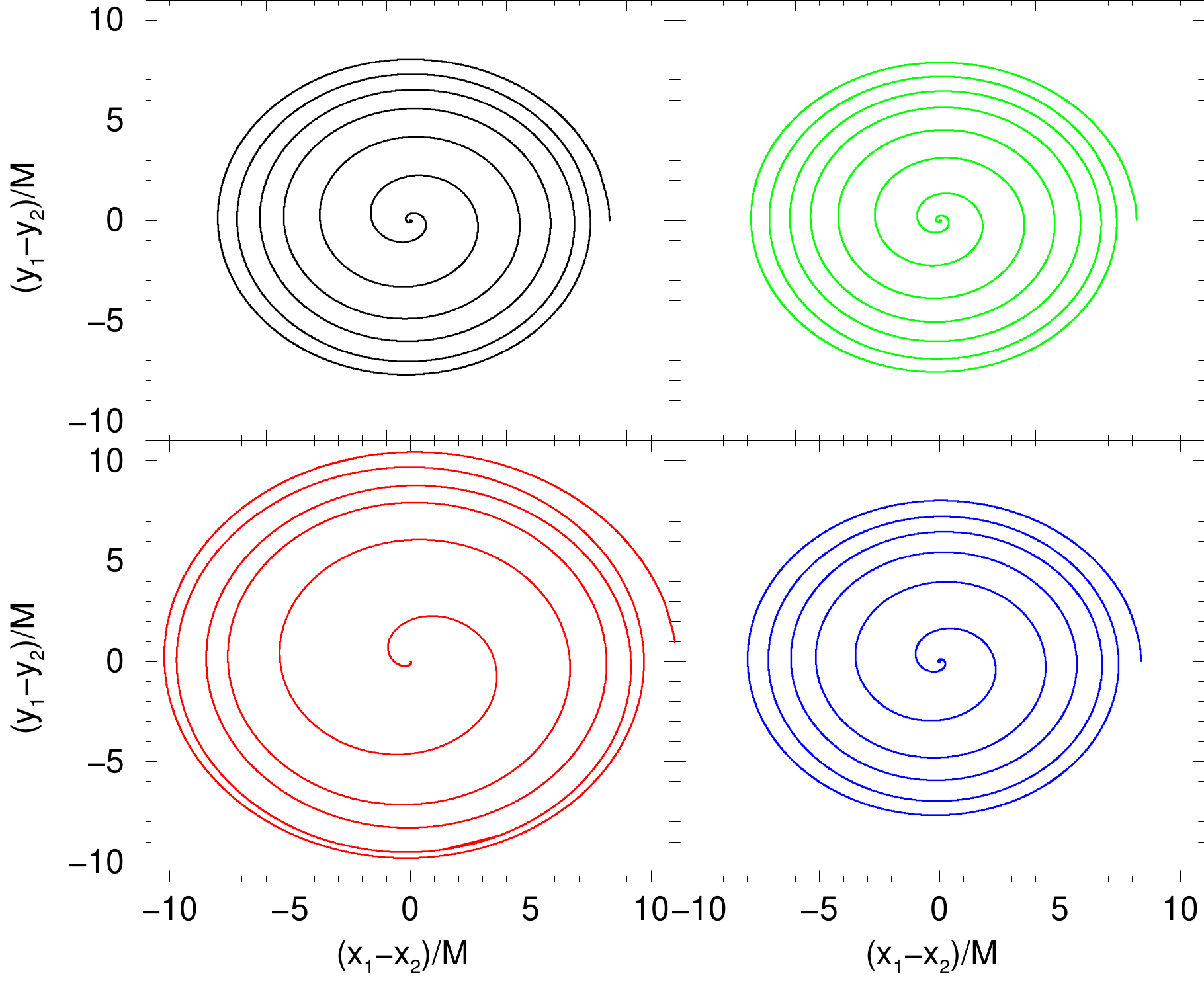}
  \caption{$xy$ plane projections of the trajectories for
various ``hangup-kick'' configurations. (Top Left) Trajectory for
the A7TH22.5PH0 configuration, (Bottom Left) trajectory for the
A7TH135PH0 configuration, (Top Right) trajectory for the
A9TH15PH0 configuration, (Bottom Right)  trajectory for the
A9TH60PH0 configuration.  The plot shows the trajectories for
configurations with the largest and smallest inclination angle for
the $\alpha=1/\sqrt{2}$ and $\alpha=0.9$ configurations. Note that the
eccentricity is larger for large $\theta$ and that the eccentricity
decreases more slowly.}
\label{fig:hangup_xytrack}
\end{figure}

\begin{figure}
  \includegraphics[width=\columnwidth]{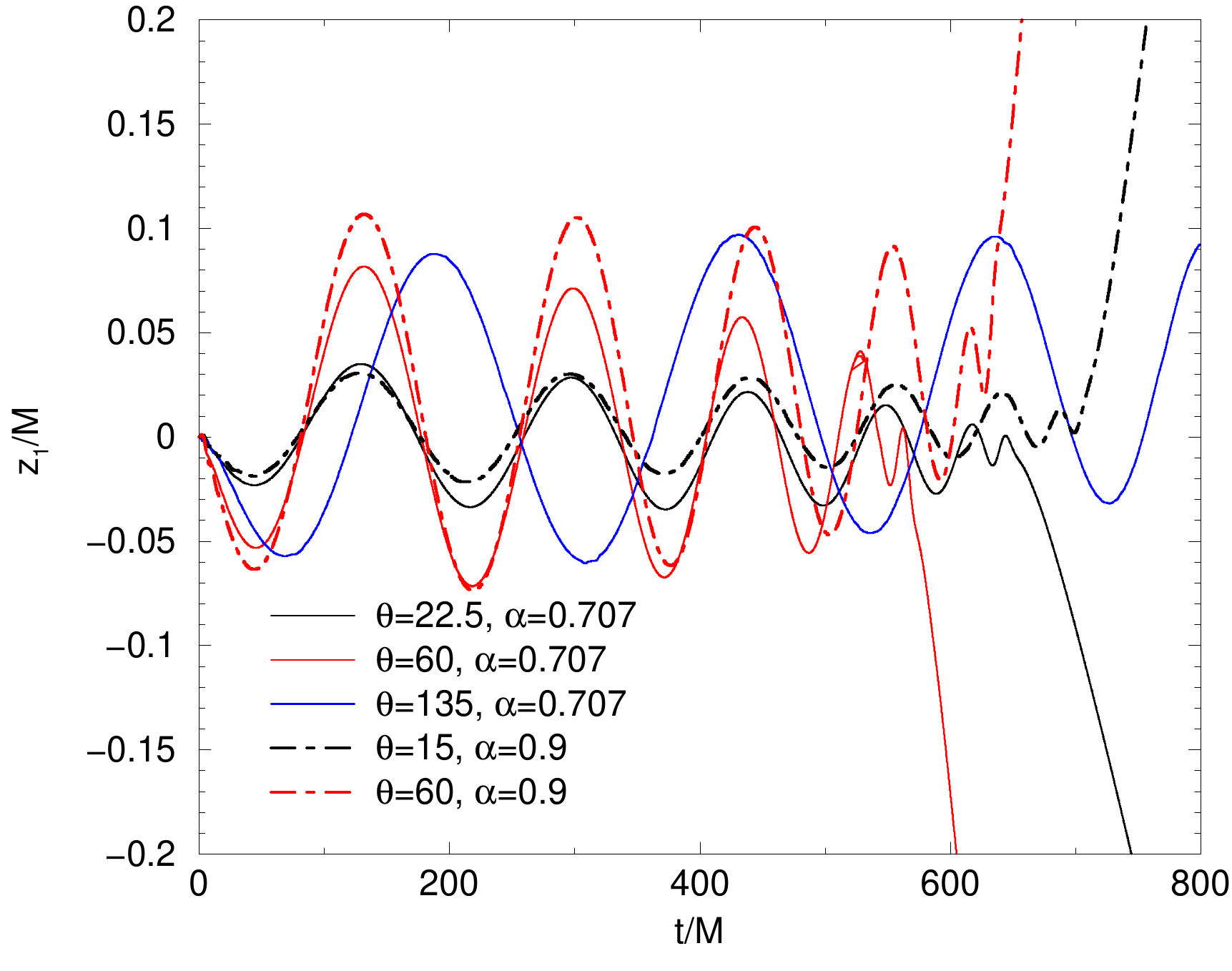}
  \caption{The elevation of the trajectory as a function of time
for the several of the ``hangup-kick'' configurations. 
Note that the ``bobbing'' amplitude
does not necessarily correspond to a large recoil. The A7TH135PH0
configuration has a factor of 2 smaller recoil than he A7TH60PH0
configuration, but a slightly larger bobbing amplitude.
}
\label{fig:hangup_ztrack}
\end{figure}

Initial data parameters for the 48 simulations are given in
Table~\ref{tab:ID}.  We denote these configurations by AsTHxxxPHyyy,
where s indicates the approximate individual  spin magnitude (7 for
$\alpha_i = 1/\sqrt{2}$ and 9 for $\alpha_i = 0.9$), xxx indicates the
angle the spin makes with respect to the $z$ axis, and yyy indicates the
angle the spin makes with respect to the $y$ axis. Here xxx and yyy are
in degrees. An eccentricity reduction procedure like those given
in~\cite{Pfeiffer:2007yz, Buonanno:2010yk} could be used to generate
configurations with very low eccentricity, but the amount of time
required to reduce the eccentricity for 48 configurations would have
been too long. We therefore chose to start from 3.5 Post-Newtonian (PN)
quasicircular orbital parameters from further separations,
such that each binary completed
5-6 orbits prior to merger, and then relied on the radiation of
angular momentum during this 5-6 orbit inspiral to reduce 
the eccentricity. The initial
separations varied between 10.16M and 8.2M, depending on the magnitude
of the hangup effect, with smaller initial separations for
configurations that exhibit larger hangups. Example trajectories for
several of these configurations are given in
Figs.~\ref{fig:hangup_xytrack}~and~\ref{fig:hangup_ztrack}.


\begin{widetext}

\begin{table}[t]
  \caption{Initial data parameters for the 48 ``hangup-kick''
configurations. In all cases the puncture masses were
chosen such that the total ADM mass of the binary was $1.0\pm10^{-6}M$. Here
the punctures are located at $\pm(x,0,0)$ with momenta $\pm(0, p,0)$ and spins
$\vec S = (\pm S_x, \pm S_y, S_z)$. The approximate initial
eccentricities,
eccentricities measured over the last orbit, and the number of orbits,
are also given.}
   \label{tab:ID}
\begin{ruledtabular}
\begin{tabular}{l|lllllll}
  CONF & $m_p/M$ & $x/M$ & $p/M$ & $S_x/M^2$ & $S_y/M^2$ & $S_z/M^2$ &
($e_{\rm init}$, $N_{\rm orbits}$, $e_{\rm merge}$) \\
\hline
A7TH22.5PH0 & 0.361001 & 4.141042 & 0.105976 & 0.000000 & 0.069426 &
0.167609 & ( 0.02, 5.5, 0.004)\\
A7TH22.5PH30 & 0.361022 & 4.141042 & 0.105976 & -0.034713 & 0.060125 & 0.167609 \\
A7TH22.5PH90 & 0.361085 & 4.141042 & 0.105976 & -0.069426 & 0.000000 & 0.167609 \\
A7TH22.5PH130 & 0.361050 & 4.141042 & 0.105976 & -0.053183 & -0.044626 & 0.167609 \\
A7TH22.5PH210 & 0.361022 & 4.141042 & 0.105976 & 0.034713 & -0.060125 & 0.167609 \\
A7TH22.5PH315 & 0.361043 & 4.141042 & 0.105976 & 0.049092 & 0.049092 & 0.167609 \\
\hline
A7TH45PH0 & 0.360775 & 4.175510 & 0.106744 & 0.000000 & 0.128222 &
0.128222 & (0.027, 5, 0.0054)\\
A7TH45PH30 & 0.360849 & 4.175510 & 0.106744 & -0.064111 & 0.111043 & 0.128222 \\
A7TH45PH90 & 0.361068 & 4.175510 & 0.106744 & -0.128222 & 0.000000 & 0.128222 \\
A7TH45PH130 & 0.360947 & 4.175510 & 0.106744 & -0.098224 & -0.082419 & 0.128222 \\
A7TH45PH210 & 0.360849 & 4.175510 & 0.106744 & 0.064111 & -0.111043 & 0.128222 \\
A7TH45PH315 & 0.360922 & 4.175510 & 0.106744 & 0.090667 & 0.090667 & 0.128222 \\
\hline
A7TH60PH0 & 0.360607 & 4.207527 & 0.107470 & 0.000000 & 0.156971 &
0.090627 & (0.022, 4.5, 0.0052)\\
A7TH60PH30 & 0.360718 & 4.207527 & 0.107470 & -0.078485 & 0.135941 & 0.090627 \\
A7TH60PH90 & 0.361052 & 4.207527 & 0.107470 & -0.156971 & 0.000000 & 0.090627 \\
A7TH60PH130 & 0.360868 & 4.207527 & 0.107470 & -0.120246 & -0.100899 & 0.090627 \\
A7TH60PH210 & 0.360718 & 4.207527 & 0.107470 & 0.078485 & -0.135941 & 0.090627 \\
A7TH60PH315 & 0.360830 & 4.207527 & 0.107470 & 0.110995 & 0.110995 & 0.090627 \\
\hline
A7TH120PH0 & 0.362448 & 5.295630 & 0.095864 & 0.000000 & 0.156161 &
-0.090160 & (0.026, 5, 0.003) \\
A7TH120PH30 & 0.362537 & 5.295630 & 0.095864 & -0.078081 & 0.135240 & -0.090160 \\
A7TH120PH90 & 0.362803 & 5.295630 & 0.095864 & -0.156161 & 0.000000 & -0.090160 \\
A7TH120PH130 & 0.362656 & 5.295630 & 0.095864 & -0.119627 & -0.100379 & -0.090160 \\
A7TH120PH210 & 0.362537 & 5.295630 & 0.095864 & 0.078081 & -0.135240 & -0.090160 \\
A7TH120PH315 & 0.362625 & 5.295630 & 0.095864 & 0.110423 & 0.110423 & -0.090160 \\
\hline
A7TH135PH0 & 0.362878 & 5.534525 & 0.093655 & 0.000000 & 0.127399 &
-0.127399 & (0.02, 5, 0.005) \\
A7TH135PH30 & 0.362934 & 5.534525 & 0.093655 & -0.063699 & 0.110331 & -0.127399 \\
A7TH135PH90 & 0.363104 & 5.534525 & 0.093655 & -0.127399 & 0.000000 & -0.127399 \\
A7TH135PH130 & 0.363011 & 5.534525 & 0.093655 & -0.097593 & -0.081890 & -0.127399 \\
A7TH135PH210 & 0.362934 & 5.534525 & 0.093655 & 0.063699 & -0.110331 & -0.127399 \\
A7TH135PH315 & 0.362991 & 5.534525 & 0.093655 & 0.090085 & 0.090085 & -0.127399 \\
\hline
A9TH15PH0 & 0.177282 & 4.094887 & 0.104887 & 0.000000 & 0.059803 &
0.223187 & (0.027, 6, 0.003)\\
A9TH15PH30 & 0.177339 & 4.094887 & 0.104887 & -0.029901 & 0.051791 & 0.223187 \\
A9TH15PH90 & 0.177509 & 4.094887 & 0.104887 & -0.059803 & 0.000000 & 0.223187 \\
A9TH15PH130 & 0.177415 & 4.094887 & 0.104887 & -0.045811 & -0.038440 & 0.223187 \\
A9TH15PH210 & 0.177339 & 4.094887 & 0.104887 & 0.029901 & -0.051791 & 0.223187 \\
A9TH15PH315 & 0.177395 & 4.094887 & 0.104887 & 0.042287 & 0.042287 & 0.223187 \\
\hline
A9TH30PH0 & 0.176649 & 4.116022 & 0.105345 & 0.000000 & 0.115496 &
0.200045 &(0.027, 5.5, 0.003)\\
A9TH30PH30 & 0.176864 & 4.116022 & 0.105345 & -0.057748 & 0.100022 & 0.200045 \\
A9TH30PH90 & 0.177505 & 4.116022 & 0.105345 & -0.115496 & 0.000000 & 0.200045 \\
A9TH30PH130 & 0.177152 & 4.116022 & 0.105345 & -0.088475 & -0.074239 & 0.200045 \\
A9TH30PH210 & 0.176864 & 4.116022 & 0.105345 & 0.057748 & -0.100022 & 0.200045 \\
A9TH30PH315 & 0.177078 & 4.116022 & 0.105345 & 0.081668 & 0.081668 & 0.200045 \\
\hline
A9TH60PH0 & 0.174838 & 4.190252 & 0.107000 & 0.000000 & 0.199841 &
0.115378 & (0.027, 5, 0.0055) \\
A9TH60PH30 & 0.175510 & 4.190252 & 0.107000 & -0.099920 & 0.173067 & 0.115378 \\
A9TH60PH90 & 0.177498 & 4.190252 & 0.107000 & -0.199841 & 0.000000 & 0.115378 \\
A9TH60PH130 & 0.176408 & 4.190252 & 0.107000 & -0.153087 & -0.128455 & 0.115378 \\
A9TH60PH210 & 0.175510 & 4.190252 & 0.107000 & 0.099920 & -0.173067 & 0.115378 \\
A9TH60PH315 & 0.176177 & 4.190252 & 0.107000 & 0.141309 & 0.141309 & 0.115378 \\
\end{tabular}
\end{ruledtabular}
\end{table}

\end{widetext}

\begin{figure}
  \includegraphics[width=\columnwidth]{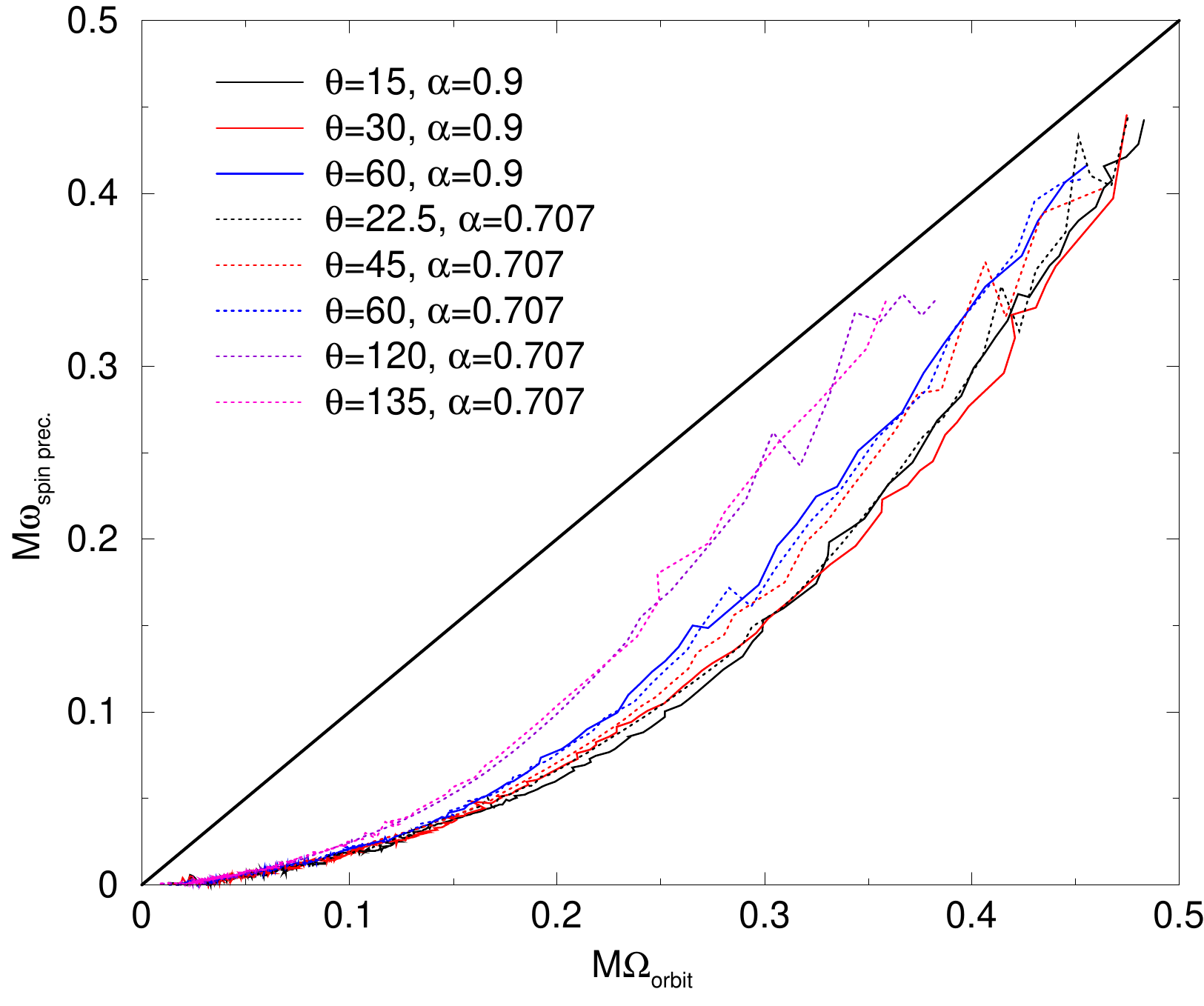}
  \caption{Spin precession frequency versus orbital frequencies for
  $\alpha=1/\sqrt{2}$ and $\alpha=0.9$ configurations. In each case,
  only the $\phi_i=0$ configuration is
shown. The trend appears to be (but see $\theta=30$ for $\alpha=0.9$)
that $\omega_{\rm spin\ prec.}$ increases as $\theta$ increases, with
very little variation with the $z$ component of the spin or even
total spin $\alpha$. Note the at late times (larger $\Omega_{\rm
orbit}$) $\omega_{\rm spin\ prec.}$ approaches $\Omega_{\rm
orbit}$, indicating that the spin processes at nearly the same rate
as the orbit during the final orbit and plunge.}
\label{fig:omega_s_v_omega_o} 
\end{figure}

The orbital motion of these BHBs has an interesting property, the
spin precession frequency and the orbital frequency agree right
near merger (in these coordinates). Initially the spin precession
frequencies are much lower than the orbital frequency, but ramp up
dramatically near merger. Figure~\ref{fig:omega_s_v_omega_o} shows
this behavior for the different $\theta$ and $\alpha$ configurations
(for clarity in the plot, we only show the $\phi_i=0$ configurations).
Interestingly, the
strongest effect seems to be due to the inclination angle $\theta$
(which is measured with respect to the orbital angular momentum axis,
i.e.\ the $z$ axis) rather than on the projected $z$ spin or total spin.
For a given $\alpha$ and $\theta$ there are variations with
$\phi_i$, but these are smaller than the variations with $\theta$.
This rapid increase of the precession frequency near merger
(as measured with the techniques of \cite{Campanelli:2006fy}),
increasing all the way up  to the orbital frequency,
is in contrast with the
much milder increase in the spin magnitude due to weak tidal
effects\cite{Campanelli:2006fg}. It also lends support to modeling
of black hole merger assuming geodetic precession \cite{Nichols:2011ih}.

In Table~\ref{tab:RAD} we compare the radiated mass and angular
momentum as calculated directly from $\psi_4$ to the corresponding
quantities derived from the remnant mass and spin. The difference
between these quantities is a better measurement of the error than
those derived from variations in the final horizon mass and spin or in
extrapolation of the waveform to infinity. That is, there are
systematic errors due to truncation error and finite extraction
radius, and the difference between these two measurements gives a lower
bound to the error.

The dimensionless spin $\alpha_{\rm merger}$, the orientation $\varphi$ of the
spin during the final orbit and plunge, as well as the remnant BH
properties, including recoil velocity,
 are given in Table~\ref{tab:REM}. Note here that the
orientation $\varphi$ is the angle that the spin of BH1 (the BH
originally located on the positive $x$ axis)
 makes with the
spin of BH1 in the corresponding AsTHxxxPH0 configuration in a rotated
frame where infall directions all coincide (see \cite{Lousto:2008dn}).
Also note that the largest measured recoil for these runs is
 $(4079.5 \pm 10.1)\ \KMS$ for the A9TH60PH30 and A9TH60PH210 configurations,
which exceeds both the largest
measured quasicircular ``superkick recoil'' of $3300\ \KMS$~\cite{Dain:2008ck} 
and even the theoretical ``superkick maximum'' recoil of $3680\pm130\
\KMS$~\cite{Lousto:2010xk}.
The values
given for the spins near merger should only be taken as an
approximation. The spin-up apparent in the A9TH60 simulations (i.e.\
the difference between $\alpha_{\rm merger}$ and the initial spin of
$\alpha=0.9$) was due
to the lower resolution used in these simulations (simulations of
A9TH45 with the same resolution as A9TH60 showed an even stronger
spin-up at late times that converged away with higher resolution).
Interestingly, these highly-spinning configurations
can exhibit both spin-up and spin-down, depending on the location of the
refinement boundaries, when not fully resolved. The other A9 runs used
higher resolution and show much
better spin conservation.

We evolve the A9TH15 and A9TH30 configurations with 10 levels of
refinement and maximum resolution of $h=M/153.6$. The width of this
level was $2\times0.35M$, while the radius of the horizons grew to
$0.24M$. Our initial explorations used grids that were smaller in
radius, but we found that using larger grids improved the
 spin conservation considerably. The
A9TH60 configuration were evolved with grids a factor of 1.2 coarser,
and consequently the spins near merger are not as accurate.  Note that these
A9TH60 runs were performed first, and based on the errors in these
simulations, we refined the grid for the other A9 runs.

\begin{widetext}

\begin{table}[t]
  \caption{A comparison of the radiated mass and angular momentum with the 
predictions based on the final remnant mass and spin and the initial
ADM mass and angular momentum for the 48 ``hangup-kick''
configurations.
}
   \label{tab:RAD}
\begin{ruledtabular}
\begin{tabular}{l|llll}
  CONF & ($M_{\rm ADM} - M_{H})/M$ & $\delta E_{\rm rad}/M$ & $(J^z_{\rm ADM} - S^z_{H})/M^2$ & $\delta J^z_{\rm rad}/M^2$  \\
\hline
A7TH22.5PH0 & $ 0.065949\pm 0.000113 $ & $ 0.063479\pm 0.000169 $ & $0.452347 \pm 0.001590$ & $ 0.440182\pm0.006183$\\                                        
A7TH22.5PH30 & $ 0.065440\pm 0.000125 $ & $ 0.063034\pm 0.000164 $ & $0.450602 \pm 0.001706$ & $ 0.438756\pm0.006006$\\                                       
A7TH22.5PH90 & $ 0.065188\pm 0.000121 $ & $ 0.062822\pm 0.000171 $ & $0.450332 \pm 0.001675$ & $ 0.439398\pm0.005161$\\                                       
A7TH22.5PH130 & $ 0.065842\pm 0.000105 $ & $ 0.063363\pm 0.000190 $ & $0.452581 \pm 0.001493$ & $ 0.440443\pm0.005901$\\                                      
A7TH22.5PH210 & $ 0.065440\pm 0.000125 $ & $ 0.063035\pm 0.000165 $ & $0.450601 \pm 0.001706$ & $ 0.438766\pm0.005996$\\                                      
A7TH22.5PH315 & $ 0.065910\pm 0.000105 $ & $ 0.063423\pm 0.000188 $ & $0.452761 \pm 0.001487$ & $ 0.440460\pm0.006077$\\                                      
\hline
A7TH45PH0 & $ 0.058764\pm 0.000006 $ & $ 0.056538\pm 0.000181 $ & $0.417697 \pm 0.000084$ & $ 0.403669\pm0.007395$\\                                          
A7TH45PH30 & $ 0.058594\pm 0.000008 $ & $ 0.056460\pm 0.000153 $ & $0.416460 \pm 0.000088$ & $ 0.403290\pm0.006741$\\                                         
A7TH45PH90 & $ 0.056266\pm 0.000010 $ & $ 0.054394\pm 0.000120 $ & $0.408936 \pm 0.000072$ & $ 0.397875\pm0.004794$\\                                         
A7TH45PH130 & $ 0.056959\pm 0.000010 $ & $ 0.054940\pm 0.000155 $ & $0.412016 \pm 0.000078$ & $ 0.399658\pm0.005982$\\                                        
A7TH45PH210 & $ 0.058594\pm 0.000008 $ & $ 0.056460\pm 0.000153 $ & $0.416459 \pm 0.000088$ & $ 0.403290\pm0.006739$\\                                        
A7TH45PH315 & $ 0.057167\pm 0.000008 $ & $ 0.055120\pm 0.000159 $ & $0.412756 \pm 0.000078$ & $ 0.400125\pm0.006224$\\                                        
\hline
A7TH60PH0 & $ 0.050887\pm 0.000004 $ & $ 0.049046\pm 0.000145 $ & $0.379033 \pm 0.000032$ & $ 0.365962\pm0.007685$\\                                          
A7TH60PH30 & $ 0.052125\pm 0.000003 $ & $ 0.050267\pm 0.000146 $ & $0.383455 \pm 0.000029$ & $ 0.371173\pm0.006948$\\                                         
A7TH60PH90 & $ 0.051234\pm 0.000005 $ & $ 0.049628\pm 0.000094 $ & $0.378829 \pm 0.000027$ & $ 0.368250\pm0.005342$\\                                         
A7TH60PH130 & $ 0.049522\pm 0.000004 $ & $ 0.047918\pm 0.000103 $ & $0.372896 \pm 0.000025$ & $ 0.361386\pm0.006580$\\                                        
A7TH60PH210 & $ 0.052124\pm 0.000003 $ & $ 0.050266\pm 0.000146 $ & $0.383456 \pm 0.000029$ & $ 0.371168\pm0.006945$\\                                        
A7TH60PH315 & $ 0.049445\pm 0.000004 $ & $ 0.047811\pm 0.000112 $ & $0.372866 \pm 0.000030$ & $ 0.360951\pm0.006697$\\                                        
\hline
A7TH120PH0 & $ 0.033193\pm 0.000002 $ & $ 0.032297\pm 0.000048 $ & $0.303865 \pm 0.000005$ & $ 0.298884\pm0.006258$\\                                         
A7TH120PH30 & $ 0.033107\pm 0.000002 $ & $ 0.032297\pm 0.000010 $ & $0.303095 \pm 0.000005$ & $ 0.295666\pm0.008881$\\                                        
A7TH120PH90 & $ 0.031686\pm 0.000002 $ & $ 0.031037\pm 0.000015 $ & $0.297018 \pm 0.000005$ & $ 0.292110\pm0.006177$\\                                        
A7TH120PH130 & $ 0.031974\pm 0.000002 $ & $ 0.031188\pm 0.000044 $ & $0.298771 \pm 0.000005$ & $ 0.297030\pm0.002584$\\                                       
A7TH120PH210 & $ 0.033107\pm 0.000002 $ & $ 0.032298\pm 0.000010 $ & $0.303097 \pm 0.000005$ & $ 0.295663\pm0.008889$\\                                       
A7TH120PH315 & $ 0.032117\pm 0.000002 $ & $ 0.031308\pm 0.000051 $ & $0.299419 \pm 0.000005$ & $ 0.297731\pm0.002485$\\                                       
\hline
A7TH135PH0 & $ 0.029675\pm 0.000004 $ & $ 0.029004\pm 0.000024 $ & $0.287481 \pm 0.000006$ & $ 0.290758\pm0.000188$\\                                         
A7TH135PH30 & $ 0.030091\pm 0.000004 $ & $ 0.029449\pm 0.000043 $ & $0.289543 \pm 0.000005$ & $ 0.293205\pm0.000038$\\                                        
A7TH135PH90 & $ 0.030000\pm 0.000004 $ & $ 0.029437\pm 0.000063 $ & $0.288696 \pm 0.000006$ & $ 0.294013\pm0.001881$\\                                        
A7TH135PH130 & $ 0.029415\pm 0.000003 $ & $ 0.028803\pm 0.000032 $ & $0.285893 \pm 0.000006$ & $ 0.290818\pm0.001668$\\                                       
A7TH135PH210 & $ 0.030085\pm 0.000004 $ & $ 0.029448\pm 0.000048 $ & $0.289513 \pm 0.000006$ & $ 0.293050\pm0.000149$\\                                       
A7TH135PH315 & $ 0.029385\pm 0.000004 $ & $ 0.028766\pm 0.000032 $ & $0.285792 \pm 0.000006$ & $ 0.290244\pm0.001187$\\                                       
\hline
A9TH15PH0 & $ 0.086926\pm 0.000422 $ & $ 0.082160\pm 0.000612 $ & $0.540966 \pm 0.004231$ & $ 0.523516\pm0.003980$\\                                          
A9TH15PH30 & $ 0.087312\pm 0.000396 $ & $ 0.082564\pm 0.000612 $ & $0.541701 \pm 0.003972$ & $ 0.524385\pm0.004228$\\                                         
A9TH15PH90 & $ 0.086773\pm 0.000341 $ & $ 0.082188\pm 0.000602 $ & $0.539095 \pm 0.003287$ & $ 0.523929\pm0.003074$\\                                         
A9TH15PH130 & $ 0.086317\pm 0.000384 $ & $ 0.081695\pm 0.000599 $ & $0.538447 \pm 0.003722$ & $ 0.522743\pm0.002983$\\                                        
A9TH15PH210 & $ 0.087315\pm 0.000394 $ & $ 0.082564\pm 0.000612 $ & $0.541724 \pm 0.003950$ & $ 0.524384\pm0.004228$\\                                        
A9TH15PH315 & $ 0.086325\pm 0.000394 $ & $ 0.081686\pm 0.000600 $ & $0.538588 \pm 0.003835$ & $ 0.522588\pm0.003187$\\                                        
\hline
A9TH30PH0 & $ 0.077615\pm 0.000339 $ & $ 0.073045\pm 0.000461 $ & $0.509243 \pm 0.005934$ & $ 0.484306\pm0.004248$\\                                          
A9TH30PH30 & $ 0.076957\pm 0.000316 $ & $ 0.072527\pm 0.000493 $ & $0.507180 \pm 0.005340$ & $ 0.483676\pm0.004906$\\                                         
A9TH30PH90 & $ 0.078900\pm 0.000166 $ & $ 0.074900\pm 0.000485 $ & $0.510059 \pm 0.002727$ & $ 0.496097\pm0.000678$\\                                         
A9TH30PH130 & $ 0.079891\pm 0.000209 $ & $ 0.075617\pm 0.000486 $ & $0.513344 \pm 0.003666$ & $ 0.496170\pm0.000939$\\
A9TH30PH210 & $ 0.076956\pm 0.000316 $ & $ 0.072556\pm 0.000469 $ & $0.507180 \pm 0.005341$ & $ 0.483678\pm0.004902$\\
A9TH30PH315 & $ 0.079810\pm 0.000224 $ & $ 0.075477\pm 0.000486 $ & $0.513380 \pm 0.003973$ & $ 0.495112\pm0.001297$\\
\hline
A9TH60PH0 & $ 0.056289\pm 0.000311 $ & $ 0.055634\pm 0.000322 $ & $0.387215 \pm 0.007605$ & $ 0.411510\pm0.009728$\\
A9TH60PH30 & $ 0.059774\pm 0.000325 $ & $ 0.058422\pm 0.000326 $ & $0.402796 \pm 0.005399$ & $ 0.422236\pm0.007268$\\
A9TH60PH90 & $ 0.054809\pm 0.000087 $ & $ 0.052769\pm 0.000184 $ & $0.394238 \pm 0.001762$ & $ 0.396247\pm0.003734$\\
A9TH60PH130 & $ 0.055249\pm 0.000168 $ & $ 0.053567\pm 0.000241 $ & $0.392036 \pm 0.003341$ & $ 0.399533\pm0.008047$\\
A9TH60PH210 & $ 0.059774\pm 0.000325 $ & $ 0.058422\pm 0.000326 $ & $0.402796 \pm 0.005399$ & $ 0.422236\pm0.007268$\\
A9TH60PH315 & $ 0.054902\pm 0.000201 $ & $ 0.053558\pm 0.000260 $ & $0.388072 \pm 0.004485$ & $ 0.400454\pm0.008515$\\

\end{tabular}
\end{ruledtabular}
\end{table}

\begin{table}[t]
  \caption{Merger and remnant BH properties of the 48 configurations.
 $S_{H}$ is the spin
angular momentum of the remnant, $M_{H}$ is the Christodoulou
mass, $V^z_{\rm recoil}$ is the recoil velocity, $\alpha_{\rm merger}$
is an approximate value of the dimensionless spin during the merger phase,
and $\varphi$ is the angle between the direction of the spin of BH1
(in the rotated frame) and the spin of BH1 in the corresponding PH0
configuration (see Section IV). }
 \label{tab:REM}
\begin{ruledtabular}
\begin{tabular}{l|lllll}
  CONF & $M_{H}/M$ & $S_{H}/M^2$ & $V^z_{\rm recoil} (\KMS)$ & $\alpha_{\rm merger}$  & $\varphi$\\
\hline
A7TH22.5PH0 & $ 0.934051\pm 0.000113 $ & $ 0.760575\pm 0.001590 $ & $-925.3 \pm 1.0$ & 0.71 & 0 \\
A7TH22.5PH30 & $ 0.934560\pm 0.000125 $ & $ 0.762320\pm 0.001706 $ & $-7.9 \pm 2.3$ & 0.71 & 31.53 \\
A7TH22.5PH90 & $ 0.934812\pm 0.000121 $ & $ 0.762590\pm 0.001675 $ & $1531.5 \pm 2.6$ & 0.71 & 91.68 \\
A7TH22.5PH130 & $ 0.934158\pm 0.000105 $ & $ 0.760341\pm 0.001493 $ & $1735.2 \pm 1.2$ & 0.71 & 131.14 \\
A7TH22.5PH210 & $ 0.934560\pm 0.000125 $ & $ 0.762321\pm 0.001706 $ & $8.0 \pm 2.3$ & 0.71 & 211.52 \\
A7TH22.5PH315 & $ 0.934089\pm 0.000105 $ & $ 0.760161\pm 0.001487 $ & $-1703.7 \pm 1.1$ & 0.71 & 315.97 \\
\hline
A7TH45PH0 & $ 0.941235\pm 0.000006 $ & $ 0.730165\pm 0.000084 $ & $2527.7 \pm 5.4$ & 0.71 & 0 \\
A7TH45PH30 & $ 0.941406\pm 0.000008 $ & $ 0.731403\pm 0.000088 $ & $1708.6 \pm 1.2$ & 0.71 & 29.33 \\
A7TH45PH90 & $ 0.943734\pm 0.000010 $ & $ 0.738926\pm 0.000072 $ & $-1199.6 \pm 4.8$ & 0.71 & 91.22 \\
A7TH45PH130 & $ 0.943041\pm 0.000010 $ & $ 0.735846\pm 0.000078 $ & $-2486.4 \pm 0.2$ & 0.71 & 131.57 \\
A7TH45PH210 & $ 0.941406\pm 0.000008 $ & $ 0.731403\pm 0.000088 $ & $-1708.2 \pm 1.3$ & 0.71 & 209.35 \\
A7TH45PH315 & $ 0.942833\pm 0.000008 $ & $ 0.735106\pm 0.000078 $ & $2569.6 \pm 0.7$ & 0.71 & 316.39 \\
\hline
A7TH60PH0 & $ 0.949113\pm 0.000004 $ & $ 0.706585\pm 0.000032 $ & $-2786.0 \pm 4.6$ & 0.71 & 0 \\
A7TH60PH30 & $ 0.947875\pm 0.000003 $ & $ 0.702163\pm 0.000029 $ & $-2886.8 \pm 2.9$ & 0.71 & 27.78 \\
A7TH60PH90 & $ 0.948766\pm 0.000005 $ & $ 0.706789\pm 0.000027 $ & $-968.8 \pm 0.4$ & 0.71 & 89.42 \\
A7TH60PH130 & $ 0.950477\pm 0.000004 $ & $ 0.712722\pm 0.000025 $ & $1167.1 \pm 4.6$ & 0.71 & 129.89 \\
A7TH60PH210 & $ 0.947876\pm 0.000003 $ & $ 0.702162\pm 0.000029 $ & $2886.7 \pm 3.0$ & 0.71 & 207.73 \\
A7TH60PH315 & $ 0.950555\pm 0.000004 $ & $ 0.712752\pm 0.000030 $ & $-1495.7 \pm 5.5$ & 0.71 & 316.71 \\
\hline
A7TH120PH0 & $ 0.966806\pm 0.000002 $ & $ 0.531135\pm 0.000005 $ & $1754.0 \pm 6.7$ & 0.71 & 0 \\
A7TH120PH30 & $ 0.966893\pm 0.000002 $ & $ 0.531905\pm 0.000005 $ & $1370.2 \pm 5.4$ & 0.71 & 29.99 \\
A7TH120PH90 & $ 0.968314\pm 0.000002 $ & $ 0.537982\pm 0.000005 $ & $-269.0 \pm 1.3$ & 0.71 & 86.11 \\
7TH120PH130 & $ 0.968026\pm 0.000002 $ & $ 0.536229\pm 0.000005 $ & $-1400.6 \pm 3.5$ & 0.71 & 129.11 \\
A7TH120PH210 & $ 0.966893\pm 0.000002 $ & $ 0.531903\pm 0.000005 $ & $-1370.7 \pm 5.4$ & 0.71 & 209.98 \\
A7TH120PH315 & $ 0.967883\pm 0.000002 $ & $ 0.535581\pm 0.000005 $ & $1495.8 \pm 4.1$ & 0.71 & 314.62 \\
\hline
A7TH135PH0 & $ 0.970326\pm 0.000004 $ & $ 0.494390\pm 0.000006 $ & $1108.0 \pm 1.1$ & 0.71 & 0 \\
A7TH135PH30 & $ 0.969909\pm 0.000004 $ & $ 0.492329\pm 0.000005 $ & $1328.0 \pm 1.9$ & 0.71 & 28.77 \\
A7TH135PH90 & $ 0.970000\pm 0.000004 $ & $ 0.493176\pm 0.000006 $ & $775.6 \pm 2.0$ & 0.71 & 90.61 \\
A7TH135PH130 & $ 0.970585\pm 0.000003 $ & $ 0.495978\pm 0.000006 $ & $-207.1 \pm 0.6$ & 0.71 & 133.14 \\
A7TH135PH210 & $ 0.969915\pm 0.000004 $ & $ 0.492359\pm 0.000006 $ & $-1326.6 \pm 1.9$ & 0.71 & 208.32 \\
A7TH135PH315 & $ 0.970615\pm 0.000004 $ & $ 0.496080\pm 0.000006 $ & $332.6 \pm 0.4$ & 0.71 & 318.33 \\
\hline
A9TH15PH0 & $ 0.913073\pm 0.000422 $ & $ 0.764409\pm 0.004231 $ & $2028.2 \pm 20.6$ & 0.90 & 0 \\
A9TH15PH30 & $ 0.912688\pm 0.000396 $ & $ 0.763674\pm 0.003972 $ & $1764.0 \pm 23.2$ & 0.90 & 30.06 \\
A9TH15PH90 & $ 0.913227\pm 0.000341 $ & $ 0.766280\pm 0.003287 $ & $22.9 \pm 8.4$ & 0.90 & 91.51 \\
A9TH15PH130 & $ 0.913683\pm 0.000384 $ & $ 0.766929\pm 0.003722 $ & $-1323.0 \pm 9.6$ & 0.90 & 130.98 \\
A9TH15PH210 & $ 0.912685\pm 0.000394 $ & $ 0.763651\pm 0.003950 $ & $-1763.9 \pm 23.2$ & 0.90 & 210.02 \\
A9TH15PH315 & $ 0.913676\pm 0.000394 $ & $ 0.766788\pm 0.003835 $ & $1455.4 \pm 11.0$ & 0.90 & 316.18 \\
\hline
A9TH30PH0 & $ 0.922385\pm 0.000339 $ & $ 0.758052\pm 0.005934 $ & $-886.8 \pm 13.6$ & 0.895 & 0 \\
A9TH30PH30 & $ 0.923043\pm 0.000316 $ & $ 0.760116\pm 0.005340 $ & $-2358.8 \pm 6.0$ & 0.895 & 28.07 \\
A9TH30PH90 & $ 0.921100\pm 0.000166 $ & $ 0.757236\pm 0.002727 $ & $-3346.5 \pm 34.1$ & 0.895 & 80.64 \\
A9TH30PH130 & $ 0.920108\pm 0.000209 $ & $ 0.753952\pm 0.003666 $ & $-2306.6 \pm 39.5$ & 0.895 & 122.75 \\
A9TH30PH210 & $ 0.923043\pm 0.000316 $ & $ 0.760115\pm 0.005341 $ & $2360.4 \pm 4.9$ & 0.895 & 208.02 \\
A9TH30PH315 & $ 0.920190\pm 0.000224 $ & $ 0.753916\pm 0.003973 $ & $2040.8 \pm 38.7$ & 0.895 & 308.95 \\
\hline
A9TH60PH0 & $ 0.943710\pm 0.000311 $ & $ 0.740252\pm 0.007605 $ & $3792.7 \pm 5.4$ & 0.92 & 0 \\
A9TH60PH30 & $ 0.940226\pm 0.000325 $ & $ 0.724671\pm 0.005399 $ & $4079.5 \pm 10.1$ & 0.92 & 36.83 \\
A9TH60PH90 & $ 0.945191\pm 0.000087 $ & $ 0.733228\pm 0.001762 $ & $-2352.1 \pm 1.6$ & 0.92 & 146.22 \\
A9TH60PH130 & $ 0.944751\pm 0.000168 $ & $ 0.735431\pm 0.003341 $ & $-3054.6 \pm 0.7$ & 0.92 & 160.22 \\
A9TH60PH210 & $ 0.940226\pm 0.000325 $ & $ 0.724671\pm 0.005399 $ & $-4079.5 \pm 10.1$ & 0.92 & 216.90 \\
A9TH60PH315 & $ 0.945098\pm 0.000201 $ & $ 0.739395\pm 0.004485 $ & $2772.5 \pm 0.8$ & 0.92 & 334.02 \\
\end{tabular}
\end{ruledtabular}
\end{table}

\end{widetext}

\section{Results and modeling of recoil velocities}\label{Sec:Results}

With the discovery of very large recoil \cite{Campanelli:2007ew}
velocities for certain configurations of merging spinning BHBs,
the need for an empirical model for the recoil velocity as a function
of the progenitor's parameters was apparent.
Our approach to provide that phenomenological formula
was based on the observation
that the recoil of spinning BHs is largely generated around the
time of merger of the two holes \cite{Lousto:2007db};
and that this nearly instantaneous burst of radiation of linear momentum 
can be modeled by a 
parametrized dependence of the leading (on spins and mass ratio) 
post-Newtonian (PN) expressions for the linear momentum 
radiated\cite{Kidder:1995zr}.

In Ref.~\cite{Lousto:2009mf} we extended our original empirical formula
for the recoil velocity imparted to the remnant of a 
BHB merger~\cite{Campanelli:2007ew, Campanelli:2007cga} to include
next-to-leading-order corrections (based on the PN work of
\cite{Racine:2008kj}), still linear in the spins
\begin{eqnarray}\label{eq:Pempirical}
\vec{V}_{\rm recoil}(q,\vec\alpha)&=&v_m\,\hat{e}_1+
v_\perp(\cos\xi\,\hat{e}_1+\sin\xi\,\hat{e}_2)+v_\|\,\hat{n}_\|,\nonumber\\
\end{eqnarray}
where
\begin{eqnarray}\label{eq:Pempiricalcont}
v_m&=&A_m\frac{\eta^2(1-q)}{(1+q)}\left[1+B_m\,\eta\right],\nonumber\\
v_\perp&=&H\frac{\eta^2}{(1+q)}\left[
(1+B_H\,\eta)\,(\alpha_2^\|-q\alpha_1^\|)\right.\nonumber\\
&&\left.+\,H_S\,\frac{(1-q)}{(1+q)^2}\,(\alpha_2^\|+q^2\alpha_1^\|)\right],\nonumber\\
v_\|&=&K\frac{\eta^2}{(1+q)}\Bigg[
(1+B_K\,\eta)
\left|\vec\alpha_2^\perp-q\vec\alpha_1^\perp\right|
\nonumber \\ && \quad \times
\cos(\phi_\Delta-\phi_1)\nonumber\\
&&+\,K_S\,\frac{(1-q)}{(1+q)^2}\,\left|\vec \alpha_2^\perp+q^2\vec \alpha_1^\perp\right|
\nonumber \\ && \quad \times
\cos(\phi_S-\phi_2)\Bigg],
\end{eqnarray}
and $\eta=q/(1+q)^2$, with $q=m_1/m_2$
the mass ratio of the smaller to larger mass hole,
$\vec{\alpha}_i=\vec{S}_{i}/m_i^2$, $m_i$ is shorthand for
$m_{H\,i}$ the mass of BH $i$, the index $\perp$ and $\|$ refer to
perpendicular and parallel to the orbital angular momentum respectively,
$\hat{e}_1,\hat{e}_2$ are
orthogonal unit vectors in the orbital plane, and $\xi$ measures the
angle between the unequal mass and spin contribution to the recoil
velocity in the orbital plane. 
 The angles $\phi_{\Delta}$ and $\phi_S$ are defined as the angle
between the in-plane component $\vec \Delta^\perp = M (\vec S_2^\perp/m_2 - \vec
S_1^\perp/m_1)$  and $\vec S^\perp=\vec S_1^\perp+\vec S_2^\perp$ respectively
and a fiducial direction at merger (see Ref.~\cite{Lousto:2008dn} for
a description of the technique). Note that $\vec \Delta =  M (\vec
S_2/m_2 - \vec S_1/m_1)$ can be expressed as $\vec \Delta =
M^2\left(\vec \alpha_2 - q\vec\alpha_1\right)/(1+q)$.
Phases $\phi_1$ and $\phi_2$ depend
on the initial separation of the holes for quasicircular orbits
(astrophysically realistic evolutions of comparable masses BHs
lead to nearly zero eccentricity mergers).

The most recent published estimates for the above parameters can be found in
\cite{Lousto:2008dn,Zlochower:2010sn} 
and references therein. The current best
estimates are: $A_m = 1.2\times 10^{4}\
\KMS$, $B_m = -0.93$, $H = (6.9\pm0.5)\times 10^{3}\ \KMS$,
$K=(5.9\pm0.1)\times 10^4\ \KMS$, and $\xi \sim 145^\circ$,
and $K_S=-4.254$. Here we set $B_H$ and $B_K$ to zero, which is
consistent with the findings in~\cite{Lousto:2009mf}, where it was found that the
uncertainties in the coefficients are of the same magnitude as the
coefficients themselves.

Although the post-Newtonian approximation fails to provide accurate 
amplitudes for each velocity component, the above parametrization and
fitting to a set of full numerical simulations has shown its predictive
power in a number of occasions; for instance by predicting the mass
ratio dependence that was later confirmed by sets of lengthy 
numerical simulations\cite{Lousto:2008dn}. 
The success of the original formula allowed the study of higher-order dependencies on the spin of the holes.
In a previous study~\cite{Lousto:2010xk},
 we found that the ``superkick'' recoil (where the two BHs have equal
mass, equal intrinsic spin magnitudes $\alpha$, and spins lying in the 
orbital plane in opposite directions)  has the
following dependence on the intrinsic spin $\alpha$ and orientation $\phi$ (the
angle between the in-plane spin vector and the infall direction
near merger),
\begin{eqnarray}
  V &=& V_1 \cos(\phi-\phi_1) + V_3 \cos(3\phi - 3\phi_3), \nonumber \\
    V_1 &=& V_{1,1} \alpha + V_{1,3}\alpha^3, \nonumber \\
    V_3 &=& V_{3,1} \alpha + V_{3,3} \alpha^3,
\label{eq:superkick}
\end{eqnarray}
where
$V_{1,3}=(-15.46\pm2.66)\ \KMS$, $V_{3,1}=(15.65\pm3.01)\ \KMS$, and
$V_{3,3}=(105.90\pm4.50)\ \KMS$, while $V_{1,1} = (3681.77\pm2.66)\ \KMS$.
From that study, it was clear that in the ``superkick'' configuration,
the dominant contribution, even at large $\alpha$, is linear
in $\alpha$ and proportional to $\cos(\phi)$. Note that because of the small contributions of $V_3$ and $V_{1,3}$, we neglect these
terms in the statistical studies below (where we take a uniform
distribution in $\phi-\phi_1$).

In Ref.~\cite{Boyle:2007ru} an alternative approach to fitting recoil
and remnant mass and spin of a merged BHB was developed. It is based
on a Taylor expansion in terms of the binary parameters and exploits
all the symmetries of the problem (note that our approach also
incorporates these symmetries because it is based on PN formulas for
the instantaneous recoil).  Using one of our previous set of six
``superkick" simulations in \cite{Campanelli:2007cga}, the authors in
\cite{Boyle:2007ru} fitted the recoil velocities to terms in
$\cos(\phi)$ and $\cos(3\phi)$ to extract the cubic (in spin)
dependence of the recoil from a single set of simulations with constant
total spin. We also modeled that cubic dependence with more
simulations including a range of spin magnitude
in~\cite{Lousto:2010xk}.  Ref.~\cite{Boyle:2007ru} also fitted the
recoil velocity as a function of the angle $\theta$ that the spins
make with the orbital angular momentum using the data from a series
of runs reported in Ref.~\cite{Herrmann:2007ex}.  However, those
results are inconclusive since they could not model the recoil as a
function of $\phi$ or model the precession of the orbital plane using
the data in Ref.~\cite{Herrmann:2007ex}.

\begin{figure}
  \includegraphics[width=1.65in]{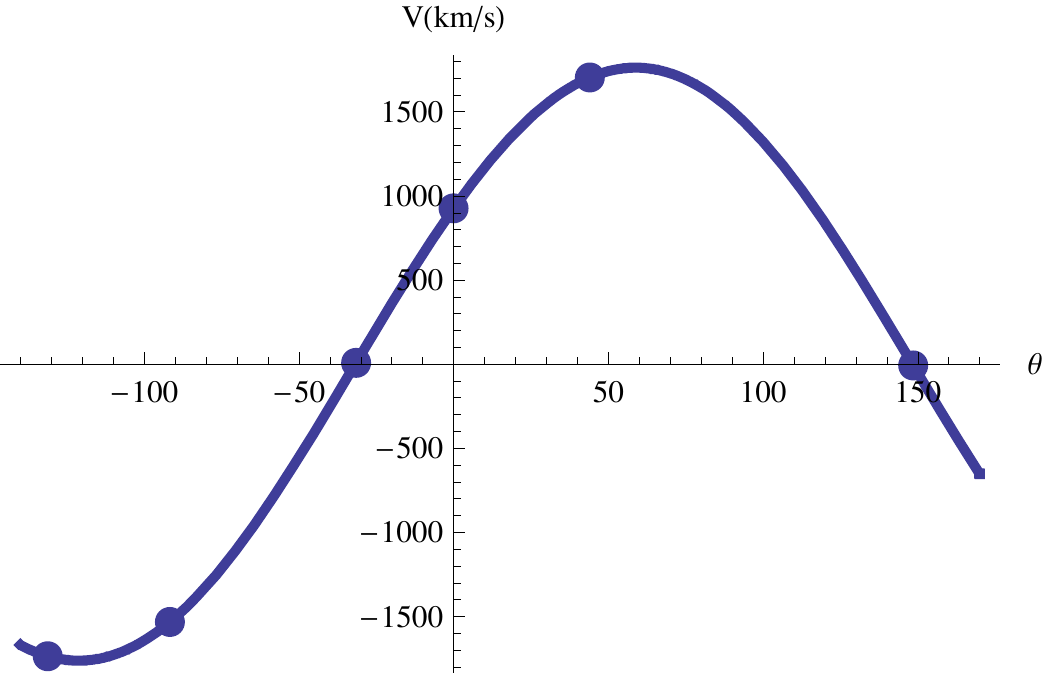}
  \includegraphics[width=1.65in]{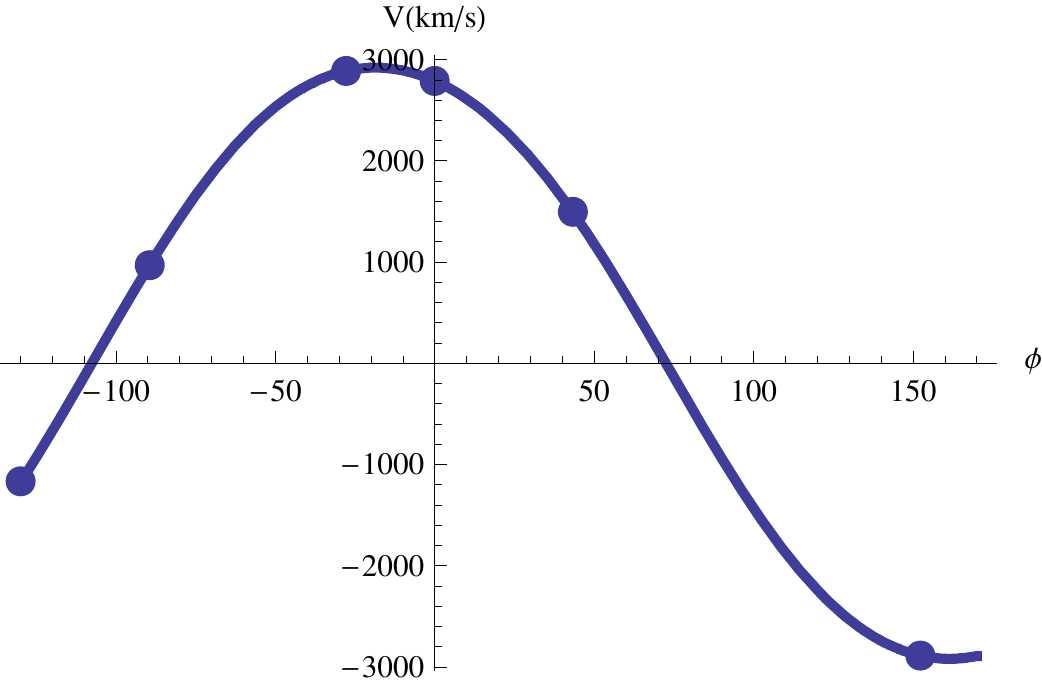}
  \caption{A fit of $V_{\rm recoil}$ versus $\varphi$ for the
A7TH22.5PHyyy (Left) and  A7TH60PHyyy (Right) configurations.}
  \label{fig:phifit7}
\end{figure}

\begin{figure}
  \includegraphics[width=1.65in]{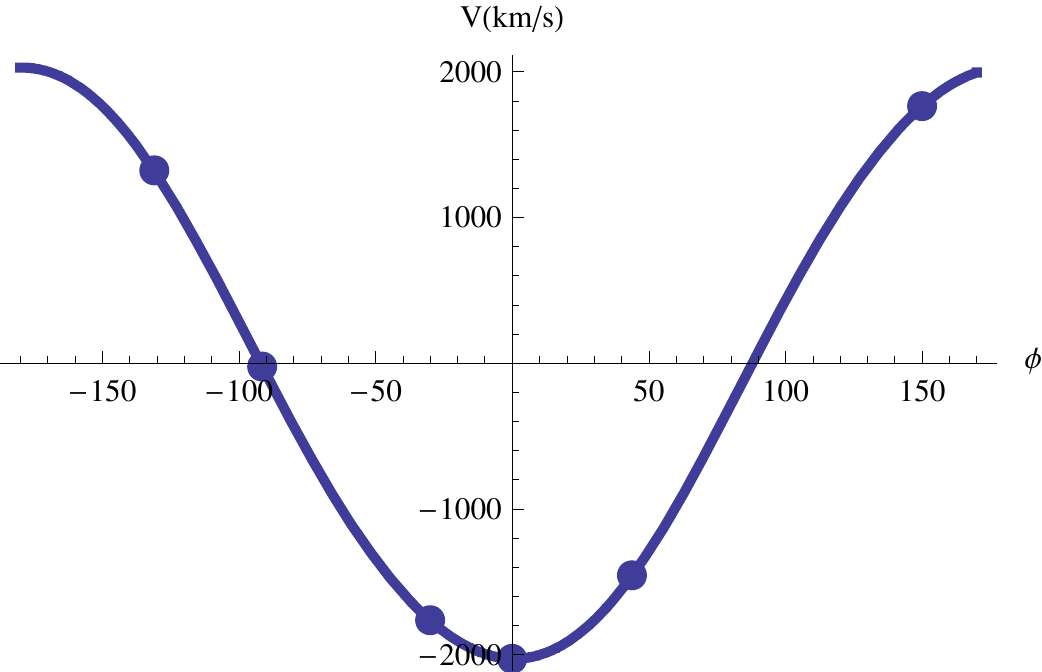}
  \includegraphics[width=1.65in]{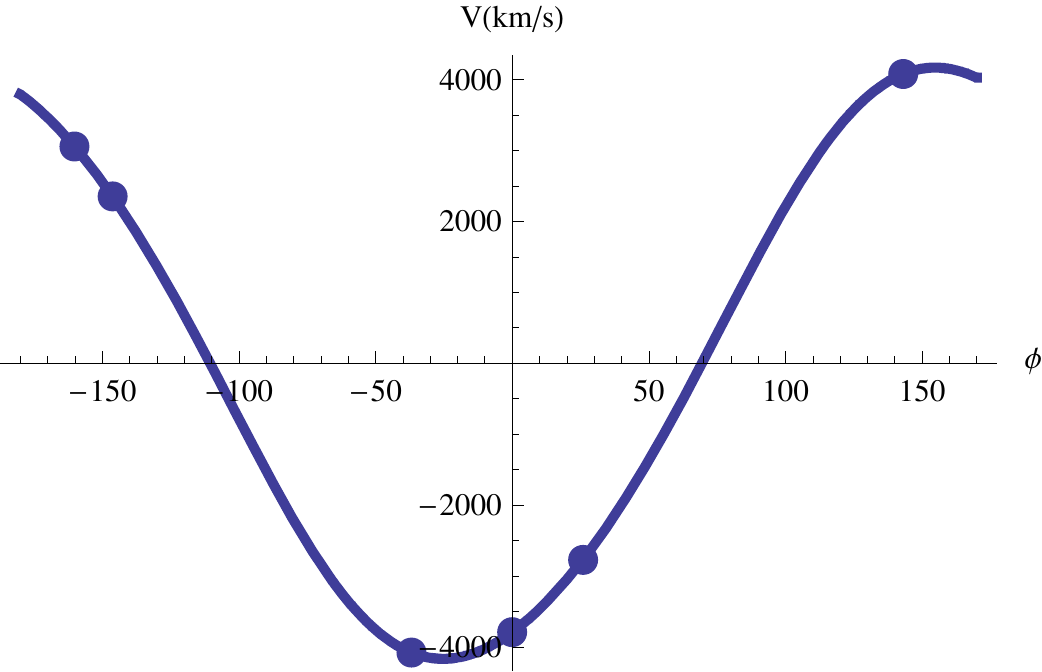}
  \caption{A fit of $V_{\rm recoil}$ versus $\varphi$ for the
A9TH15PHyyy (Left) and  A9TH60PHyyy (Right) configurations.}
  \label{fig:phifit9}
\end{figure}

\begin{figure}
  \includegraphics[width=\columnwidth]{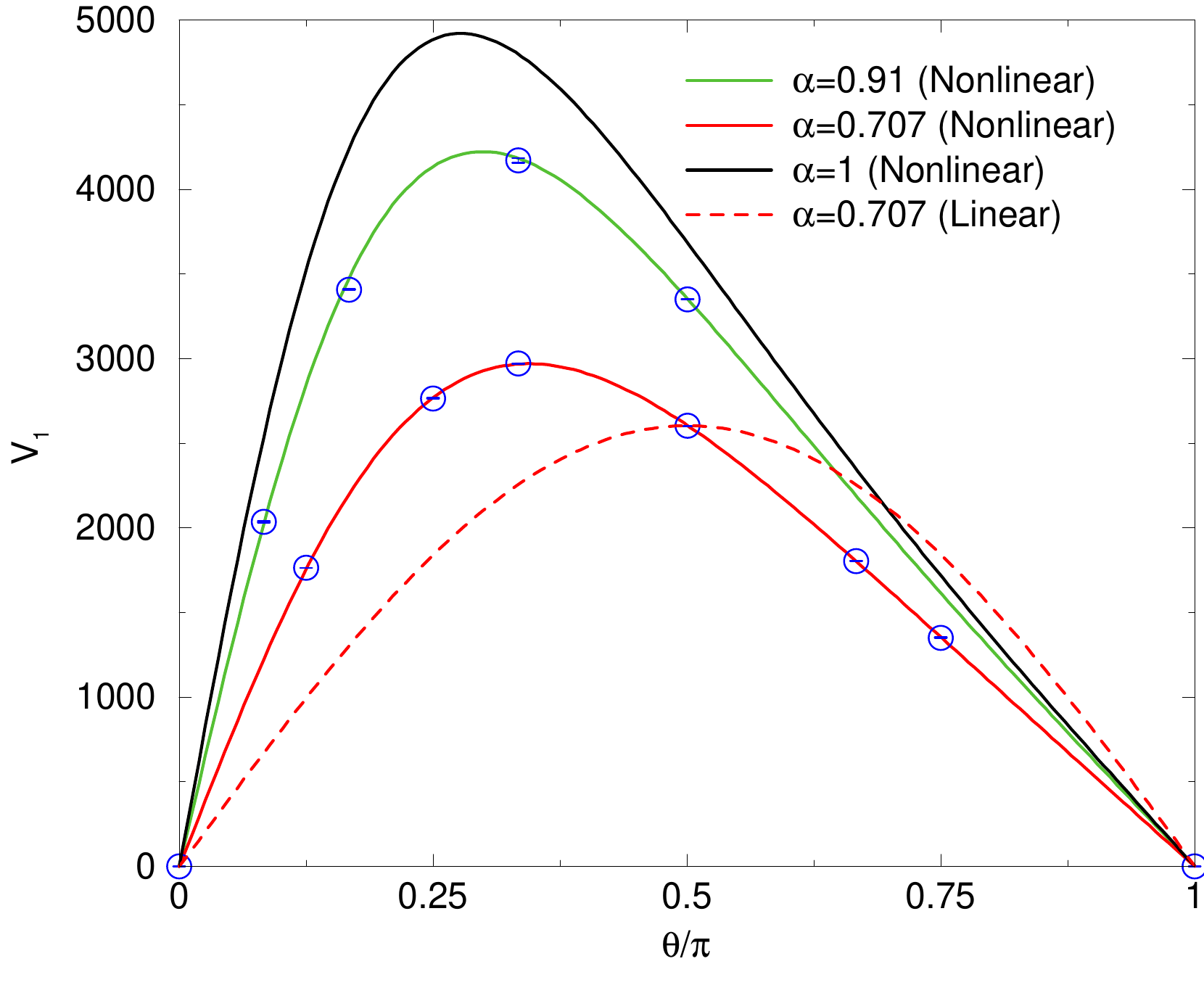}
  \caption{ fit of the recoil ($V_1$) to the form
Eq.~(\ref{eq:pade}) for the $\alpha=1/\sqrt{2}$ configurations,
and predictions (based on this fitting) for the
$\alpha=0.91$ recoils. Note how well the $\alpha=0.91$ curve matches
the four measured values.
For reference, curves corresponding to the original empirical
formula prediction (which only had terms linear
in $\Delta$) for $\alpha=1/\sqrt{2}$ and the new
formula for $\alpha=1$ are also included. Note the skew in the
velocity profile compared to the linear predictions.}
  \label{fig:kick_fit}
\end{figure}

In order to analyze the results of the present simulations, we use the
techniques developed in~\cite{Lousto:2008dn}. Briefly, we rotate each
configuration such that the trajectories near merger overlap. We then
calculate the spins in this rotated frame. The angle $\varphi$ is then
defined to be the angle between the AsTHxxPHyyy spin  of BH1 (the BH
originally located on the positive $x$ axis) and the spin of BH1 in
the corresponding AsTHxxPH0 configuration.  Note that, for a given
family of fixed spin and spin inclination angle $\theta$, the angle
$\varphi$ and $\phi$ differ by a constant, which can be absorbed in
the fitting constants $\phi_1$ and $\phi_3$.  We then fit the recoil
to the form
\begin{equation}
V_{\rm rec} = V_{1}
\cos(\varphi - \phi_1) + V_{3} \cos(3 \varphi - 3 \phi_3)
  \label{eq:phifit}
\end{equation}
 for each
set of configurations with the same spin and $\theta$, and then fit
the dominant $V_{1}$ coefficient as a function of $\varphi$.  Results
from these fits are given in Table~\ref{tab:fit} and
Figs.~\ref{fig:phifit7}-\ref{fig:kick_fit}.
Note that A9TH60 runs show the largest discrepancies in the fit,
consistent with the larger errors in these simulations due to a coarser
global resolution (see Sec.~\ref{sec:Runs} above).

Based on the ``superkick'' formula~(\ref{eq:superkick}), we expected that
the recoil would have the form
\begin{eqnarray}
  V_1 =&& \left(V_{1,1} + V_A \alpha  \cos \theta
+ V_B \alpha^2 \cos^2\theta  + V_C \alpha^3 \cos^3\theta
\right)\times\nonumber \\ 
  &&\ \alpha \sin \theta,
  \label{eq:FS}
\end{eqnarray}
where $V_1$ is the component of the recoil proportional to $\cos
\phi$,  $V_{1,1}$ arises from the ``superkick'' formula, and the remaining
terms are proportional to linear, quadratic, and higher orders in
$S_z/m^2=\alpha \cos \theta$ (the spin component in the direction of
the orbital angular momentum). Here, we do not consider terms
higher-order in the in-plane component of $\vec \Delta\propto \vec
\alpha_2 - q \vec \alpha_1$ denoted by $\Delta^\perp$
($\Delta^\perp\propto\alpha \sin \theta$ here), where $q$ is
the mass ratio, because our previous studies showed that these terms
were small at $\theta=90^\circ$. A fit to this ansatz~(\ref{eq:FS})
showed that the truncated series appears to converge very slowly
with coefficients $V_{1,1}=(3677.76\pm15.17)\ \KMS$,
$V_A=(2481.21\pm67.09)\ \KMS$, $V_B=(1792.45\pm92.98)\ \KMS$,
$V_C=(1506.52\pm286.61)\ \KMS$ that have
relatively large uncertainties.  In addition, we propose the
modification
\begin{equation}
  V_1 = \left(\frac{1 + E \alpha \cos
\theta}{1+F \alpha \cos \theta}\right)\,\, D \alpha \sin \theta
\label{eq:pade}
\end{equation}
which can be thought of as a resummation of Eq.~(\ref{eq:FS}) with an
additional term $E \alpha \cos \theta$,
and fit to $D$, $E$, $F$ (where we used the prediction
of~\cite{Lousto:2010xk} to model the  $V_1$ for $\theta=90^\circ$)
and find $D=(3684.73\pm5.67)\ \KMS$, $E=0.0705\pm0.0127$, and
$F=-0.6238\pm0.0098$. Note that $E$ is approximately $1/10$ of $F$,
indicating that coefficients in this series get progressively smaller
faster than in Eq.~(\ref{eq:FS}). Interestingly, a fit to just
$$V_1 = \left(\frac{1 }{1+F \alpha \cos \theta}\right)\,\, D \alpha
\sin \theta$$ failed to produce sensible results (a badly conditioned
matrix was encountered).
The two
formulas (\ref{eq:FS}) and (\ref{eq:pade}) give very similar results
for a broad range of $\alpha$ (see Tables~\ref{tab:a7fitcomp},
 \ref{tab:a9fitcomp}, and \ref{tab:maxtheta}).
 We then use Eq.~(\ref{eq:pade}) to predict the
recoil for higher spin $\alpha=0.9$ and test this formula for three angles
$\theta=90^\circ$, $\theta=60^\circ$, and $\theta=15^\circ$, with very good
agreement (see Fig.~\ref{fig:kick_fit}). In actuality,
both Eq.~(\ref{eq:pade}) and Eq.~(\ref{eq:FS}) provide
accurate predictions for our measured recoils at $\alpha=0.9$.
We show the errors in the predictions for the $\alpha=0.9$
configurations in Table~\ref{tab:a9fitcomp}.

We also tried fits to
\begin{equation}
  V_1 = \left(\frac{1 + E \alpha \cos\theta 
+ G \alpha^2 \cos^2 \theta}{1+F \alpha \cos \theta}\right)\,\, D
\alpha \sin \theta,
\label{eq:pade2}
\end{equation}
but found that the coefficients were not well determined. In this
case, we found $D = 3686.34\pm8.87$, $E=0.055\pm0.056$,
$F=-0.638\pm0.051$, $G=-0.014\pm0.050$. The errors in both $E$ and $G$
in this case are larger than the values of the coefficients
themselves. We therefore do not use Eq.~(\ref{eq:pade2}) in the
analysis below. Similarly large uncertainties are encountered if the
quadratic $F$ correction is in the denominator of Eq.~(\ref{eq:pade2})
rather than the numerator.

\begin{table}[t]
  \caption{Fits of recoil velocities as a function of $\varphi$ 
for each family of
configurations with fixed $\alpha$ and $\theta$ to
the form Eq.~(\ref{eq:superkick}).}
   \label{tab:fit}
\begin{ruledtabular}
\begin{tabular}{l|llll}
CONF & $V_{1} (\KMS)$ & $V_{3} (\KMS)$ & $\phi_1$ & $\phi_3$ \\
\hline
A7TH22.5 & $1764\pm 1$ & $4.6\pm 0.1$ & $58.36\pm0.01$ & $280\pm1$ \\
A7TH45 & $2766\pm 1$ & $41.6\pm 0.9$ & $203.55\pm0.02$ & $83 \pm 2$\\
A7TH60 & $2972 \pm 3$ & $54\pm3$ & $342.25\pm0.07$ & $141\pm4$\\
A7TH120 & $1806\pm1$ & $27.4\pm0.1$ & $191.78\pm0.01$ & $62\pm1$\\
A7TH135 & $1352\pm1$ & $16.6\pm 0.7$ & $145.21\pm0.04$ & $277 \pm 4$\\
A9TH15 & $2038\pm2$ & $27\pm2$ & $178.55\pm0.09$ & $291 \pm 7$ \\
A9TH30 & $3408\pm2$ & $42\pm2$ & $284.91 \pm 0.03$ & $285 \pm3$ \\
A9TH60 & $4171\pm14$ & $87\pm10$ & $158.3\pm0.7$ & $17\pm17$ \\
\end{tabular}
\end{ruledtabular}
\end{table}

\begin{table}[t]
  \caption{Predictions for $\alpha=1/\sqrt{2}$ simulations based on Eq.~(\ref{eq:pade}) (denoted by pade) 
and Eq.~(\ref{eq:FS}) (denoted by FS), as well as the measured $V_1$.
Note that the $\theta=90^\circ $ measured value comes from Ref.~\cite{Lousto:2010xk}. Velocities are in units of $\KMS$.}
   \label{tab:a7fitcomp}
\begin{ruledtabular}
\begin{tabular}{l|lll}
CONF & $V_{1,{\rm pade}}$ & $V_{1, {\rm FS}}$ & $V_{1, {\rm meas}}$ \\
\hline
TH22.5 & 1760.46 & 1754.47 & 1764 \\
TH45 &  2771.92 & 2777.4 & 2766 \\
TH60 &  2967.09 & 2967.33 & 2972 \\
TH90* &  2605.5 & 2600.57 & 2603.4 \\
TH120 & 1802.6 & 1811.4 & 1806 \\
TH135 &  1354.82 & 1348.48 & 1352\\
\end{tabular}
\end{ruledtabular}
\end{table}

\begin{table}[t]
  \caption{
Predictions for $\alpha=0.9$ simulations based on Eq.~(\ref{eq:pade}) (denoted by pade) 
and Eq.~(\ref{eq:FS}) (denoted by FS), as well as the measured $V_1$.
Note that the $\theta=90^\circ $ measured value comes from Ref.~\cite{Lousto:2010xk}. When applying Eqs.~(\ref{eq:pade}) and (\ref{eq:FS})  we use
$\alpha_{\rm merger}$.  The relative error quoted here is the relative error in the prediction based on Eq.~(\ref{eq:pade}). Velocities are in units of
\ $\KMS$.
}
   \label{tab:a9fitcomp}
\begin{ruledtabular}
\begin{tabular}{l|llll}
CONF & $V_{1,{\rm pade}}$ & $V_{1,{\rm FS}}$ & $V_{1,{\rm
meas}}$ & Rel. Error \\
\hline
TH15 & 1990 & 1905.23 & 2038 & -2.4\% \\
TH30 & 3367 & 3302.23 & 3408 & -1.2\%\\
TH60 & 4251 & 4258.62 & 4171 & 1.9\% \\
TH90* & 3353.1 & 3346.76 & 3350.41 & 0.1\%  \\
\end{tabular}
\end{ruledtabular}
\end{table}
\begin{table}[t]
  \caption{
Angle $\theta$ that maximizes the recoil and the maximum recoil
 as a function of $\alpha$ for
Eq.~(\ref{eq:pade}) and Eq.~(\ref{eq:FS}). Velocities are in units of\ $\KMS$
while angles are measured in degrees.
}
   \label{tab:maxtheta}
\begin{ruledtabular}
\begin{tabular}{l|llll}
$\alpha$ & $\theta_{\rm pade}$ &  $V_{\rm pade}$ ($\KMS$) &
$\theta_{\rm FS}$ & $V_{\rm FS}$ ($\KMS$) \\
\hline
0.1 & $86^\circ$ & 369 & $86^\circ$ & 369 \\
0.5 & $70^\circ$ & 1961 & $70^\circ$ & 1955 \\
0.707 & $62^\circ$ & 2969 & $61^\circ$ & 2968 \\
0.91 & $54^\circ$ & 4225  & $54^\circ$ & 4232 \\
1 & $50^\circ$ & 4926 & $51^\circ$ & 4915 \\
\end{tabular}
\end{ruledtabular}
\end{table}

\section{Black Hole spin evolution in gas rich galaxy mergers}\label{Sec:Accretion}

Full numerical simulations of BHBs typically start when the BHs
are at distances of the order of $10M$ from each other. There are
good reasons for this. Numerical runs are still extremely expensive, they
need to run  on hundreds of nodes for weeks at a time to obtain
accurate computations of the gravitational radiation. Those few runs allow
us to infer generic behaviors of the remnants, e.g.\ the modeling of
the recoils by the phenomenological Eq.~(\ref{eq:Pempirical}). 
While these initial separations are extremely close by astrophysical 
standards,  most of the nonlinear general relativistic effects take 
place at these and closer separations.
However, if one wants to study 
statistical distributions of recoils by astrophysical seeds one would 
like to input realistic spin and mass ratio distributions for the merging 
BHB. In a first study of such systems we assume an isotropic
distribution of the spin direction of the BHs \cite{Lousto:2009ka}. 
This
could represent ``dry'' binary mergers. We point out below the
relevance of pre-merger accretion to partially align the BH 
spins with the orbital 
angular momentum. We then perform a preliminary study 
of an extended recoil formula 
(\ref{eq:newPempirical})
to see the differences between the predictions of the
recoil formula based on only linear
terms in the spins and the new updated formula \cite{Lousto:2011kp}.

The spins of massive BHs binding in binaries as a result of galaxy
mergers can be deeply affected by gas accretion during the last stages
of their orbital decay (for separations $\lsim 100$ pc). This is due
to the gas overdensities that the galaxy mergers are expected to
convey into the nuclei of the galaxy remnants. Such dense gas
structures have a disk like morphology (``circumnuclear disks"),
reminiscent of the initial net angular momentum of the inflowing
material, as observed in high resolution simulations
\citep[e.g.][]{mayer07,hopkins10} as well as in real merging systems
\citep[e.g.][]{downes98,daviesA04,daviesB04}.

Bogdanovic et al~\cite{Bogdanovic:2007hp} proposed a physical process
that could {\it align} the BH spins with the angular momentum of the
nuclear disk in which the binary orbit is embedded, thus leading to
slow recoils for the BH remnant.  The evolution of the spin directions
is due to the torques exerted by the gas accreting onto the BHs. Since
this process happens on a timescale shorter than the orbital decay of
the BHs in the remnant nucleus, the spins tend to align before the two
BHs bind in a binary. As a consequence, the evolution of the spin of
each BH in this earlier phase can be studied independently, neglecting
the presence of the second BH.

The evolution of spin direction and magnitude is governed on small
scales (milli-pc, much smaller than the circumnuclear disk within
which BHs, and their accretion disks, are embedded).  As shown by
\cite{bardeen75}, if the orbital angular momentum of an accretion
disk around the BH is misaligned with respect to the BH spin, the
coupled action of viscosity and relativistic Lense-Thirring precession
(`inertial frame dragging') causes important changes in the structure
of an accretion disk, warping the disk. The equilibrium profile of the
warped disk can be computed by solving the equation:
\begin{eqnarray} \label{eqn:angular momentum}
\frac{1}{R}\frac{\partial}{\partial R}(R \vec{L} v_{\rm R})=
\frac{1}{R}\frac{\partial}{\partial R}\left(\nu_1 \Sigma R^3 \frac{d\Omega}{dR}~ \hat{l} \right)+ \nonumber \\
+\frac{1}{R}\frac{\partial}{\partial R}\left(\frac{1}{2}\nu_2 R L \frac{\partial \hat{l}}{\partial R} \right) 
+ \frac{2G}{c^2} \frac{\vec{S}_{\rm BH} \times \vec{L}} {R^3} 
\end{eqnarray}
\citep[see][]{Perego:2009cw}, where $R$ is the distance from the BH,
$v_R$ is the radial drift velocity, $\Sigma$ is the surface density,
$\Omega$ is the Keplerian angular velocity of the gas, and $\nu_1$
($\nu_2$) is the radial (vertical) viscosity.  $\vec{L}$ is the local
angular momentum surface density of the disk, defined by its modulus
$L$ and the unit vector $\hat{l}$ that defines its direction. The disk
profile that is a solution of Eq.~(\ref{eqn:angular momentum}) is
composed of three regions, whose relative importance depends on the
values of the specific disk parameters. In the outermost region the
angular momentum of the gas is unperturbed by any relativistic effect,
and therefore the direction of the disk's angular momentum is
independent of the BH spin.  In the inner region, the fluid is forced
to rotate in the equatorial plane of the rotating BH, on either
prograde or retrograde orbits. Therefore in this inner region the disk
is either completely aligned or completely antialigned with respect to
the BH spin. Finally, between the inner and outer regions,
characterized by different directions of their angular momenta, there
exists a transition region, centered at $\sim 100-1000$ gravitational
radii, where the disk is warped connecting the inner and outer parts
of the disk, misaligned one with respect to the other.

The spin of a BH embedded in a warped disk evolves under the influence
of the disk itself.  The BH spin evolution is described by the
equation~\citep{Perego:2009cw}:
\begin{equation} \label{eqn:jbh precession-disk}
\frac{d\vec{S}_{\rm BH}}{dt} =\dot{M}\Lambda(R_{\rm ISO})\hat
     {l}(R_{\rm ISO}) + \frac{4\pi G}{c^2}\int_{\rm
       disk}\frac{\vec{L} \times \vec{S}_{\rm BH}}{R^2}dR.
\end{equation}
The first term in Eq.~(\ref{eqn:jbh precession-disk}) accounts for the
angular momentum deposited onto the BH by the accreted gas at the
innermost stable orbit (ISO), where $\Lambda(R_{\rm ISO})$ denotes the
angular momentum per unit mass \citep[Eq. 12.7.18 in][]{Shapiro83}
evaluated at $R_{\rm ISO}$ and $\hat {l}(R_{\rm ISO})$ the local disk
angular momentum direction, which is parallel to $\vec{S}_{\rm BH}$ as
discussed above.  The second term describes the interaction of the BH
spin with the warped disk. It is responsible for the evolution of the
BH spin direction, and tends to align the direction of the BH spin
with the angular momentum of the outer regions of the accretion disk.

The efficiency of the alignment depends on the dynamics of the
inflowing material that fuels the small scale accretion disks.  It is
particularly relevant to determine whether the accretion flow
maintains a nearly constant direction of the angular momentum over the
growth episode (i.e. the accretion is ``coherent''), or not. Only a
substantial amount of gas ($1 \sim 10\%$ of the BH mass) accreting
from the same plane (for both the BHs) can significantly align the two
spins. In order to constrain the degree of coherency of the gas
accreting onto the BHs, and to predict the spin configurations in BH
binaries, Dotti et al~\cite{Dotti:2009vz} performed numerical
simulations of BH pairs in large scale nuclear disks with the
N--Body/SPH code {\small GADGET} \citep{GADGET}, upgraded to include
the accretion physics.  Here we give a short summary of the initial
conditions for the different runs. For a more detailed discussion, we
defer the reader to Refs.~\cite{dotti09,Dotti:2009vz}.

The two BHs are placed in the plane of a massive circumnuclear gaseous
disk, embedded in a larger stellar spheroid.  The disk is modeled with
$\approx 2 \times 10^6$ gas particles, has a total mass
$M_{\rm{Disk}}=10^8 \msun$, and follows a Mestel surface density
profile $\Sigma(R) \propto R^{-1},$ where $R$ is the radial distance
projected into the disk plane.  Dotti et al. truncated the disk at an
outer radius of 100 pc.  The massive disk is rotationally supported in
$R$ and has a vertical thickness of 8 pc.  Gas is evolved assuming a
polytropic equation of state with index $\gamma=5/3$ or $\gamma=7/5$.
In the former case, the disk is termed ``hot'' as the temperature is
proportional to a higher power of density than in the latter class of
models (``cold'' cases).  The cold case has been shown to provide a
good approximation to a gas of solar metallicity heated by a starburst
\citep{spaans00,klessen07}. The hot case instead corresponds to an
adiabatic monatomic gas, as if radiative cooling were completely
suppressed during the merger, for example as a result of radiative
heating after gas accretion onto the BHs \citep{mayer07}. The
spheroidal component (bulge) is modeled with $10^5$ collisionless
particles, initially distributed as a Plummer sphere with a total mass
$\mbulge(=6.98 \times M_{\rm{Disk}})$.  The mass of the bulge within
$100$ pc is five times the mass of the disk, as suggested by
\cite{downes98}.  The BHs are equal in mass ($m_{\rm BH}=4\times
10^6\,\msun$), and their initial separation is 50 pc. A BH is placed
at rest at the center of the circumnuclear disk, while the other is
moving on an initially eccentric ($e_0\simeq 0.7$) counter-rotating
(retrograde BH) or corotating (prograde BH) orbit with respect to the
circumnuclear disk.  Given the large masses of the disk and the bulge,
the dynamics of the moving BH (secondary) is unaffected by the
presence of the primary until the BHs form a gravitationally bound
system. Furthermore, the gravitational interaction between the
orbiting BH and the rotating gas forces the BH to corotate on almost
circular orbits with the disk \citep{dotti06,dotti07}, before the BHs
bind in a binary. As a consequence, the initial orbital configurations
of the BHs do not influence the final degree of alignment, that, as
will be discussed in the following, depends only on the
thermodynamical state of the disk.

To follow the evolution of the BH spins, it is necessary to track the dynamics
of the gas accreting onto the two central objects. In the simulations
discussed in \cite{Dotti:2009vz} a gas particle can be accreted by a BH if
the following two criteria are fulfilled:\\
 \noindent
$\bullet$ the sum of the kinetic and internal energy of the gas
particle is lower than $b$-times the modulus of its
gravitational energy (all the energies are computed with respect to
each BH);\\ 
\noindent $\bullet$ the total mass accreted per unit time onto the BH
every timestep is lower than the accretion rate corresponding to the
Eddington luminosity computed assuming a radiative
efficiency of 10\%.\\ The parameter $b$ is a constant
that defines the degree to which a particle is bound to the BH in
order to be accreted. Dotti et al~\cite{Dotti:2009vz} set $b=0.3$. Note that due to the
nature of the above criteria, the gas particles can accrete onto the
BHs only if the time-varying Bondi-Hoyle-Lyttleton radius is resolved
in the simulations.
Such a small radius can be resolved only by performing very high
resolution simulations. The gravitational softening parameter of the BHs is 0.1
pc. The gravitational softening of the gas particles is set to the
same value, in order to prevent numerical errors. This is also the
spatial resolution of the hydrodynamical force in the highest density
regions\footnote{The code computes the density of each SPH particle
  averaging over $N_{\rm neigh}=32$ neighbors.}.

The simulations discussed in Dotti et al~\cite{Dotti:2009vz} cannot follow the
dynamics of the accreting gas on unresolved scales.  Dotti and
collaborators assume that, below the spatial resolution of the runs,
gas settles on standard geometrically thin/optically thick $\alpha$-disks
\citep{shakura73}.
The properties of those two unresolved disks (one surrounding each of the two BHs of the binary) 
embedded in the larger scale circumnuclear disk, are determined by the properties of the accreting
material.  Each gas particle accreted by the BH carries with it mass and angular
momentum. These are data Dotti et al. used to model the unresolved accretion discs
around the two BHs, becoming the outer boundary conditions for Eq.~(\ref{eqn:angular
  momentum})~and~(\ref{eqn:jbh precession-disk}). 
Specifically, Dotti and collaborators define $\hat{l}_{\rm
  edge}$ as the unit vector defining the direction of the angular
momentum of the accretion disk in its outermost region, i.e. where it
is unaffected by any general relativistic effect.  
In a warped  $\alpha$-disk the two viscosities (radial, $\nu_1$ and vertical, $\nu_2$) can be described in
terms of two different dimensionless viscosity parameters, $\alpha_1$
and $\alpha_2$, through the relations $\nu_{1,2}=\alpha_{1,2}Hc_{\rm
  s}$, where $H$ is the disk vertical scale height and $c_{\rm s}$ is
the sound speed of the gas in the accretion disk. Additionally, 
$\alpha_2=f_2 / (2 \alpha_1)$, with $\alpha_1 = 0.1$ and $f_2 = 0.6$
\citep{lodato07}.  Further details on the procedure used to evolve the
BH spins can be found in \cite{Perego:2009cw}.

\begin{figure}
 \includegraphics[width=\columnwidth]{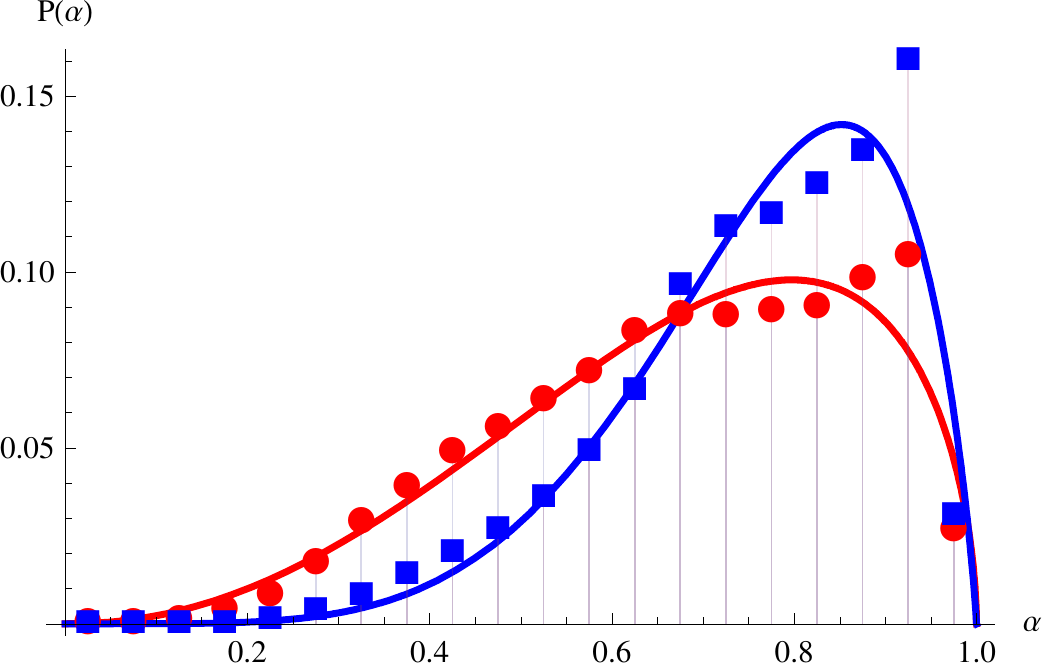}
\caption{The probability that the dimensionless spin of a BH in a merging binary has
a given magnitude $\alpha$
for BHs in cold disks (squares) and in hot disks (circles). The
fits to Beta functions are reasonably good.}
\label{fig:spin_dist_fits}
\end{figure}

\begin{figure}
 \includegraphics[width=\columnwidth]{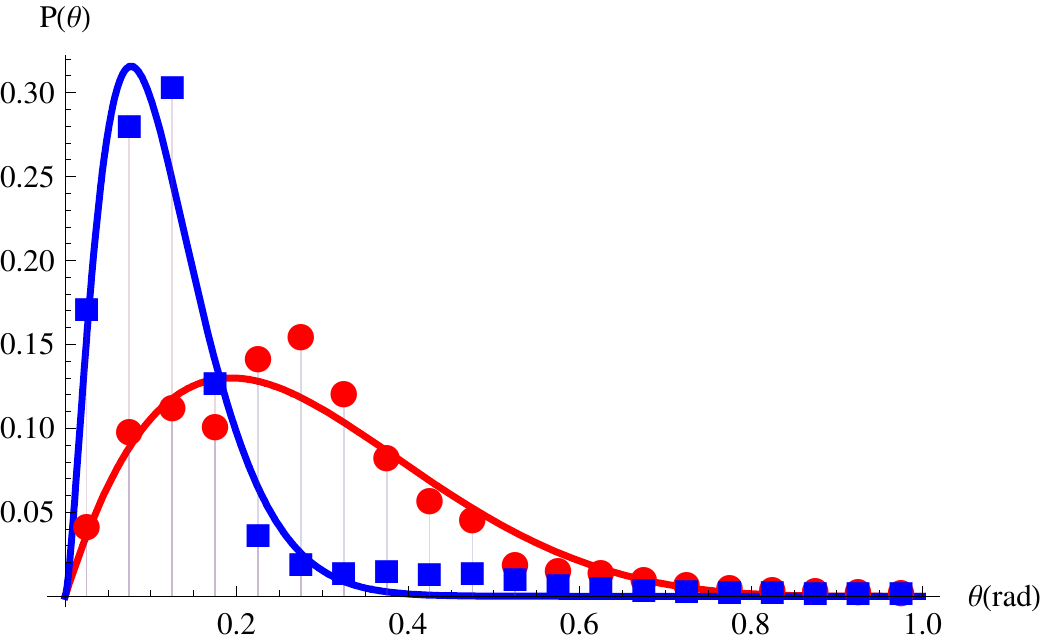}
\caption{The probability that the spin of a BH in a merging binary is
at an inclination angle $\theta$ with respect to the orbital angular
momentum for cold  (squares) and hot  (circles) circumnuclear disks. The
fits to Beta functions are reasonably good, but miss the small tail in
the cold distribution.}
\label{fig:theta_dist_fits}
\end{figure}

The resulting distributions of spin magnitudes and inclinations with
respect to the angular momentum of the newly formed binary are shown
in Figs.~\ref{fig:spin_dist_fits}~and~\ref{fig:theta_dist_fits},
 respectively. Red circles
refer to BHs embedded in hot disks, blue squares to BHs in cold
disks. In both the cases, the well defined angular momentum of the
large scale nuclear disk results in coherent accretion flows onto the
two BHs, and, as a consequence, in high spins strongly aligned with
the angular momentum of the BH binary. Note that, in absence of any
alignment, the distributions in Fig.~\ref{fig:theta_dist_fits} should be
$\propto$ sin$(\theta)$ in the whole interval $[0,\pi]$.  As discussed
in \cite{Dotti:2009vz}, a ``hotter'' disk, with a stiffer equation of
state, is more pressure supported in the center, and, as a
consequence, the degree of alignment is lower. Because of this
additional support, the accretion rates onto the BHs in the hot runs
are lower, corresponding to spin distributions less skewed towards
high spin values.

In order to perform the statistical studies below, we fit the spin
magnitude and inclination
angle distributions above to Beta distributions, which have the form
$P(x) \propto (1-x)^{(b-1)}  x^{(a-1)}$.
Fits to the dimensionless spin magnitudes for BHs in hot and cold
gaseous environments give
$a=3.212\pm0.258$, $b=1.563\pm0.093$, and $a=5.935\pm0.642$,
$b=1.856\pm0.146$, respectively. A comparison of these fits with the
measured probabilities of a BH having a given spin magnitude is given
in Fig.~\ref{fig:spin_dist_fits}.
Fits to the inclination angle for the hot and cold cases angular
distributions give $a=2.018\pm0.181$, $b=5.244\pm0.604$, and
$a=2.544\pm 0.198$, $b=19.527\pm2.075$, respectively. Note that these
distributions $P(\theta)$ are for $\theta$ in radians. The Beta
distribution is not defined for $\theta>1$, but the data are
consistent with near zero probabilities for angles larger than 1
radian. 
A comparison of these fits with the
measured probabilities of a BH having a given spin direction is given
in Fig.~\ref{fig:theta_dist_fits}.

\section{Extending the hangup-kick formula}

While accretion will tend to align the spins of the two BHs in a
BHB with the orbital angular momentum, it will not align the
in-plane components of the spins. Additionally, the expected
distribution of mass ratios~\cite{Yu:2011vp, Stewart:2008ep,
Hopkins:2009yy} indicates that equal-mass mergers
are rare. We therefore need a way to extend Eq.~(\ref{eq:FS})
 to
generic BHBs.

Using the same post-Newtonian analysis~\cite{Racine:2008kj} as in~\cite{Lousto:2009mf},
we can extend
formula~(\ref{eq:FS}) to less
symmetric configurations by replacing $\alpha \sin\theta$ by
$|\vec \alpha_2^\perp - q \alpha_1^\perp|/(1+q)$ and $\alpha \cos \theta$
by $2[\alpha_2^z + q^2 \alpha_1^z]/(1+q)^2$.
Importantly, we are assuming that terms
proportional to $|\vec \alpha_2^\perp - q \vec \alpha_1^\perp]^n$ (for $n>1$)
are negligible.
 If this is not the case, then our expansion,
which can be thought of as a Fourier sine series, would still converge, but our
extension would contain errors that may not be small. For example,
if a term like $(\alpha^\perp)^2 \alpha^z$ were present,
this would contribute to all even components of the
Fourier sine series and when extending the series, we would have
to take this into account.
 This would change the behavior
of kick even in more symmetric configurations.
This degeneracy in the
interpretation of the sine series can be broken by examining
configurations with constant $\alpha^z$ (while varying
$\alpha^\perp$) and constant $\alpha^\perp$ (while
varying $\alpha^z$). These, and other configurations, will be the
subject of an upcoming paper.
Our justification for not including these terms is that
the higher-order $\alpha^\perp$ terms are small in the ``superkick''
configuration. Furthermore, the accuracy with which
formula~(\ref{eq:FS}) predicts the
results of our $\alpha=0.91$ simulations supports
the conclusion that these terms remain small.
 This can be verified by confirming that
formula~(\ref{eq:FS}) is accurate for
all $\theta$ and  $\alpha$ (a subject of our ongoing analysis that
will be reported in a forthcoming paper).
We emphasize that the proposed extension is an ansatz, that while
reasonable as a starting point for the modeling, needs to be
thoroughly tested and refined.

Our new ansatz for the recoil velocity 
 modifies Eq.~(\ref{eq:Pempirical}) by changing the
``superkick'' $v_\|$ term. The ansatz has the form (after dropping
terms that previous studies indicated were small~\cite{Lousto:2009mf}):
\begin{eqnarray}\label{eq:newPempirical}
&& \vec{V}_{\rm recoil}(q,\vec\alpha)=v_m\,\hat{e}_1+
v_\perp(\cos\xi\,\hat{e}_1+\sin\xi\,\hat{e}_2)+v_\|\,\hat{n}_\|,\nonumber\\
&&v_m=A_m\frac{\eta^2(1-q)}{(1+q)}\left[1+B_m\,\eta\right],\nonumber\\
&&v_\perp=H\frac{\eta^2}{(1+q)}\left[
\,(\alpha_2^\|-q\alpha_1^\|) \right],\nonumber\\
&&v_\|= 16 \eta^2/(1+q)\Bigg[V_{1,1} + V_A \tilde S_z
+ V_B \tilde S_z^2 + V_C \tilde S_z^3\Bigg]\nonumber\\
&& \mbox{\ \ \ \
}\times\left|\vec \alpha_2^\perp-q\vec \alpha_1^\perp\right|\cos(\phi_\Delta-\phi_1),\nonumber\\
\end{eqnarray}
where $\vec{\tilde{S}} = 2 (\vec \alpha_2 + q^2\vec \alpha_1)/(1+q)^2$,
and the coefficients we use in the statistical studies below are
$H=6.9\times10^3$~\cite{Lousto:2007db}, $A_m=1.2\times10^4$,
$B_m=-0.93$~\cite{Gonzalez:2006md}, and the remaining coefficients are
obtained from Eq.~(\ref{eq:FS}) above.

\section{Statistical Studies}\label{Sec:Statistics}

\begin{figure}
 \includegraphics[width=\columnwidth]{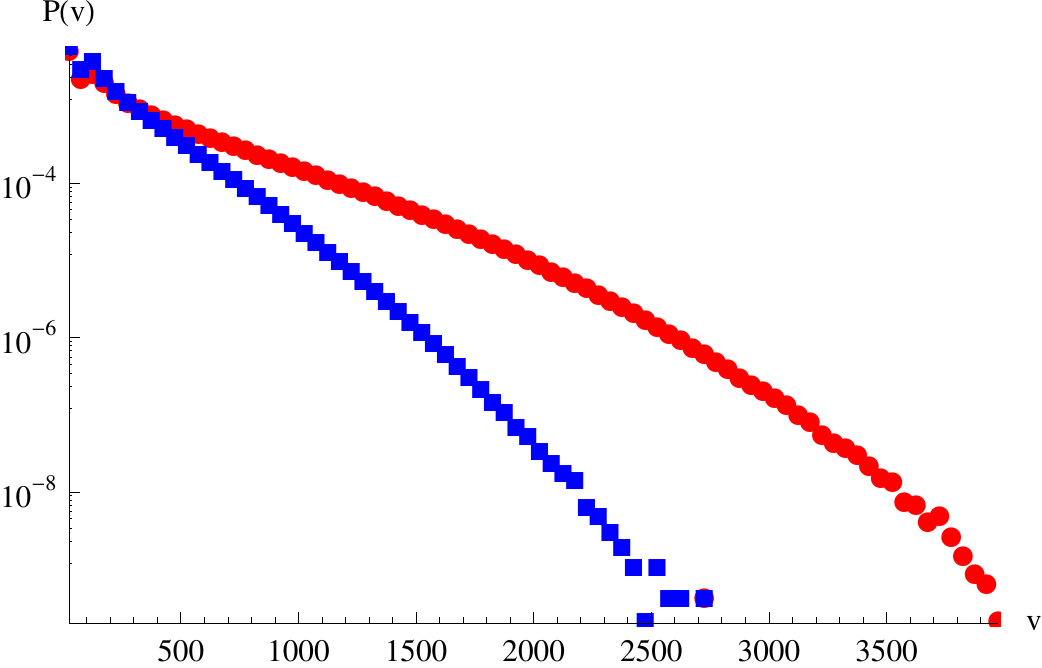}
 \caption{Probability distribution $P(v)$ of the recoil magnitude for BHBs alignment configurations in 
  hot (top curve, red circles) and cold disks  BHBs (lower
curve, blue squares).
The velocity is in units of $\KMS$.}
  \label{fig:v_dist}
\end{figure}

\begin{figure}
 \includegraphics[width=\columnwidth]{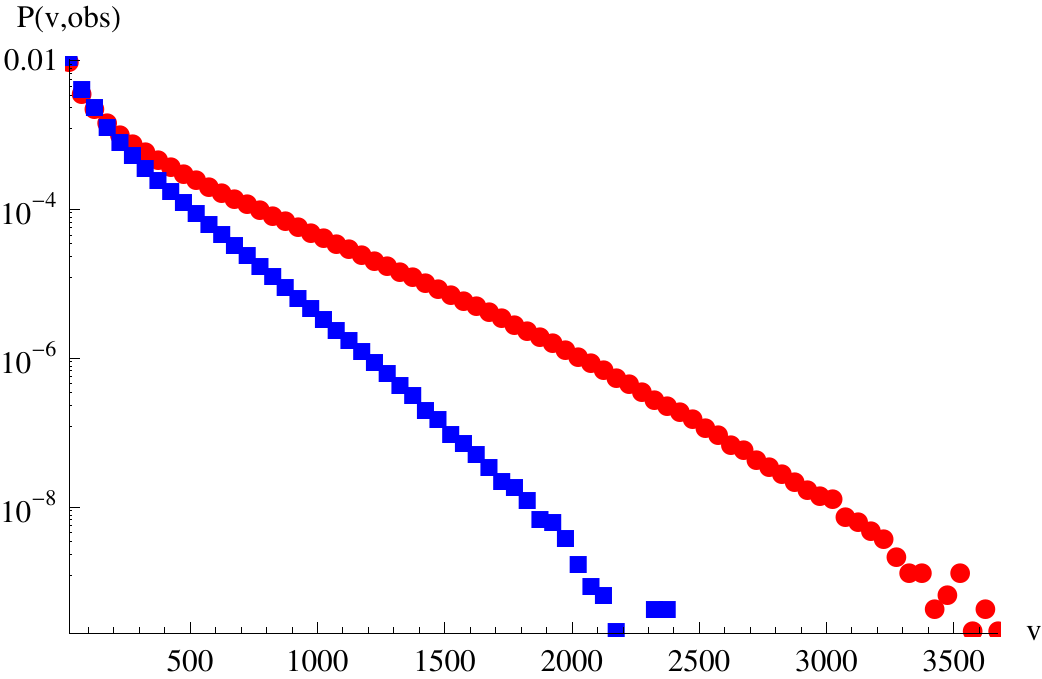}
 \caption{Probability distribution $P(v)$ of the recoil magnitude 
along the line of
sight for BHBs in
  hot (top curve, red circles) and cold disks   (lower
curve, blue squares). The
velocity is in units of $\KMS$.}
  \label{fig:ov_dist}
\end{figure}

Using Eq.~(\ref{eq:newPempirical}) and the above fitted spin magnitude
$(\alpha)$,
and direction distributions ($\theta$), the mass ratio
distribution suggested in~\cite{Yu:2011vp, Stewart:2008ep,
Hopkins:2009yy},  $P(q) \propto q^{-0.3}
(1-q)$,
and assuming that the two BHs can have arbitrary
orientations for the in-plane component of the spin 
(i.e.\ uniform probability in the range
$0\leq\phi\leq2\pi$),
we obtain probabilities for the recoil
velocity magnitude and direction. To perform our statistical studies,
we choose $10^8$ configurations ($2\times10^8$ configurations in
total) 
randomly chosen based on the above
probability distributions and examine the predicted recoil magnitude
and direction. Our results are summarized
in Table~\ref{tab:stats} and
Figs.~\ref{fig:v_dist}~and~\ref{fig:ov_dist}.
Although we include the pure unequal mass recoil [$v_m$ in
Eq.~(\ref{eq:newPempirical})], we note that the modeling of
the angle $\xi$ as a function of the binary's parameter is incomplete.
However $v_m$ has normally a nonleading effect.
We therefore chose a constant $\xi=145^\circ$, as suggested in
our previous study~\cite{Lousto:2007db}.
 
\begin{figure}
 \includegraphics[width=\columnwidth]{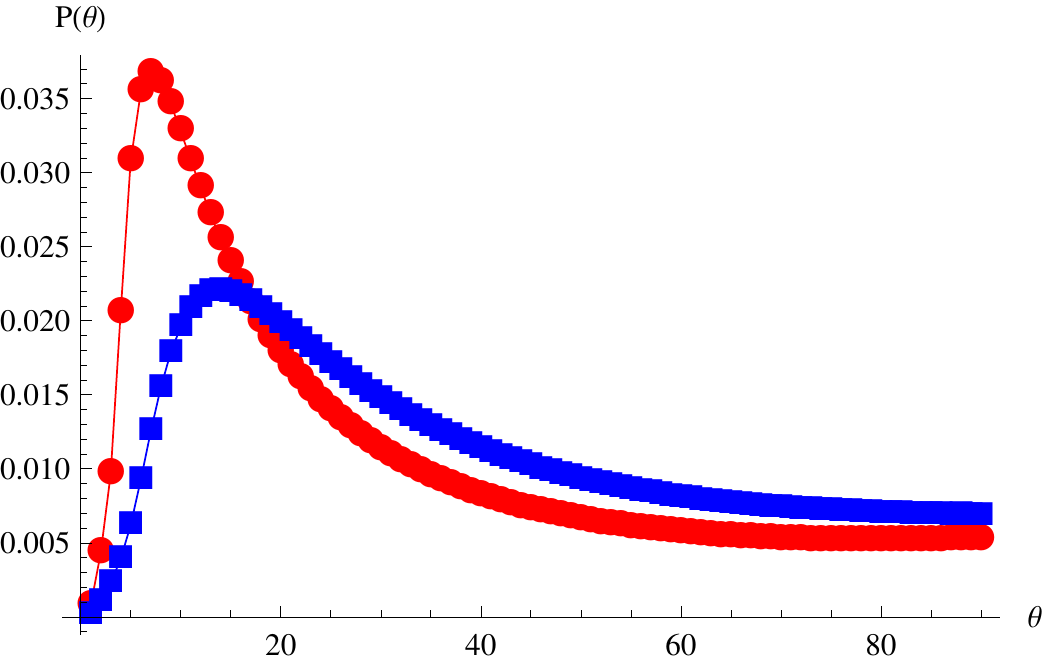}
  \caption{The probability distribution of the inclination angle
$\theta$ of the recoil
(measured with respect to the axis of the angular momentum)
 for hot (narrower
distribution, red circles) and cold environments (blue squares). Angles are measured in degrees. Note
that $P(180^\circ-\theta) = P(\theta)$. These distributions were
created by mapping $\theta\to 180^\circ-\theta$ for $\theta>90^\circ$. }
  \label{fig:kick_inc}
\end{figure}

\begin{table}[t]
  \caption{Recoil velocity probabilities (in percent) 
for BHs in hot and cold disks
aligned binaries and the probabilities for the recoil along the
line-of-sight having the given magnitude range (denoted by Obs.). 
For the hot case, there is a nontrivial probability of observing
a recoil larger than $2000\ \KMS$, but for cold disks, such
recoils are suppressed. Velocities are in units of $\KMS$.}
   \label{tab:stats}
\begin{ruledtabular}
\begin{tabular}{l|llll}
  Vel. ($\KMS)$ & (Hot) & Obs. (Hot) & (Cold) &
Obs. (Cold)\\ \hline
0-100     &34.2593  \%& 60.1847  \%& 41.4482  \%& 71.2967 \%  \\
100-200   &21.1364  \%& 16.9736  \%& 28.3502  \%& 16.8471 \%  \\
200-300   &11.6901  \%&  8.1110  \%& 12.503  \%&   6.1508 \%  \\
300-400   & 7.8400  \%&  4.8108  \%&  7.0967  \%&  2.8281 \%  \\
400-500   & 5.7590  \%&  3.0913  \%&  4.2490  \%&  1.3973 \%  \\
500-1000  &14.0283  \%&  5.6593  \%&  5.9309  \%&  1.4258 \%  \\
1000-1500 & 4.0183  \%&  0.9809  \%&  0.4030  \%&  0.0526 \%  \\
1500-2000 & 1.0309  \%&  0.1638  \%&  0.0185  \%&  0.0015 \%  \\
2000-2500 & 0.2047  \%&  0.0223  \%&  0.0005  \%&  $2\times10^{-5}\%$\\
2500-3000 & 0.0296  \%&  0.0023  \%&  $1\times10^{-5}\%$& 0.\% \\
3000-3500 & 0.0032  \%&  0.0002  \%& 0. \%& 0.\% \\
3500-4000 & 0.0002  \%& $4.\times 10^{-6} $ \%& 0.\%  \%& 0.\%  \\
\end{tabular}
\end{ruledtabular}
\end{table}
In Fig.~\ref{fig:kick_inc} we show the probabilities that the recoil has
a given inclination angle (angle with respect to the orbital angular
momentum) for hot and cold disks.
Because of the
$\theta\to 180^\circ-\theta$ symmetry, we map all recoil angles to the
interval $0^\circ\leq\theta\leq90^\circ$.
Here $P(\theta)$ is the probability integrating over all possible recoil magnitudes, i.e. $P(\theta) = \int_0^\infty P(\theta, v) dv$. 
The distribution is normalized such that $\int_0^{90} P(\theta)
d\theta=1$.
The angular distribution is
broader for accretion in cold disks, since that tends to suppress the
``hangup-kick'' and ``superkick'', while the distribution is more sharply
peaked near $\theta=0$ for hot disks.

\begin{table}[t]
  \caption{Recoil velocity direction ($\theta$)
 probabilities for the hot and cold cases. Recoils with $\theta>90^\circ$ have been
remapped using the symmetry $\theta\to180^\circ - \theta$.
}
   \label{tab:stats_theta}
\begin{ruledtabular}
\begin{tabular}{l|ll}
  Range & (Hot) & (Cold) \\ \hline
$0^\circ-30^\circ$     &  61.6\% &  47.0\% \\
$30^\circ-60^\circ$    &  22.6\% & 31.4\% \\
$60^\circ-90^\circ$    &  15.7\% & 21.6\%\\
\end{tabular}
\end{ruledtabular}
\end{table}

This strong angular dependence of the recoil has particular relevance
for the studies of the observational consequences of merging and
kicked BHs surrounded by preexisting gas disks 
\cite{Milosavljevic:2004cg,Schnittman:2008ez,Lippai:2008fx,Shields:2008va,Rossi:2009nk,Corrales:2009nv,Ponce:2011kv}. 
The merger of a
BHB resulting in the remnant BH moving across the matter that
surrounded the original BHB would greatly affect the dynamics of the
gas and its thermodynamic state. This translates into distinctive
electromagnetic signatures that could reveal the presence of recoiling
BHs. The effect is very pronounced when the BH recoils in the orbital
plane with a large magnitude.  
The strong preference for large recoils along the axis of the disk
over those recoiling along the disk itself can strongly suppress the
magnitude of such signatures.

\begin{figure}
  \includegraphics[width=1.65in]{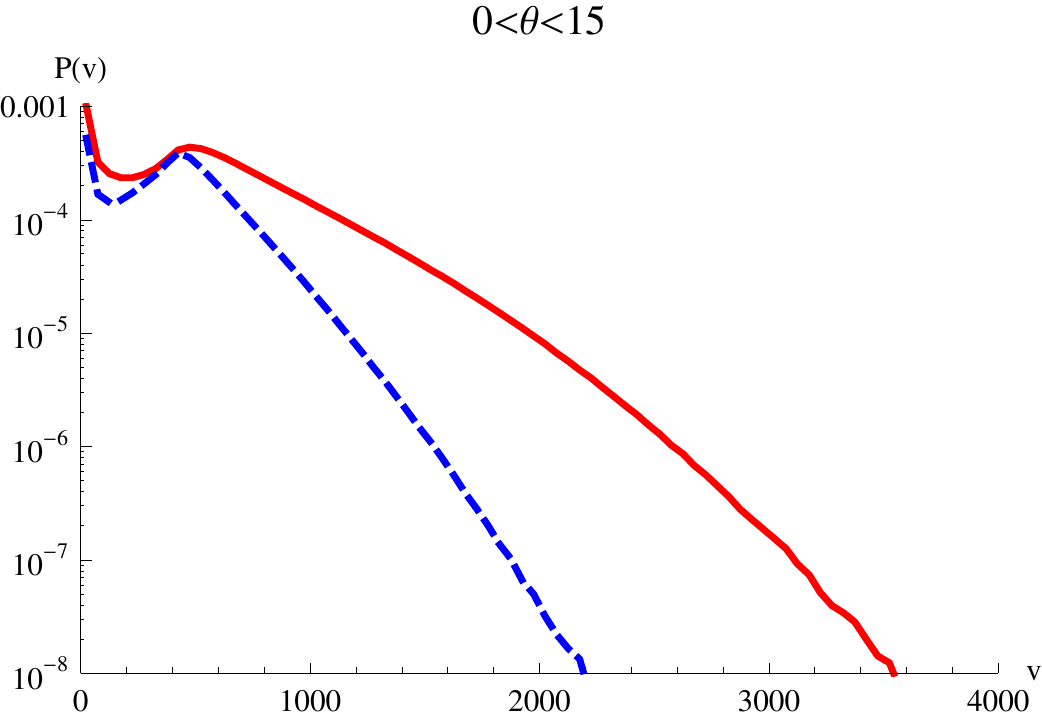}
  \includegraphics[width=1.65in]{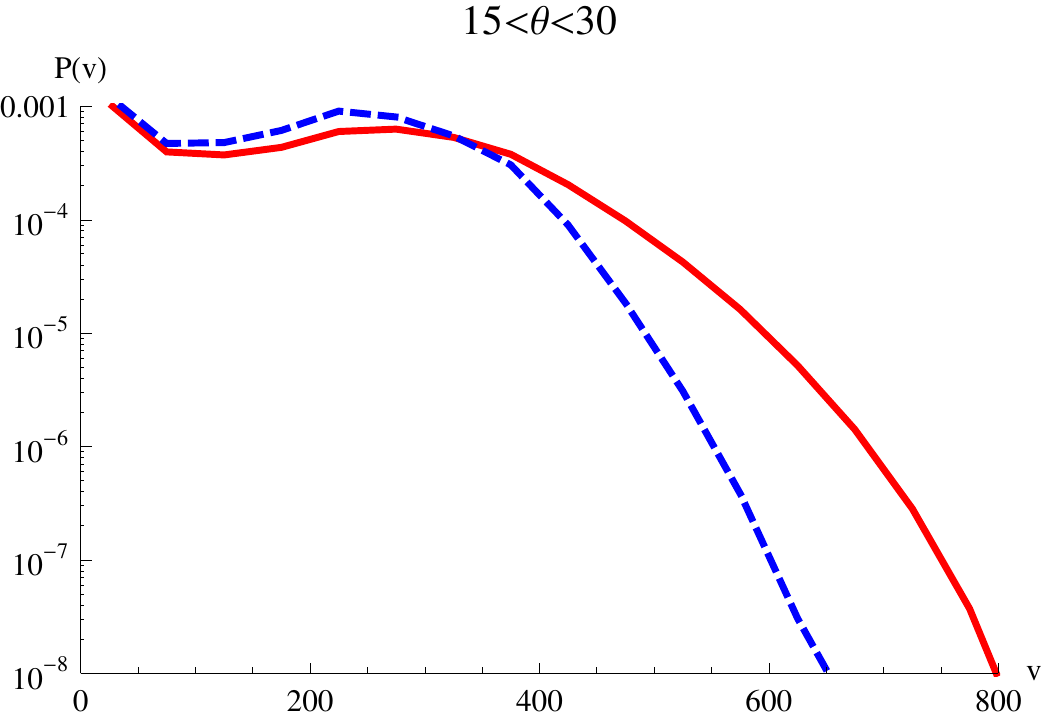}
  \includegraphics[width=1.65in]{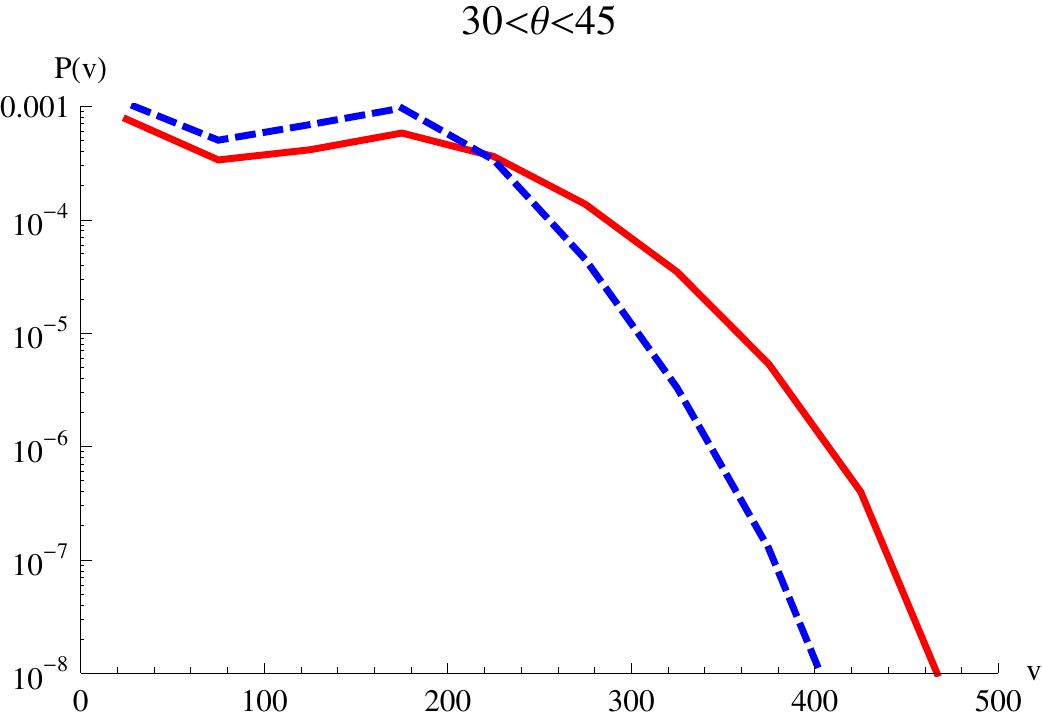}
  \includegraphics[width=1.65in]{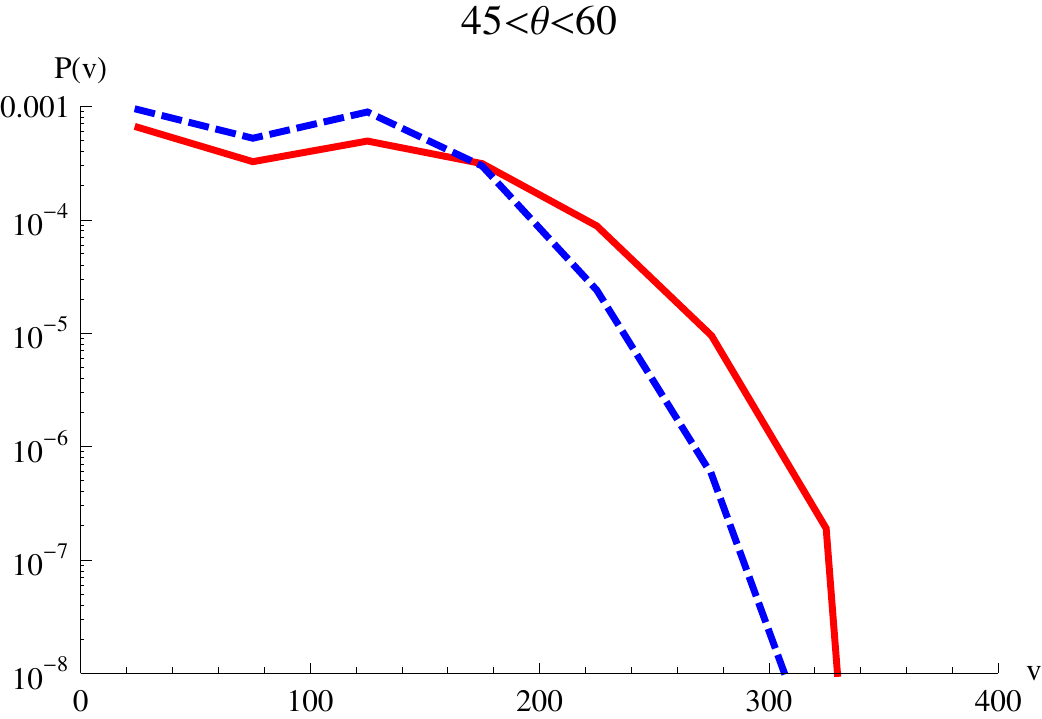}
  \includegraphics[width=1.65in]{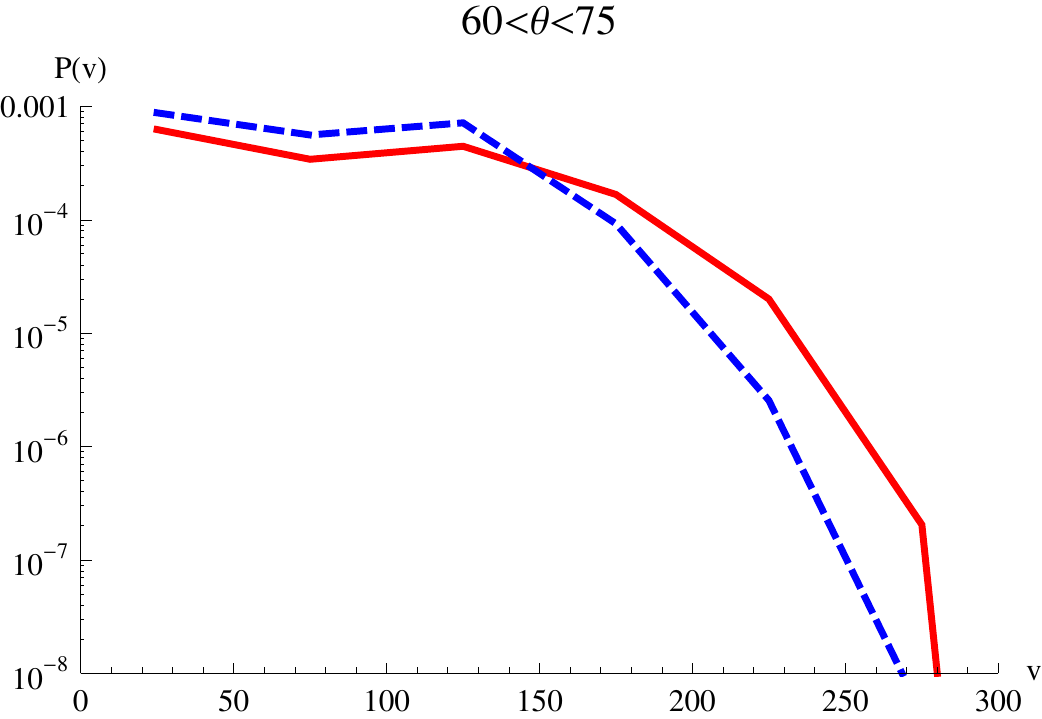}
  \includegraphics[width=1.65in]{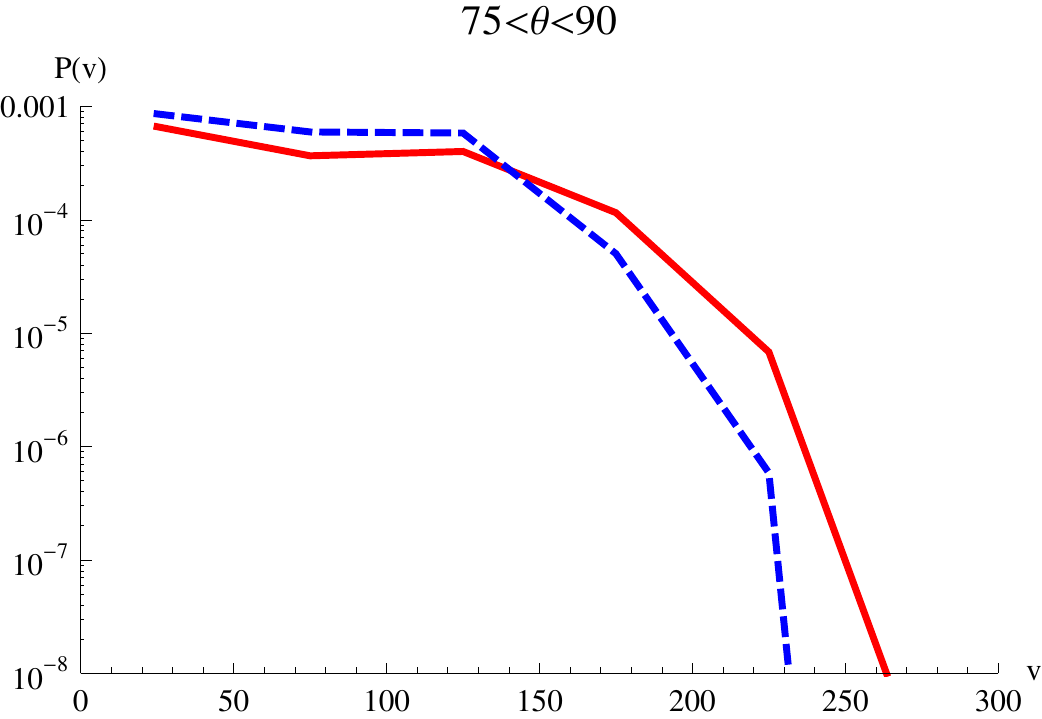}

  \caption{Recoil velocity distributions for hot (red solid
lines) and cold (blue dashed lines) disks
   for recoils in angular intervals 15 degrees wide
(Eq.~(\ref{eq:newPempirical}) predicts equal probabilities for recoils
at an angle $\theta$ and $180^\circ-\theta$). Note how rapidly
the maximum recoil decreases as a function of $\theta$. Recoils as
large as $1000\ \KMS$ must have $\theta<15^\circ$ and in-plane
recoils are less than $270\ \KMS$ (but see comment about unequal-mass
recoils in the text).
  }
  \label{fig:p_v_theta_15}
\end{figure}

\begin{figure}
  \includegraphics[width=\columnwidth]{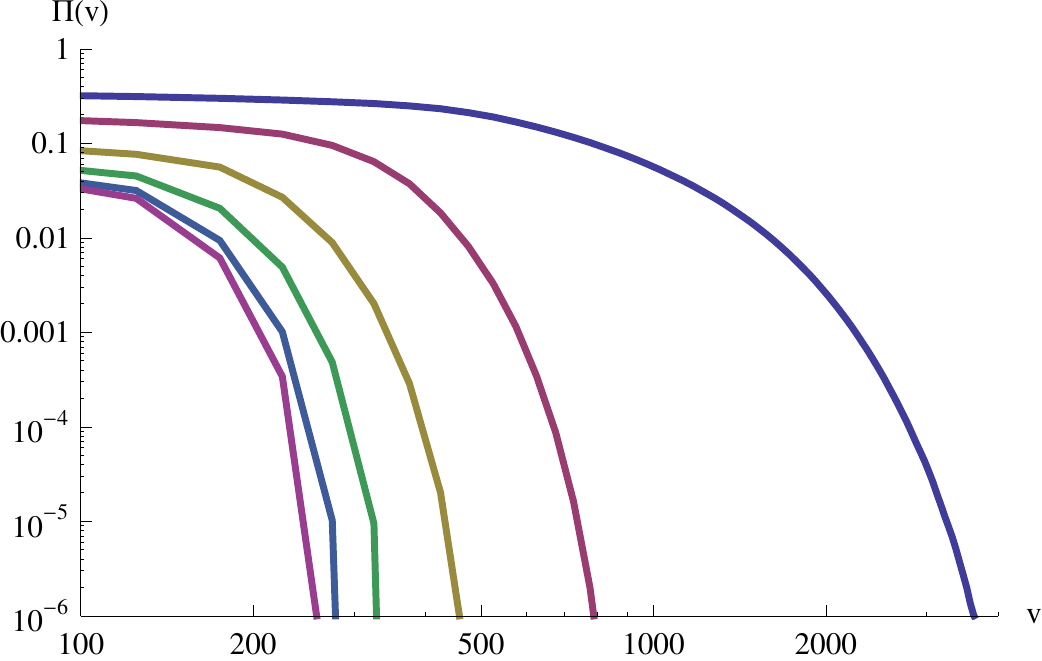}
  \includegraphics[width=\columnwidth]{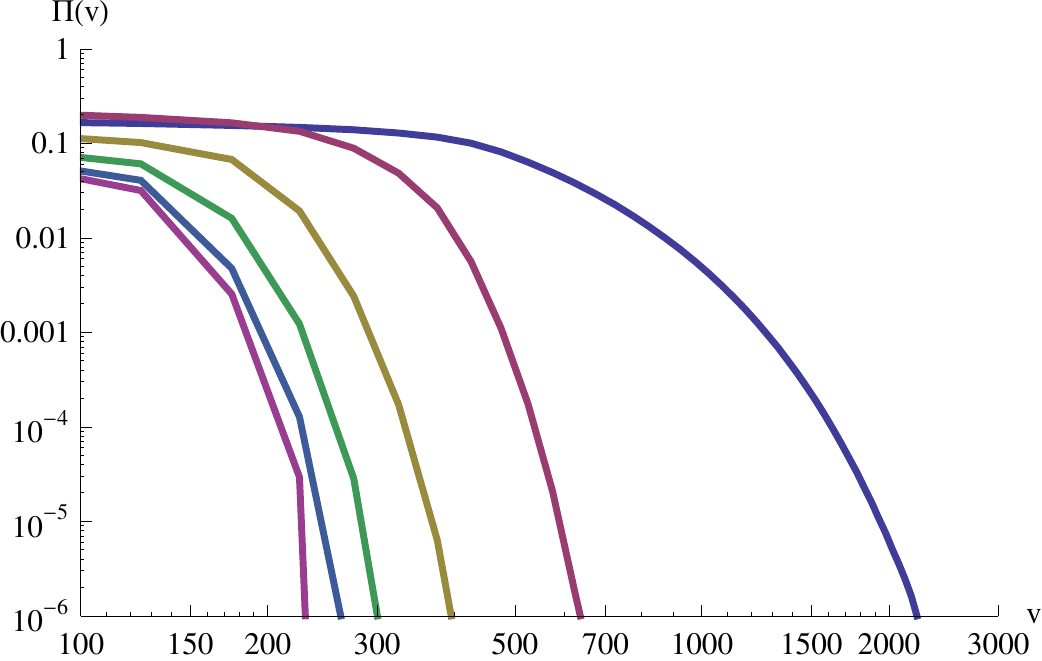}

  \caption{The integrated probability $\Pi(v)$ of a recoil having
velocity $v$ or larger [$\Pi(v) = \int_v^\infty P(\nu) d\nu$] for
hot (top) and cold (bottom) accretion disks for recoils in the ranges
$0^\circ <\theta<15^\circ$, $15^\circ<\theta<30^\circ$, $\cdots$,
$75^\circ<\theta<90^\circ$. In both cases, for recoils larger than
$200\ \KMS$, the recoil probabilities
are smaller for larger values of $\theta$. In the cold case,
low-velocity ($<200\ \KMS)$ recoils with an angle  
$0^\circ < \theta < 15^\circ$ are less
probable than recoils with angle $15^\circ < \theta < 30^\circ$. }
  \label{fig:prob_v_theta_15}
\end{figure}

In Fig.~\ref{fig:p_v_theta_15} we show the recoil velocity distribution
for hot and cold disks integrated over $15^\circ$
intervals of the inclination angle $\theta$ (in Figs.~\ref{fig:v_dist}~and~\ref{fig:ov_dist} above, the integration is
over $0^\circ\leq\theta\leq 180^\circ$). While in
Fig.~\ref{fig:prob_v_theta_15}, we show the integrated probability,
$\Pi(v)$ of a recoil having
velocity $v$ or larger [$\Pi(v) = \int_v^\infty P(\nu) d\nu$].
If we consider recoils
within $15^\circ$ of the orbital axis, we see that velocities up to
$900\ \KMS$ are
likely (i.e.\ about $1\%$ probability) 
for cold  disks and up to $1600\ \KMS$ for hot  disks (see also
Fig.~\ref{fig:prob_v_theta_15}).
If we look at angles between $15^\circ$ and $30^\circ$ we see 
the recoils are limited to less than $800\ \KMS$ (even for hot disks,
a recoil of $600\ \KMS$ has $<0.1\%$  probability).
At larger $\theta$ angles, the maximum recoil drops below $270\ \KMS$.

Since the most striking effects are likely due to BHs recoiling with large
magnitudes through the plane of the disk, it is interesting
to examine the probabilities
of such events occurring. Even with integrated probabilities of $\lesssim 10^{-4}$,
such events may be observed in large surveys of galaxies, e.g.\ 
the Sloan Digital Sky Survey DR7 contains $\sim9\times10^{5}$
galaxies if the observable effects last long enough.
We can see in Figs.~\ref{fig:p_v_theta_15}~and~\ref{fig:prob_v_theta_15} that while
in hot disks we can observe recoils of nearly $3000\ \KMS$,
cold disks limit the maximum observable recoil 
to $2000\ \KMS$. On the other end, if we require the recoiling hole to
be  within $30^\circ$ of the orbital plane (i.e.\ $\theta>60^\circ$)
we cannot observe recoils larger than $ 250\ \KMS$, while at
intermediate angles velocities seem limited to $400\ \KMS$ for both
cold and hot  disks.

\begin{figure}
  \includegraphics[width=.49\columnwidth]{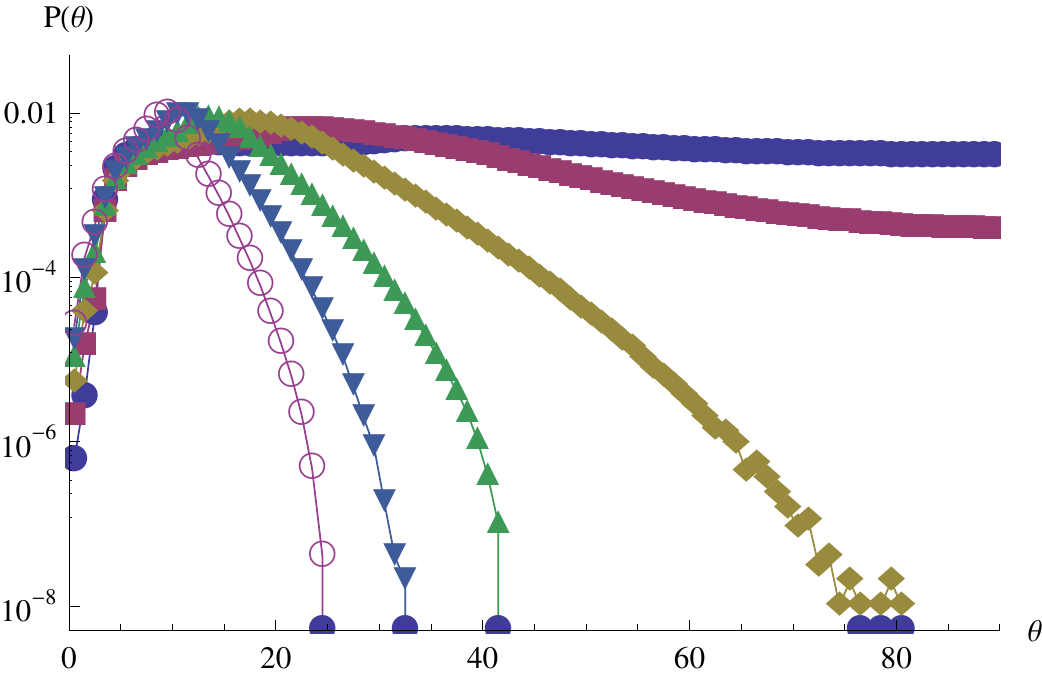}
  \includegraphics[width=.49\columnwidth]{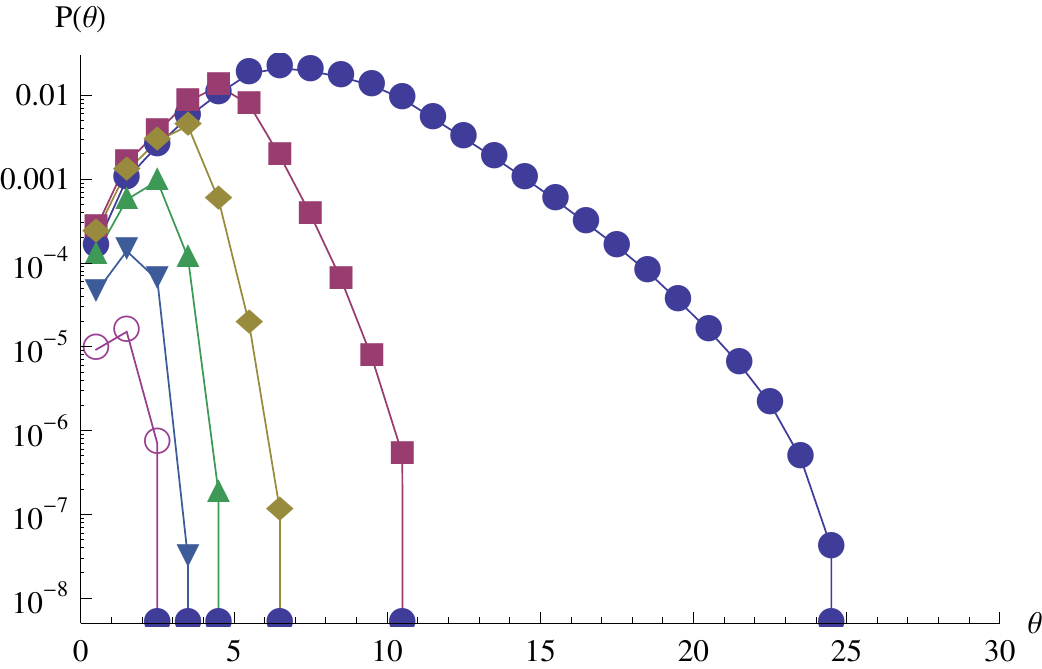}

  \includegraphics[width=.49\columnwidth]{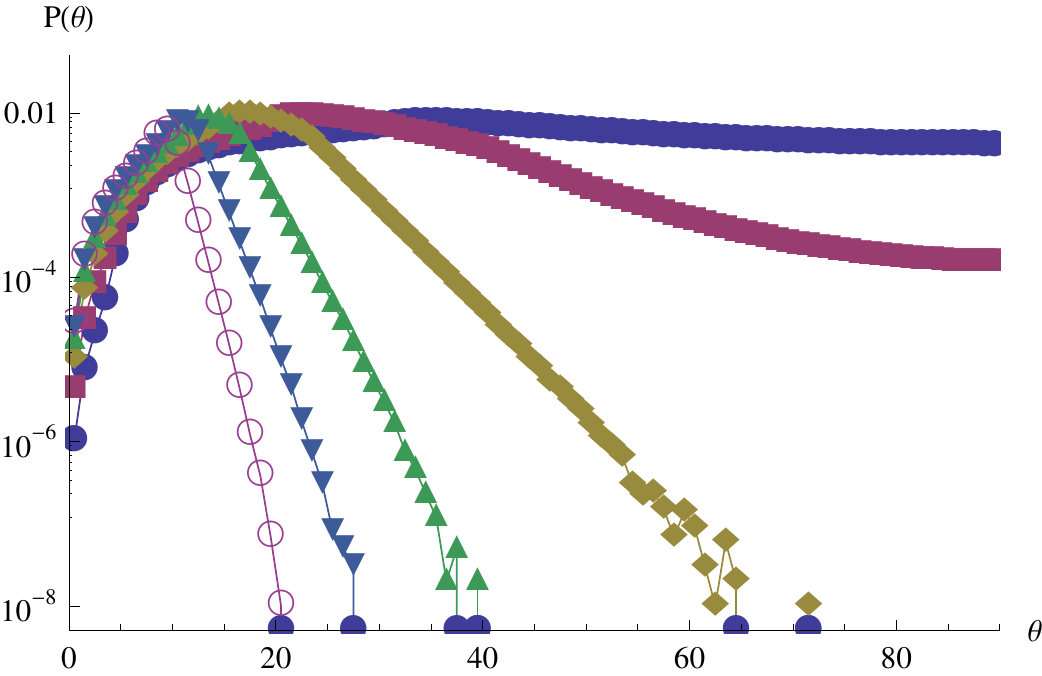}
  \includegraphics[width=.49\columnwidth]{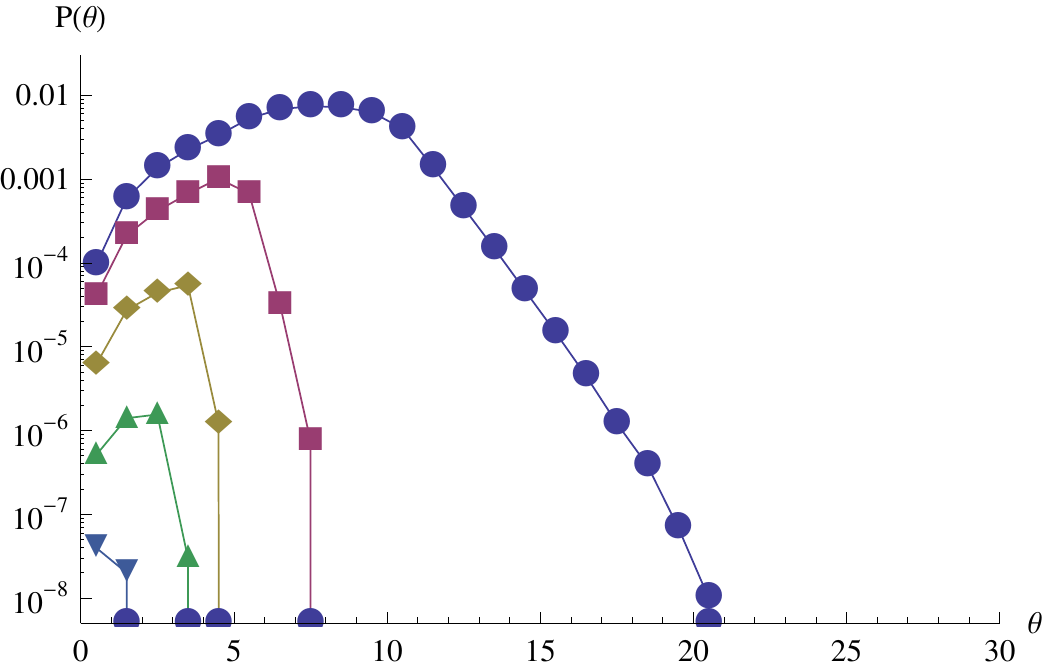}
  \caption{Recoil angle probabilities $P(\theta)$ for hot (top) and
 cold (bottom) disks aligned
binaries. The plots on the left show angular probabilities for
velocities in the ranges $0-100\ \KMS$, $100-200\ \KMS$,
$\cdots$, $500-600\ \KMS$. The plots on the right show 
$P(\theta)$ for velocity ranges of $500-1000\ \KMS$, $\cdots$,
$3000-3500\ \KMS$. Probabilities for $\theta$ and $180^\circ-\theta$ are
equal. In the plots, closed circles correspond to the smallest range,
followed by squares, diamonds, triangles (vertex up), triangles
(vertex down), and open circles. The circles on the axis are an
artifact of the visualization tool. }
\label{fig:p_theta_v_range}
\end{figure}

In Fig.~\ref{fig:p_theta_v_range},
 we show the angular distribution for
recoils in given velocity ranges. Again, because of the 
$\theta\to180^\circ-\theta$ symmetry, we map all recoil angles to the
interval $0\leq\theta\leq90^\circ$. For convenience, we plot the
probabilities in degrees rather than radians. For these plots, we used
$10^8$ randomly chosen binaries consistent with the above
distributions for spin magnitude, spin direction, and mass ratio.
The maximum angle the recoil can make with the orbital angular
momentum axis is very restrictive for large velocities, as is
shown in Table~\ref{tab:max_theta_v_range}.
\begin{table}[t]
  \caption{Maximum recoil angle $\theta$ (angle with respect to the
orbital angular momentum axis) for given recoil velocity ranges. Note
here that $\theta_{\rm max} < \delta$ means that $\theta$ must smaller
than $\delta$ or larger than $180^\circ-\delta$.}
   \label{tab:max_theta_v_range}
\begin{ruledtabular}
\begin{tabular}{l|ll}
  Range & $\theta_{\rm max}$ (Hot) & $\theta_{\rm max}$ (Cold) \\ \hline
$0-100\ \KMS$  & $90^\circ$ & $90^\circ$ \\
$100-200\ \KMS$ & $90^\circ$ & $90^\circ$ \\
$200-300\ \KMS$ & $<80^\circ$  & $<70^\circ$ \\
$300-400\ \KMS$ & $<45^\circ$ & $<40^\circ$ \\
$400-500\ \KMS$ & $<33^\circ$ & $<30^\circ$ \\
$500-600\ \KMS$ & $<25^\circ$ &  $<21^\circ$ \\
$500-1000\ \KMS$ & $<25^\circ$ &  $<21^\circ$ \\
$1000-1500\ \KMS$ & $<11^\circ$ & $<8^\circ$ \\
$1500-2000\ \KMS$ & $<7^\circ$ & $ <5^\circ$  \\
$2000-2500\ \KMS$ &  $<5^\circ$ & $<4^\circ$ \\
$2500-3000\ \KMS$ & $ < 4^\circ$ & $<2^\circ$ \\
$3000-3500\ \KMS$ & $ < 3^\circ $ & *** \\
\end{tabular}
\end{ruledtabular}
\end{table}

\section{Discussion}\label{Sec:Discussion}

We studied in detail a family of BHB configurations with full
numerical relativity that allowed us to single out the
``hangup-kick''. This effect is expected to be relevant in generic BBH
mergers since it arises from a combination of generic properties of
the orbital dynamics of spinning BHBs, namely the ``orbital-hangup
effect''~\cite{Campanelli:2006uy} and the ``superkick
effect''~\cite{Campanelli:2007cga}.  We present evidence that this effect
increases the maximum recoil velocity achievable from the merger of
two orbiting BHs by up to $1200\ \KMS$ with respect to previous
estimates, approaching nearly $5000\ \KMS$. Even more importantly this
maximum recoil is reached for spins at angles near $50^\circ$ with
respect to the orbital momentum of the binary system. We have also
shown evidence that accretion in the premerger stage of the
binary tends to align spins with the angular momentum of the system,
leading to distributions that favor the ``hangup-kick''
configurations with respect to the purely in-plane ``superkick'' ones.
Due to depletion of nearby matter, the merger itself occurs in
a ``dry'' regime where accretion no longer affects the BH spins.

In an attempt to estimate the probability of observing such large
recoils
in
real astronomical systems, we assumed accretion driven distributions
for the spin magnitudes and directions, based on the two extreme scenarios of
cold and hot  disks, and assumed a mass ratio distribution
based on independent estimates, to obtain non negligible probabilities
of observing
recoils of several thousand $\KMS$. In particular, 
the results in Table \ref{tab:stats}
indicate that surveys exploring peculiar differential
radial velocities should observe at least one case of a
``line-of-sight'' velocity above $2000\ \KMS$ out of
four thousand merged galaxies (assuming ``hot'' disks). 
The probability that a remnant BH receives a recoil exceeding the
escape velocity (in any direction) of giant galaxies ($2000\ \KMS$) is
ten times larger.
 Probabilities of recoils exceeding the escape velocity
quickly rise to 5\% 
for galaxies with escape velocities of $1000\ \KMS$and nearly
20\% for galaxies with escape velocities of
$500\ \KMS$.
These numbers indicate that recoil velocities and modeling the accretion
of the supermassive BHs in centers of galaxies should be important
ingredients in understanding the growth of supermassive BHs and
large scale structure formation in the universe. 

Our initial study showed the relevance of recoil and accretion modeling
in order to better understand  how BHs evolve and grow in the universe.
There are several aspects that deserve further study. The recoil formula needs
to be further tested and developed in the intermediate mass ratio regime and 
for fully precessing BHBs. Accretion needs to be modeled at
even smaller scales, i.e. at sub-milli-parsecs. In between the accretion regime
governed by Newtonian physics and the fully nonlinear regime, a slow adiabatic inspiral occurs. This
intermediate regime can be described by semianalytic methods, such as the 
post-Newtonian approximations. It has been pointed out that resonances for
certain mass ratios can lead to further spin alignment
\cite{Schnittman:2004vq,Kesden:2010ji} 
and hence change the initial spin distributions used in the applications of
the recoil formula (\ref{eq:newPempirical}). 
Further theoretical and observational explorations
of the recoil phenomena are well worth being pursued since they could
represent the first prediction and verification of General Relativity
in its most highly dynamical and nonlinear regime.

\acknowledgments 

The authors thank M.Favata and V.Paschalidis
for pointing out Ref.~\cite{Nichols:2011ih} and T. Bogdanovic and C.
Miller for careful reading of the manuscript.
CL and YZ gratefully acknowledge the NSF for financial support from Grants
AST-1028087, PHY-0929114, PHY-0969855, PHY-0903782, OCI-0832606, and
DRL-1136221,  and NASA for financial support from NASA Grant No.
07-ATFP07-0158. Computational resources were provided by the Ranger
system at the Texas Advance Computing Center (Teragrid allocation
TG-PHY060027N), which is supported in part by the NSF, and by
NewHorizons at Rochester Institute of Technology, which was supported
by NSF grant No. PHY-0722703, DMS-0820923 and AST-1028087.
MV acknowledges the NSF for financial support from 
NSF award AST-1107675.

\bibliographystyle{apsrev}
\bibliography{../../../../Bibtex/references}

\begin{thebibliography}{83}
\expandafter\ifx\csname natexlab\endcsname\relax\def\natexlab#1{#1}\fi
\expandafter\ifx\csname bibnamefont\endcsname\relax
  \def\bibnamefont#1{#1}\fi
\expandafter\ifx\csname bibfnamefont\endcsname\relax
  \def\bibfnamefont#1{#1}\fi
\expandafter\ifx\csname citenamefont\endcsname\relax
  \def\citenamefont#1{#1}\fi
\expandafter\ifx\csname url\endcsname\relax
  \def\url#1{\texttt{#1}}\fi
\expandafter\ifx\csname urlprefix\endcsname\relax\def\urlprefix{URL }\fi
\providecommand{\bibinfo}[2]{#2}
\providecommand{\eprint}[2][]{\url{#2}}

\bibitem[{\citenamefont{{Begelman} et~al.}(1984)\citenamefont{{Begelman},
  {Blandford}, and {Rees}}}]{1984RvMP...56..255B}
\bibinfo{author}{\bibfnamefont{M.~C.} \bibnamefont{{Begelman}}},
  \bibinfo{author}{\bibfnamefont{R.~D.} \bibnamefont{{Blandford}}},
  \bibnamefont{and} \bibinfo{author}{\bibfnamefont{M.~J.}
  \bibnamefont{{Rees}}}, \bibinfo{journal}{Reviews of Modern Physics}
  \textbf{\bibinfo{volume}{56}}, \bibinfo{pages}{255} (\bibinfo{year}{1984}).

\bibitem[{\citenamefont{{Redmount} and {Rees}}(1989)}]{Redmount:1989}
\bibinfo{author}{\bibfnamefont{I.~H.} \bibnamefont{{Redmount}}}
  \bibnamefont{and} \bibinfo{author}{\bibfnamefont{M.~J.}
  \bibnamefont{{Rees}}}, \bibinfo{journal}{Comments on Astrophysics}
  \textbf{\bibinfo{volume}{14}}, \bibinfo{pages}{165} (\bibinfo{year}{1989}).

\bibitem[{\citenamefont{{Blecha} et~al.}(2011)\citenamefont{{Blecha}, {Cox},
  {Loeb}, and {Hernquist}}}]{Blecha:2011}
\bibinfo{author}{\bibfnamefont{L.}~\bibnamefont{{Blecha}}},
  \bibinfo{author}{\bibfnamefont{T.~J.} \bibnamefont{{Cox}}},
  \bibinfo{author}{\bibfnamefont{A.}~\bibnamefont{{Loeb}}}, \bibnamefont{and}
  \bibinfo{author}{\bibfnamefont{L.}~\bibnamefont{{Hernquist}}},
  \bibinfo{journal}{MNRAS} \textbf{\bibinfo{volume}{412}},
  \bibinfo{pages}{2154} (\bibinfo{year}{2011}), \eprint{1009.4940}.

\bibitem[{\citenamefont{{Fitchett}}(1983)}]{1983MNRAS.203.1049F}
\bibinfo{author}{\bibfnamefont{M.~J.} \bibnamefont{{Fitchett}}},
  \bibinfo{journal}{MNRAS} \textbf{\bibinfo{volume}{203}},
  \bibinfo{pages}{1049} (\bibinfo{year}{1983}).

\bibitem[{\citenamefont{{Fitchett} and
  {Detweiler}}(1984)}]{1984MNRAS.211..933F}
\bibinfo{author}{\bibfnamefont{M.~J.} \bibnamefont{{Fitchett}}}
  \bibnamefont{and}
  \bibinfo{author}{\bibfnamefont{S.}~\bibnamefont{{Detweiler}}},
  \bibinfo{journal}{Mon. Not. R. astr. Soc.} \textbf{\bibinfo{volume}{211}},
  \bibinfo{pages}{933} (\bibinfo{year}{1984}).

\bibitem[{\citenamefont{Blanchet et~al.}(2005)\citenamefont{Blanchet, Qusailah,
  and Will}}]{Blanchet:2005rj}
\bibinfo{author}{\bibfnamefont{L.}~\bibnamefont{Blanchet}},
  \bibinfo{author}{\bibfnamefont{M.~S.~S.} \bibnamefont{Qusailah}},
  \bibnamefont{and} \bibinfo{author}{\bibfnamefont{C.~M.} \bibnamefont{Will}},
  \bibinfo{journal}{Astrophys. J.} \textbf{\bibinfo{volume}{635}},
  \bibinfo{pages}{508} (\bibinfo{year}{2005}), \eprint{astro-ph/0507692}.

\bibitem[{\citenamefont{Le~Tiec et~al.}(2010)\citenamefont{Le~Tiec, Blanchet,
  and Will}}]{LeTiec:2009yg}
\bibinfo{author}{\bibfnamefont{A.}~\bibnamefont{Le~Tiec}},
  \bibinfo{author}{\bibfnamefont{L.}~\bibnamefont{Blanchet}}, \bibnamefont{and}
  \bibinfo{author}{\bibfnamefont{C.~M.} \bibnamefont{Will}},
  \bibinfo{journal}{Class. Quant. Grav.} \textbf{\bibinfo{volume}{27}},
  \bibinfo{pages}{012001} (\bibinfo{year}{2010}), \eprint{0910.4594}.

\bibitem[{\citenamefont{Campanelli}(2005)}]{Campanelli:2004zw}
\bibinfo{author}{\bibfnamefont{M.}~\bibnamefont{Campanelli}},
  \bibinfo{journal}{Class. Quant. Grav.} \textbf{\bibinfo{volume}{22}},
  \bibinfo{pages}{S387} (\bibinfo{year}{2005}), \eprint{astro-ph/0411744}.

\bibitem[{\citenamefont{Pretorius}(2005)}]{Pretorius:2005gq}
\bibinfo{author}{\bibfnamefont{F.}~\bibnamefont{Pretorius}},
  \bibinfo{journal}{Phys. Rev. Lett.} \textbf{\bibinfo{volume}{95}},
  \bibinfo{pages}{121101} (\bibinfo{year}{2005}), \eprint{gr-qc/0507014}.

\bibitem[{\citenamefont{Campanelli
  et~al.}(2006{\natexlab{a}})\citenamefont{Campanelli, Lousto, Marronetti, and
  Zlochower}}]{Campanelli:2005dd}
\bibinfo{author}{\bibfnamefont{M.}~\bibnamefont{Campanelli}},
  \bibinfo{author}{\bibfnamefont{C.~O.} \bibnamefont{Lousto}},
  \bibinfo{author}{\bibfnamefont{P.}~\bibnamefont{Marronetti}},
  \bibnamefont{and}
  \bibinfo{author}{\bibfnamefont{Y.}~\bibnamefont{Zlochower}},
  \bibinfo{journal}{Phys. Rev. Lett.} \textbf{\bibinfo{volume}{96}},
  \bibinfo{pages}{111101} (\bibinfo{year}{2006}{\natexlab{a}}),
  \eprint{gr-qc/0511048}.

\bibitem[{\citenamefont{Baker et~al.}(2006)\citenamefont{Baker, Centrella,
  Choi, Koppitz, and van Meter}}]{Baker:2005vv}
\bibinfo{author}{\bibfnamefont{J.~G.} \bibnamefont{Baker}},
  \bibinfo{author}{\bibfnamefont{J.}~\bibnamefont{Centrella}},
  \bibinfo{author}{\bibfnamefont{D.-I.} \bibnamefont{Choi}},
  \bibinfo{author}{\bibfnamefont{M.}~\bibnamefont{Koppitz}}, \bibnamefont{and}
  \bibinfo{author}{\bibfnamefont{J.}~\bibnamefont{van Meter}},
  \bibinfo{journal}{Phys. Rev. Lett.} \textbf{\bibinfo{volume}{96}},
  \bibinfo{pages}{111102} (\bibinfo{year}{2006}), \eprint{gr-qc/0511103}.

\bibitem[{\citenamefont{Gonz\'alez et~al.}(2007)\citenamefont{Gonz\'alez,
  Sperhake, Brugmann, Hannam, and Husa}}]{Gonzalez:2006md}
\bibinfo{author}{\bibfnamefont{J.~A.} \bibnamefont{Gonz\'alez}},
  \bibinfo{author}{\bibfnamefont{U.}~\bibnamefont{Sperhake}},
  \bibinfo{author}{\bibfnamefont{B.}~\bibnamefont{Brugmann}},
  \bibinfo{author}{\bibfnamefont{M.}~\bibnamefont{Hannam}}, \bibnamefont{and}
  \bibinfo{author}{\bibfnamefont{S.}~\bibnamefont{Husa}},
  \bibinfo{journal}{Phys. Rev. Lett.} \textbf{\bibinfo{volume}{98}},
  \bibinfo{pages}{091101} (\bibinfo{year}{2007}), \eprint{gr-qc/0610154}.

\bibitem[{\citenamefont{Herrmann
  et~al.}(2007{\natexlab{a}})\citenamefont{Herrmann, Hinder, Shoemaker, Laguna,
  and Matzner}}]{Herrmann:2007ac}
\bibinfo{author}{\bibfnamefont{F.}~\bibnamefont{Herrmann}},
  \bibinfo{author}{\bibfnamefont{I.}~\bibnamefont{Hinder}},
  \bibinfo{author}{\bibfnamefont{D.}~\bibnamefont{Shoemaker}},
  \bibinfo{author}{\bibfnamefont{P.}~\bibnamefont{Laguna}}, \bibnamefont{and}
  \bibinfo{author}{\bibfnamefont{R.~A.} \bibnamefont{Matzner}},
  \bibinfo{journal}{Astrophys. J.} \textbf{\bibinfo{volume}{661}},
  \bibinfo{pages}{430} (\bibinfo{year}{2007}{\natexlab{a}}),
  \eprint{gr-qc/0701143}.

\bibitem[{\citenamefont{Koppitz et~al.}(2007)\citenamefont{Koppitz, Pollney,
  Reisswig, Rezzolla, Thornburg et~al.}}]{Koppitz:2007ev}
\bibinfo{author}{\bibfnamefont{M.}~\bibnamefont{Koppitz}},
  \bibinfo{author}{\bibfnamefont{D.}~\bibnamefont{Pollney}},
  \bibinfo{author}{\bibfnamefont{C.}~\bibnamefont{Reisswig}},
  \bibinfo{author}{\bibfnamefont{L.}~\bibnamefont{Rezzolla}},
  \bibinfo{author}{\bibfnamefont{J.}~\bibnamefont{Thornburg}},
  \bibnamefont{et~al.}, \bibinfo{journal}{Phys. Rev. Lett.}
  \textbf{\bibinfo{volume}{99}}, \bibinfo{pages}{041102}
  (\bibinfo{year}{2007}), \eprint{gr-qc/0701163}.

\bibitem[{\citenamefont{Campanelli
  et~al.}(2007{\natexlab{a}})\citenamefont{Campanelli, Lousto, Zlochower, and
  Merritt}}]{Campanelli:2007ew}
\bibinfo{author}{\bibfnamefont{M.}~\bibnamefont{Campanelli}},
  \bibinfo{author}{\bibfnamefont{C.~O.} \bibnamefont{Lousto}},
  \bibinfo{author}{\bibfnamefont{Y.}~\bibnamefont{Zlochower}},
  \bibnamefont{and} \bibinfo{author}{\bibfnamefont{D.}~\bibnamefont{Merritt}},
  \bibinfo{journal}{Astrophys. J.} \textbf{\bibinfo{volume}{659}},
  \bibinfo{pages}{L5} (\bibinfo{year}{2007}{\natexlab{a}}),
  \eprint{gr-qc/0701164}.

\bibitem[{\citenamefont{Campanelli
  et~al.}(2007{\natexlab{b}})\citenamefont{Campanelli, Lousto, Zlochower, and
  Merritt}}]{Campanelli:2007cga}
\bibinfo{author}{\bibfnamefont{M.}~\bibnamefont{Campanelli}},
  \bibinfo{author}{\bibfnamefont{C.~O.} \bibnamefont{Lousto}},
  \bibinfo{author}{\bibfnamefont{Y.}~\bibnamefont{Zlochower}},
  \bibnamefont{and} \bibinfo{author}{\bibfnamefont{D.}~\bibnamefont{Merritt}},
  \bibinfo{journal}{Phys. Rev. Lett.} \textbf{\bibinfo{volume}{98}},
  \bibinfo{pages}{231102} (\bibinfo{year}{2007}{\natexlab{b}}),
  \eprint{gr-qc/0702133}.

\bibitem[{\citenamefont{Lousto and
  Zlochower}(2011{\natexlab{a}})}]{Lousto:2011kp}
\bibinfo{author}{\bibfnamefont{C.~O.} \bibnamefont{Lousto}} \bibnamefont{and}
  \bibinfo{author}{\bibfnamefont{Y.}~\bibnamefont{Zlochower}},
  \bibinfo{journal}{Phys. Rev. Lett.} \textbf{\bibinfo{volume}{107}},
  \bibinfo{pages}{231102} (\bibinfo{year}{2011}{\natexlab{a}}),
  \eprint{1108.2009}.

\bibitem[{\citenamefont{Komossa et~al.}(2008)\citenamefont{Komossa, Zhou, and
  Lu}}]{Komossa:2008qd}
\bibinfo{author}{\bibfnamefont{S.}~\bibnamefont{Komossa}},
  \bibinfo{author}{\bibfnamefont{H.}~\bibnamefont{Zhou}}, \bibnamefont{and}
  \bibinfo{author}{\bibfnamefont{H.}~\bibnamefont{Lu}},
  \bibinfo{journal}{Astrop. J. Letters} \textbf{\bibinfo{volume}{678}},
  \bibinfo{pages}{L81} (\bibinfo{year}{2008}), \eprint{0804.4585}.

\bibitem[{\citenamefont{{Shields} and {Bonning}}(2008)}]{Shields:2008va}
\bibinfo{author}{\bibfnamefont{G.~A.} \bibnamefont{{Shields}}}
  \bibnamefont{and} \bibinfo{author}{\bibfnamefont{E.~W.}
  \bibnamefont{{Bonning}}}, \bibinfo{journal}{Astrophys. J.}
  \textbf{\bibinfo{volume}{682}}, \bibinfo{pages}{758} (\bibinfo{year}{2008}),
  \eprint{0802.3873}.

\bibitem[{\citenamefont{Bogdanovic et~al.}(2009)\citenamefont{Bogdanovic,
  Eracleous, and Sigurdsson}}]{Bogdanovic:2008uz}
\bibinfo{author}{\bibfnamefont{T.}~\bibnamefont{Bogdanovic}},
  \bibinfo{author}{\bibfnamefont{M.}~\bibnamefont{Eracleous}},
  \bibnamefont{and}
  \bibinfo{author}{\bibfnamefont{S.}~\bibnamefont{Sigurdsson}},
  \bibinfo{journal}{Astrophys. J.} \textbf{\bibinfo{volume}{697}},
  \bibinfo{pages}{288} (\bibinfo{year}{2009}), \eprint{0809.3262}.

\bibitem[{\citenamefont{Civano et~al.}(2010)}]{Civano:2010es}
\bibinfo{author}{\bibfnamefont{F.}~\bibnamefont{Civano}} \bibnamefont{et~al.},
  \bibinfo{journal}{Astrophys. J.} \textbf{\bibinfo{volume}{717}},
  \bibinfo{pages}{209} (\bibinfo{year}{2010}), \eprint{1003.0020}.

\bibitem[{\citenamefont{Eracleous et~al.}(2011)\citenamefont{Eracleous,
  Boroson, Halpern, and Liu}}]{Eracleous:2011ua}
\bibinfo{author}{\bibfnamefont{M.}~\bibnamefont{Eracleous}},
  \bibinfo{author}{\bibfnamefont{T.~A.} \bibnamefont{Boroson}},
  \bibinfo{author}{\bibfnamefont{J.~P.} \bibnamefont{Halpern}},
  \bibnamefont{and} \bibinfo{author}{\bibfnamefont{J.}~\bibnamefont{Liu}}
  (\bibinfo{year}{2011}), \eprint{1106.2952}.

\bibitem[{\citenamefont{Tsalmantza et~al.}(2011)\citenamefont{Tsalmantza,
  Decarli, Dotti, and Hogg}}]{Tsalmantza:2011ju}
\bibinfo{author}{\bibfnamefont{P.}~\bibnamefont{Tsalmantza}},
  \bibinfo{author}{\bibfnamefont{R.}~\bibnamefont{Decarli}},
  \bibinfo{author}{\bibfnamefont{M.}~\bibnamefont{Dotti}}, \bibnamefont{and}
  \bibinfo{author}{\bibfnamefont{D.~W.} \bibnamefont{Hogg}},
  \bibinfo{journal}{Astrophys. J.} \textbf{\bibinfo{volume}{738}},
  \bibinfo{pages}{20} (\bibinfo{year}{2011}), \eprint{1106.1180}.

\bibitem[{\citenamefont{Bogdanovic et~al.}(2007)\citenamefont{Bogdanovic,
  Reynolds, and Miller}}]{Bogdanovic:2007hp}
\bibinfo{author}{\bibfnamefont{T.}~\bibnamefont{Bogdanovic}},
  \bibinfo{author}{\bibfnamefont{C.~S.} \bibnamefont{Reynolds}},
  \bibnamefont{and} \bibinfo{author}{\bibfnamefont{M.~C.}
  \bibnamefont{Miller}}, \bibinfo{journal}{Astrophys. J.}
  \textbf{\bibinfo{volume}{661}}, \bibinfo{pages}{L147} (\bibinfo{year}{2007}),
  \eprint{astro-ph/0703054}.

\bibitem[{\citenamefont{{Dotti} et~al.}(2010)\citenamefont{{Dotti},
  {Volonteri}, {Perego}, {Colpi}, {Ruszkowski}, and {Haardt}}}]{Dotti:2009vz}
\bibinfo{author}{\bibfnamefont{M.}~\bibnamefont{{Dotti}}},
  \bibinfo{author}{\bibfnamefont{M.}~\bibnamefont{{Volonteri}}},
  \bibinfo{author}{\bibfnamefont{A.}~\bibnamefont{{Perego}}},
  \bibinfo{author}{\bibfnamefont{M.}~\bibnamefont{{Colpi}}},
  \bibinfo{author}{\bibfnamefont{M.}~\bibnamefont{{Ruszkowski}}},
  \bibnamefont{and} \bibinfo{author}{\bibfnamefont{F.}~\bibnamefont{{Haardt}}},
  \bibinfo{journal}{mnras} \textbf{\bibinfo{volume}{402}}, \bibinfo{pages}{682}
  (\bibinfo{year}{2010}), \eprint{0910.5729}.

\bibitem[{\citenamefont{Ansorg et~al.}(2004)\citenamefont{Ansorg, Br\"ugmann,
  and Tichy}}]{Ansorg:2004ds}
\bibinfo{author}{\bibfnamefont{M.}~\bibnamefont{Ansorg}},
  \bibinfo{author}{\bibfnamefont{B.}~\bibnamefont{Br\"ugmann}},
  \bibnamefont{and} \bibinfo{author}{\bibfnamefont{W.}~\bibnamefont{Tichy}},
  \bibinfo{journal}{Phys. Rev.} \textbf{\bibinfo{volume}{D70}},
  \bibinfo{pages}{064011} (\bibinfo{year}{2004}), \eprint{gr-qc/0404056}.

\bibitem[{\citenamefont{Brandt and Br{\"u}gmann}(1997)}]{Brandt97b}
\bibinfo{author}{\bibfnamefont{S.}~\bibnamefont{Brandt}} \bibnamefont{and}
  \bibinfo{author}{\bibfnamefont{B.}~\bibnamefont{Br{\"u}gmann}},
  \bibinfo{journal}{Phys. Rev. Lett.} \textbf{\bibinfo{volume}{78}},
  \bibinfo{pages}{3606} (\bibinfo{year}{1997}), \eprint{gr-qc/9703066}.

\bibitem[{\citenamefont{Zlochower et~al.}(2005)\citenamefont{Zlochower, Baker,
  Campanelli, and Lousto}}]{Zlochower:2005bj}
\bibinfo{author}{\bibfnamefont{Y.}~\bibnamefont{Zlochower}},
  \bibinfo{author}{\bibfnamefont{J.~G.} \bibnamefont{Baker}},
  \bibinfo{author}{\bibfnamefont{M.}~\bibnamefont{Campanelli}},
  \bibnamefont{and} \bibinfo{author}{\bibfnamefont{C.~O.}
  \bibnamefont{Lousto}}, \bibinfo{journal}{Phys. Rev.}
  \textbf{\bibinfo{volume}{D72}}, \bibinfo{pages}{024021}
  (\bibinfo{year}{2005}), \eprint{gr-qc/0505055}.

\bibitem[{\citenamefont{Marronetti et~al.}(2008)\citenamefont{Marronetti,
  Tichy, Br{\"u}gmann, Gonzalez, and Sperhake}}]{Marronetti:2007wz}
\bibinfo{author}{\bibfnamefont{P.}~\bibnamefont{Marronetti}},
  \bibinfo{author}{\bibfnamefont{W.}~\bibnamefont{Tichy}},
  \bibinfo{author}{\bibfnamefont{B.}~\bibnamefont{Br{\"u}gmann}},
  \bibinfo{author}{\bibfnamefont{J.}~\bibnamefont{Gonzalez}}, \bibnamefont{and}
  \bibinfo{author}{\bibfnamefont{U.}~\bibnamefont{Sperhake}},
  \bibinfo{journal}{Phys. Rev.} \textbf{\bibinfo{volume}{D77}},
  \bibinfo{pages}{064010} (\bibinfo{year}{2008}), \eprint{0709.2160}.

\bibitem[{\citenamefont{Lousto and
  Zlochower}(2008{\natexlab{a}})}]{Lousto:2007rj}
\bibinfo{author}{\bibfnamefont{C.~O.} \bibnamefont{Lousto}} \bibnamefont{and}
  \bibinfo{author}{\bibfnamefont{Y.}~\bibnamefont{Zlochower}},
  \bibinfo{journal}{Phys. Rev.} \textbf{\bibinfo{volume}{D77}},
  \bibinfo{pages}{024034} (\bibinfo{year}{2008}{\natexlab{a}}),
  \eprint{0711.1165}.

\bibitem[{cac()}]{cactus_web}
\bibinfo{note}{Cactus Computational Toolkit home page: {\tt
  http://cactuscode.org}}.

\bibitem[{ein()}]{einsteintoolkit}
\bibinfo{note}{Einstein Toolkit home page: {\tt http://einsteintoolkit.org}}.

\bibitem[{\citenamefont{Schnetter et~al.}(2004)\citenamefont{Schnetter, Hawley,
  and Hawke}}]{Schnetter-etal-03b}
\bibinfo{author}{\bibfnamefont{E.}~\bibnamefont{Schnetter}},
  \bibinfo{author}{\bibfnamefont{S.~H.} \bibnamefont{Hawley}},
  \bibnamefont{and} \bibinfo{author}{\bibfnamefont{I.}~\bibnamefont{Hawke}},
  \bibinfo{journal}{Class. Quantum Grav.} \textbf{\bibinfo{volume}{21}},
  \bibinfo{pages}{1465} (\bibinfo{year}{2004}), \eprint{gr-qc/0310042}.

\bibitem[{\citenamefont{Alcubierre et~al.}(2003)\citenamefont{Alcubierre,
  Br\"ugmann, Diener, Koppitz, Pollney, Seidel, and Takahashi}}]{Alcubierre02a}
\bibinfo{author}{\bibfnamefont{M.}~\bibnamefont{Alcubierre}},
  \bibinfo{author}{\bibfnamefont{B.}~\bibnamefont{Br\"ugmann}},
  \bibinfo{author}{\bibfnamefont{P.}~\bibnamefont{Diener}},
  \bibinfo{author}{\bibfnamefont{M.}~\bibnamefont{Koppitz}},
  \bibinfo{author}{\bibfnamefont{D.}~\bibnamefont{Pollney}},
  \bibinfo{author}{\bibfnamefont{E.}~\bibnamefont{Seidel}}, \bibnamefont{and}
  \bibinfo{author}{\bibfnamefont{R.}~\bibnamefont{Takahashi}},
  \bibinfo{journal}{Phys. Rev.} \textbf{\bibinfo{volume}{D67}},
  \bibinfo{pages}{084023} (\bibinfo{year}{2003}), \eprint{gr-qc/0206072}.

\bibitem[{\citenamefont{van Meter et~al.}(2006)\citenamefont{van Meter, Baker,
  Koppitz, and Choi}}]{vanMeter:2006vi}
\bibinfo{author}{\bibfnamefont{J.~R.} \bibnamefont{van Meter}},
  \bibinfo{author}{\bibfnamefont{J.~G.} \bibnamefont{Baker}},
  \bibinfo{author}{\bibfnamefont{M.}~\bibnamefont{Koppitz}}, \bibnamefont{and}
  \bibinfo{author}{\bibfnamefont{D.-I.} \bibnamefont{Choi}},
  \bibinfo{journal}{Phys. Rev.} \textbf{\bibinfo{volume}{D73}},
  \bibinfo{pages}{124011} (\bibinfo{year}{2006}), \eprint{gr-qc/0605030}.

\bibitem[{\citenamefont{Thornburg}(2004)}]{Thornburg2003:AH-finding}
\bibinfo{author}{\bibfnamefont{J.}~\bibnamefont{Thornburg}},
  \bibinfo{journal}{Class. Quant. Grav.} \textbf{\bibinfo{volume}{21}},
  \bibinfo{pages}{743} (\bibinfo{year}{2004}), \eprint{gr-qc/0306056}.

\bibitem[{\citenamefont{Dreyer et~al.}(2003)\citenamefont{Dreyer, Krishnan,
  Shoemaker, and Schnetter}}]{Dreyer02a}
\bibinfo{author}{\bibfnamefont{O.}~\bibnamefont{Dreyer}},
  \bibinfo{author}{\bibfnamefont{B.}~\bibnamefont{Krishnan}},
  \bibinfo{author}{\bibfnamefont{D.}~\bibnamefont{Shoemaker}},
  \bibnamefont{and}
  \bibinfo{author}{\bibfnamefont{E.}~\bibnamefont{Schnetter}},
  \bibinfo{journal}{Phys. Rev.} \textbf{\bibinfo{volume}{D67}},
  \bibinfo{pages}{024018} (\bibinfo{year}{2003}), \eprint{gr-qc/0206008}.

\bibitem[{\citenamefont{Campanelli and Lousto}(1999)}]{Campanelli:1998jv}
\bibinfo{author}{\bibfnamefont{M.}~\bibnamefont{Campanelli}} \bibnamefont{and}
  \bibinfo{author}{\bibfnamefont{C.~O.} \bibnamefont{Lousto}},
  \bibinfo{journal}{Phys. Rev.} \textbf{\bibinfo{volume}{D59}},
  \bibinfo{pages}{124022} (\bibinfo{year}{1999}), \eprint{gr-qc/9811019}.

\bibitem[{\citenamefont{Lousto and Zlochower}(2007)}]{Lousto:2007mh}
\bibinfo{author}{\bibfnamefont{C.~O.} \bibnamefont{Lousto}} \bibnamefont{and}
  \bibinfo{author}{\bibfnamefont{Y.}~\bibnamefont{Zlochower}},
  \bibinfo{journal}{Phys. Rev.} \textbf{\bibinfo{volume}{D76}},
  \bibinfo{pages}{041502(R)} (\bibinfo{year}{2007}), \eprint{gr-qc/0703061}.

\bibitem[{\citenamefont{Lousto and
  Zlochower}(2011{\natexlab{b}})}]{Lousto:2010xk}
\bibinfo{author}{\bibfnamefont{C.~O.} \bibnamefont{Lousto}} \bibnamefont{and}
  \bibinfo{author}{\bibfnamefont{Y.}~\bibnamefont{Zlochower}},
  \bibinfo{journal}{Phys. Rev.} \textbf{\bibinfo{volume}{D83}},
  \bibinfo{pages}{024003} (\bibinfo{year}{2011}{\natexlab{b}}),
  \eprint{1011.0593}.

\bibitem[{\citenamefont{Pfeiffer et~al.}(2007)}]{Pfeiffer:2007yz}
\bibinfo{author}{\bibfnamefont{H.~P.} \bibnamefont{Pfeiffer}}
  \bibnamefont{et~al.}, \bibinfo{journal}{Class. Quant. Grav.}
  \textbf{\bibinfo{volume}{24}}, \bibinfo{pages}{S59} (\bibinfo{year}{2007}),
  \eprint{gr-qc/0702106}.

\bibitem[{\citenamefont{Buonanno et~al.}(2011)\citenamefont{Buonanno, Kidder,
  Mroue, Pfeiffer, and Taracchini}}]{Buonanno:2010yk}
\bibinfo{author}{\bibfnamefont{A.}~\bibnamefont{Buonanno}},
  \bibinfo{author}{\bibfnamefont{L.~E.} \bibnamefont{Kidder}},
  \bibinfo{author}{\bibfnamefont{A.~H.} \bibnamefont{Mroue}},
  \bibinfo{author}{\bibfnamefont{H.~P.} \bibnamefont{Pfeiffer}},
  \bibnamefont{and}
  \bibinfo{author}{\bibfnamefont{A.}~\bibnamefont{Taracchini}},
  \bibinfo{journal}{Phys. Rev.} \textbf{\bibinfo{volume}{D83}},
  \bibinfo{pages}{104034} (\bibinfo{year}{2011}), \eprint{1012.1549}.

\bibitem[{\citenamefont{Campanelli
  et~al.}(2007{\natexlab{c}})\citenamefont{Campanelli, Lousto, Zlochower,
  Krishnan, and Merritt}}]{Campanelli:2006fy}
\bibinfo{author}{\bibfnamefont{M.}~\bibnamefont{Campanelli}},
  \bibinfo{author}{\bibfnamefont{C.~O.} \bibnamefont{Lousto}},
  \bibinfo{author}{\bibfnamefont{Y.}~\bibnamefont{Zlochower}},
  \bibinfo{author}{\bibfnamefont{B.}~\bibnamefont{Krishnan}}, \bibnamefont{and}
  \bibinfo{author}{\bibfnamefont{D.}~\bibnamefont{Merritt}},
  \bibinfo{journal}{Phys. Rev.} \textbf{\bibinfo{volume}{D75}},
  \bibinfo{pages}{064030} (\bibinfo{year}{2007}{\natexlab{c}}),
  \eprint{gr-qc/0612076}.

\bibitem[{\citenamefont{Campanelli
  et~al.}(2006{\natexlab{b}})\citenamefont{Campanelli, Lousto, and
  Zlochower}}]{Campanelli:2006fg}
\bibinfo{author}{\bibfnamefont{M.}~\bibnamefont{Campanelli}},
  \bibinfo{author}{\bibfnamefont{C.~O.} \bibnamefont{Lousto}},
  \bibnamefont{and}
  \bibinfo{author}{\bibfnamefont{Y.}~\bibnamefont{Zlochower}},
  \bibinfo{journal}{Phys. Rev.} \textbf{\bibinfo{volume}{D74}},
  \bibinfo{pages}{084023} (\bibinfo{year}{2006}{\natexlab{b}}),
  \eprint{astro-ph/0608275}.

\bibitem[{\citenamefont{Nichols and Chen}(2011)}]{Nichols:2011ih}
\bibinfo{author}{\bibfnamefont{D.~A.} \bibnamefont{Nichols}} \bibnamefont{and}
  \bibinfo{author}{\bibfnamefont{Y.}~\bibnamefont{Chen}},
  \bibinfo{journal}{Phys. Rev. D}  (\bibinfo{year}{2011}), \eprint{1109.0081}.

\bibitem[{\citenamefont{Lousto and Zlochower}(2009)}]{Lousto:2008dn}
\bibinfo{author}{\bibfnamefont{C.~O.} \bibnamefont{Lousto}} \bibnamefont{and}
  \bibinfo{author}{\bibfnamefont{Y.}~\bibnamefont{Zlochower}},
  \bibinfo{journal}{Phys. Rev.} \textbf{\bibinfo{volume}{D79}},
  \bibinfo{pages}{064018} (\bibinfo{year}{2009}), \eprint{0805.0159}.

\bibitem[{\citenamefont{Dain et~al.}(2008)\citenamefont{Dain, Lousto, and
  Zlochower}}]{Dain:2008ck}
\bibinfo{author}{\bibfnamefont{S.}~\bibnamefont{Dain}},
  \bibinfo{author}{\bibfnamefont{C.~O.} \bibnamefont{Lousto}},
  \bibnamefont{and}
  \bibinfo{author}{\bibfnamefont{Y.}~\bibnamefont{Zlochower}},
  \bibinfo{journal}{Phys. Rev.} \textbf{\bibinfo{volume}{D78}},
  \bibinfo{pages}{024039} (\bibinfo{year}{2008}), \eprint{0803.0351}.

\bibitem[{\citenamefont{Lousto and
  Zlochower}(2008{\natexlab{b}})}]{Lousto:2007db}
\bibinfo{author}{\bibfnamefont{C.~O.} \bibnamefont{Lousto}} \bibnamefont{and}
  \bibinfo{author}{\bibfnamefont{Y.}~\bibnamefont{Zlochower}},
  \bibinfo{journal}{Phys. Rev.} \textbf{\bibinfo{volume}{D77}},
  \bibinfo{pages}{044028} (\bibinfo{year}{2008}{\natexlab{b}}),
  \eprint{0708.4048}.

\bibitem[{\citenamefont{Kidder}(1995)}]{Kidder:1995zr}
\bibinfo{author}{\bibfnamefont{L.~E.} \bibnamefont{Kidder}},
  \bibinfo{journal}{Phys. Rev.} \textbf{\bibinfo{volume}{D52}},
  \bibinfo{pages}{821} (\bibinfo{year}{1995}), \eprint{gr-qc/9506022}.

\bibitem[{\citenamefont{Lousto et~al.}(2010{\natexlab{a}})\citenamefont{Lousto,
  Campanelli, Zlochower, and Nakano}}]{Lousto:2009mf}
\bibinfo{author}{\bibfnamefont{C.~O.} \bibnamefont{Lousto}},
  \bibinfo{author}{\bibfnamefont{M.}~\bibnamefont{Campanelli}},
  \bibinfo{author}{\bibfnamefont{Y.}~\bibnamefont{Zlochower}},
  \bibnamefont{and} \bibinfo{author}{\bibfnamefont{H.}~\bibnamefont{Nakano}},
  \bibinfo{journal}{Class. Quant. Grav.} \textbf{\bibinfo{volume}{27}},
  \bibinfo{pages}{114006} (\bibinfo{year}{2010}{\natexlab{a}}),
  \eprint{0904.3541}.

\bibitem[{\citenamefont{Racine et~al.}(2009)\citenamefont{Racine, Buonanno, and
  Kidder}}]{Racine:2008kj}
\bibinfo{author}{\bibfnamefont{E.}~\bibnamefont{Racine}},
  \bibinfo{author}{\bibfnamefont{A.}~\bibnamefont{Buonanno}}, \bibnamefont{and}
  \bibinfo{author}{\bibfnamefont{L.~E.} \bibnamefont{Kidder}},
  \bibinfo{journal}{Phys. Rev.} \textbf{\bibinfo{volume}{D80}},
  \bibinfo{pages}{044010} (\bibinfo{year}{2009}), \eprint{0812.4413}.

\bibitem[{\citenamefont{Zlochower et~al.}(2011)\citenamefont{Zlochower,
  Campanelli, and Lousto}}]{Zlochower:2010sn}
\bibinfo{author}{\bibfnamefont{Y.}~\bibnamefont{Zlochower}},
  \bibinfo{author}{\bibfnamefont{M.}~\bibnamefont{Campanelli}},
  \bibnamefont{and} \bibinfo{author}{\bibfnamefont{C.~O.}
  \bibnamefont{Lousto}}, \bibinfo{journal}{Class. Quant. Grav.}
  \textbf{\bibinfo{volume}{28}}, \bibinfo{pages}{114015}
  (\bibinfo{year}{2011}), \eprint{1011.2210}.

\bibitem[{\citenamefont{Boyle and Kesden}(2008)}]{Boyle:2007ru}
\bibinfo{author}{\bibfnamefont{L.}~\bibnamefont{Boyle}} \bibnamefont{and}
  \bibinfo{author}{\bibfnamefont{M.}~\bibnamefont{Kesden}},
  \bibinfo{journal}{Phys. Rev.} \textbf{\bibinfo{volume}{D78}},
  \bibinfo{pages}{024017} (\bibinfo{year}{2008}), \eprint{0712.2819}.

\bibitem[{\citenamefont{Herrmann
  et~al.}(2007{\natexlab{b}})\citenamefont{Herrmann, Hinder, Shoemaker, Laguna,
  and Matzner}}]{Herrmann:2007ex}
\bibinfo{author}{\bibfnamefont{F.}~\bibnamefont{Herrmann}},
  \bibinfo{author}{\bibfnamefont{I.}~\bibnamefont{Hinder}},
  \bibinfo{author}{\bibfnamefont{D.~M.} \bibnamefont{Shoemaker}},
  \bibinfo{author}{\bibfnamefont{P.}~\bibnamefont{Laguna}}, \bibnamefont{and}
  \bibinfo{author}{\bibfnamefont{R.~A.} \bibnamefont{Matzner}},
  \bibinfo{journal}{Phys. Rev.} \textbf{\bibinfo{volume}{D76}},
  \bibinfo{pages}{084032} (\bibinfo{year}{2007}{\natexlab{b}}),
  \eprint{0706.2541}.

\bibitem[{\citenamefont{Lousto et~al.}(2010{\natexlab{b}})\citenamefont{Lousto,
  Nakano, Zlochower, and Campanelli}}]{Lousto:2009ka}
\bibinfo{author}{\bibfnamefont{C.~O.} \bibnamefont{Lousto}},
  \bibinfo{author}{\bibfnamefont{H.}~\bibnamefont{Nakano}},
  \bibinfo{author}{\bibfnamefont{Y.}~\bibnamefont{Zlochower}},
  \bibnamefont{and}
  \bibinfo{author}{\bibfnamefont{M.}~\bibnamefont{Campanelli}},
  \bibinfo{journal}{Phys. Rev.} \textbf{\bibinfo{volume}{D81}},
  \bibinfo{pages}{084023} (\bibinfo{year}{2010}{\natexlab{b}}),
  \eprint{0910.3197}.

\bibitem[{\citenamefont{{Mayer} et~al.}(2007)\citenamefont{{Mayer},
  {Kazantzidis}, {Madau}, {Colpi}, {Quinn}, and {Wadsley}}}]{mayer07}
\bibinfo{author}{\bibfnamefont{L.}~\bibnamefont{{Mayer}}},
  \bibinfo{author}{\bibfnamefont{S.}~\bibnamefont{{Kazantzidis}}},
  \bibinfo{author}{\bibfnamefont{P.}~\bibnamefont{{Madau}}},
  \bibinfo{author}{\bibfnamefont{M.}~\bibnamefont{{Colpi}}},
  \bibinfo{author}{\bibfnamefont{T.}~\bibnamefont{{Quinn}}}, \bibnamefont{and}
  \bibinfo{author}{\bibfnamefont{J.}~\bibnamefont{{Wadsley}}},
  \bibinfo{journal}{Science} \textbf{\bibinfo{volume}{316}},
  \bibinfo{pages}{1874} (\bibinfo{year}{2007}), \eprint{0706.1562}.

\bibitem[{\citenamefont{{Hopkins} and {Quataert}}(2010)}]{hopkins10}
\bibinfo{author}{\bibfnamefont{P.~F.} \bibnamefont{{Hopkins}}}
  \bibnamefont{and}
  \bibinfo{author}{\bibfnamefont{E.}~\bibnamefont{{Quataert}}},
  \bibinfo{journal}{MNRAS} \textbf{\bibinfo{volume}{407}},
  \bibinfo{pages}{1529} (\bibinfo{year}{2010}), \eprint{0912.3257}.

\bibitem[{\citenamefont{{Downes} and {Solomon}}(1998)}]{downes98}
\bibinfo{author}{\bibfnamefont{D.}~\bibnamefont{{Downes}}} \bibnamefont{and}
  \bibinfo{author}{\bibfnamefont{P.~M.} \bibnamefont{{Solomon}}},
  \bibinfo{journal}{ApJ} \textbf{\bibinfo{volume}{507}}, \bibinfo{pages}{615}
  (\bibinfo{year}{1998}), \eprint{arXiv:astro-ph/9806377}.

\bibitem[{\citenamefont{{Davies}
  et~al.}(2004{\natexlab{a}})\citenamefont{{Davies}, {Tacconi}, and
  {Genzel}}}]{daviesA04}
\bibinfo{author}{\bibfnamefont{R.~I.} \bibnamefont{{Davies}}},
  \bibinfo{author}{\bibfnamefont{L.~J.} \bibnamefont{{Tacconi}}},
  \bibnamefont{and} \bibinfo{author}{\bibfnamefont{R.}~\bibnamefont{{Genzel}}},
  \bibinfo{journal}{ApJ} \textbf{\bibinfo{volume}{613}}, \bibinfo{pages}{781}
  (\bibinfo{year}{2004}{\natexlab{a}}), \eprint{arXiv:astro-ph/0406342}.

\bibitem[{\citenamefont{{Davies}
  et~al.}(2004{\natexlab{b}})\citenamefont{{Davies}, {Tacconi}, and
  {Genzel}}}]{daviesB04}
\bibinfo{author}{\bibfnamefont{R.~I.} \bibnamefont{{Davies}}},
  \bibinfo{author}{\bibfnamefont{L.~J.} \bibnamefont{{Tacconi}}},
  \bibnamefont{and} \bibinfo{author}{\bibfnamefont{R.}~\bibnamefont{{Genzel}}},
  \bibinfo{journal}{ApJ} \textbf{\bibinfo{volume}{602}}, \bibinfo{pages}{148}
  (\bibinfo{year}{2004}{\natexlab{b}}), \eprint{arXiv:astro-ph/0310681}.

\bibitem[{\citenamefont{{Bardeen} and {Petterson}}(1975)}]{bardeen75}
\bibinfo{author}{\bibfnamefont{J.~M.} \bibnamefont{{Bardeen}}}
  \bibnamefont{and} \bibinfo{author}{\bibfnamefont{J.~A.}
  \bibnamefont{{Petterson}}}, \bibinfo{journal}{ApJ}
  \textbf{\bibinfo{volume}{195}}, \bibinfo{pages}{L65+} (\bibinfo{year}{1975}).

\bibitem[{\citenamefont{Perego et~al.}(2009)\citenamefont{Perego, Dotti, Colpi,
  and Volonteri}}]{Perego:2009cw}
\bibinfo{author}{\bibfnamefont{A.}~\bibnamefont{Perego}},
  \bibinfo{author}{\bibfnamefont{M.}~\bibnamefont{Dotti}},
  \bibinfo{author}{\bibfnamefont{M.}~\bibnamefont{Colpi}}, \bibnamefont{and}
  \bibinfo{author}{\bibfnamefont{M.}~\bibnamefont{Volonteri}},
  \bibinfo{journal}{mnras} \textbf{\bibinfo{volume}{399}},
  \bibinfo{pages}{2249} (\bibinfo{year}{2009}), \eprint{0907.3742}.

\bibitem[{\citenamefont{Shapiro and {T}eukolsky}(1983)}]{Shapiro83}
\bibinfo{author}{\bibfnamefont{S.~L.} \bibnamefont{Shapiro}} \bibnamefont{and}
  \bibinfo{author}{\bibfnamefont{S.~A.} \bibnamefont{{T}eukolsky}},
  \emph{\bibinfo{title}{Black Holes, White Dwarfs, and Neutron Stars}}
  (\bibinfo{publisher}{John Wiley \& Sons}, \bibinfo{address}{New York},
  \bibinfo{year}{1983}).

\bibitem[{\citenamefont{{Springel} et~al.}(2001)\citenamefont{{Springel},
  {Yoshida}, and {White}}}]{GADGET}
\bibinfo{author}{\bibfnamefont{V.}~\bibnamefont{{Springel}}},
  \bibinfo{author}{\bibfnamefont{N.}~\bibnamefont{{Yoshida}}},
  \bibnamefont{and} \bibinfo{author}{\bibfnamefont{S.~D.~M.}
  \bibnamefont{{White}}}, \bibinfo{journal}{New Astronomy}
  \textbf{\bibinfo{volume}{6}}, \bibinfo{pages}{79} (\bibinfo{year}{2001}),
  \eprint{astro-ph/0003162}.

\bibitem[{\citenamefont{{Dotti} et~al.}(2009)\citenamefont{{Dotti},
  {Ruszkowski}, {Paredi}, {Colpi}, {Volonteri}, and {Haardt}}}]{dotti09}
\bibinfo{author}{\bibfnamefont{M.}~\bibnamefont{{Dotti}}},
  \bibinfo{author}{\bibfnamefont{M.}~\bibnamefont{{Ruszkowski}}},
  \bibinfo{author}{\bibfnamefont{L.}~\bibnamefont{{Paredi}}},
  \bibinfo{author}{\bibfnamefont{M.}~\bibnamefont{{Colpi}}},
  \bibinfo{author}{\bibfnamefont{M.}~\bibnamefont{{Volonteri}}},
  \bibnamefont{and} \bibinfo{author}{\bibfnamefont{F.}~\bibnamefont{{Haardt}}},
  \bibinfo{journal}{MNRAS} \textbf{\bibinfo{volume}{396}},
  \bibinfo{pages}{1640} (\bibinfo{year}{2009}), \eprint{0902.1525}.

\bibitem[{\citenamefont{{Spaans} and {Silk}}(2000)}]{spaans00}
\bibinfo{author}{\bibfnamefont{M.}~\bibnamefont{{Spaans}}} \bibnamefont{and}
  \bibinfo{author}{\bibfnamefont{J.}~\bibnamefont{{Silk}}},
  \bibinfo{journal}{ApJ} \textbf{\bibinfo{volume}{538}}, \bibinfo{pages}{115}
  (\bibinfo{year}{2000}), \eprint{arXiv:astro-ph/0002483}.

\bibitem[{\citenamefont{{Klessen} et~al.}(2007)\citenamefont{{Klessen},
  {Spaans}, and {Jappsen}}}]{klessen07}
\bibinfo{author}{\bibfnamefont{R.~S.} \bibnamefont{{Klessen}}},
  \bibinfo{author}{\bibfnamefont{M.}~\bibnamefont{{Spaans}}}, \bibnamefont{and}
  \bibinfo{author}{\bibfnamefont{A.}~\bibnamefont{{Jappsen}}},
  \bibinfo{journal}{MNRAS} \textbf{\bibinfo{volume}{374}}, \bibinfo{pages}{L29}
  (\bibinfo{year}{2007}), \eprint{arXiv:astro-ph/0610557}.

\bibitem[{\citenamefont{{Dotti} et~al.}(2006)\citenamefont{{Dotti}, {Colpi},
  and {Haardt}}}]{dotti06}
\bibinfo{author}{\bibfnamefont{M.}~\bibnamefont{{Dotti}}},
  \bibinfo{author}{\bibfnamefont{M.}~\bibnamefont{{Colpi}}}, \bibnamefont{and}
  \bibinfo{author}{\bibfnamefont{F.}~\bibnamefont{{Haardt}}},
  \bibinfo{journal}{MNRAS} \textbf{\bibinfo{volume}{367}}, \bibinfo{pages}{103}
  (\bibinfo{year}{2006}).

\bibitem[{\citenamefont{{Dotti} et~al.}(2007)\citenamefont{{Dotti}, {Colpi},
  {Haardt}, and {Mayer}}}]{dotti07}
\bibinfo{author}{\bibfnamefont{M.}~\bibnamefont{{Dotti}}},
  \bibinfo{author}{\bibfnamefont{M.}~\bibnamefont{{Colpi}}},
  \bibinfo{author}{\bibfnamefont{F.}~\bibnamefont{{Haardt}}}, \bibnamefont{and}
  \bibinfo{author}{\bibfnamefont{L.}~\bibnamefont{{Mayer}}},
  \bibinfo{journal}{MNRAS} \textbf{\bibinfo{volume}{379}}, \bibinfo{pages}{956}
  (\bibinfo{year}{2007}), \eprint{arXiv:astro-ph/0612505}.

\bibitem[{\citenamefont{{Shakura} and {Sunyaev}}(1973)}]{shakura73}
\bibinfo{author}{\bibfnamefont{N.~I.} \bibnamefont{{Shakura}}}
  \bibnamefont{and} \bibinfo{author}{\bibfnamefont{R.~A.}
  \bibnamefont{{Sunyaev}}}, \bibinfo{journal}{A\&A}
  \textbf{\bibinfo{volume}{24}}, \bibinfo{pages}{337} (\bibinfo{year}{1973}).

\bibitem[{\citenamefont{{Lodato} and {Pringle}}(2007)}]{lodato07}
\bibinfo{author}{\bibfnamefont{G.}~\bibnamefont{{Lodato}}} \bibnamefont{and}
  \bibinfo{author}{\bibfnamefont{J.~E.} \bibnamefont{{Pringle}}},
  \bibinfo{journal}{MNRAS} \textbf{\bibinfo{volume}{381}},
  \bibinfo{pages}{1287} (\bibinfo{year}{2007}), \eprint{0708.1124}.

\bibitem[{\citenamefont{Yu et~al.}(2011)\citenamefont{Yu, Lu, Mohayaee, and
  Colin}}]{Yu:2011vp}
\bibinfo{author}{\bibfnamefont{Q.}~\bibnamefont{Yu}},
  \bibinfo{author}{\bibfnamefont{Y.}~\bibnamefont{Lu}},
  \bibinfo{author}{\bibfnamefont{R.}~\bibnamefont{Mohayaee}}, \bibnamefont{and}
  \bibinfo{author}{\bibfnamefont{J.}~\bibnamefont{Colin}},
  \bibinfo{journal}{Astrophys. J.} \textbf{\bibinfo{volume}{738}},
  \bibinfo{pages}{92} (\bibinfo{year}{2011}), \eprint{1105.1963}.

\bibitem[{\citenamefont{Stewart et~al.}(2009)\citenamefont{Stewart, Bullock,
  Barton, and Wechsler}}]{Stewart:2008ep}
\bibinfo{author}{\bibfnamefont{K.~R.} \bibnamefont{Stewart}},
  \bibinfo{author}{\bibfnamefont{J.~S.} \bibnamefont{Bullock}},
  \bibinfo{author}{\bibfnamefont{E.~J.} \bibnamefont{Barton}},
  \bibnamefont{and} \bibinfo{author}{\bibfnamefont{R.~H.}
  \bibnamefont{Wechsler}}, \bibinfo{journal}{Astrophys. J.}
  \textbf{\bibinfo{volume}{702}}, \bibinfo{pages}{1005} (\bibinfo{year}{2009}),
  \eprint{0811.1218}.

\bibitem[{\citenamefont{Hopkins et~al.}(2010)\citenamefont{Hopkins, Bundy,
  Croton, Hernquist, Keres et~al.}}]{Hopkins:2009yy}
\bibinfo{author}{\bibfnamefont{P.~F.} \bibnamefont{Hopkins}},
  \bibinfo{author}{\bibfnamefont{K.}~\bibnamefont{Bundy}},
  \bibinfo{author}{\bibfnamefont{D.}~\bibnamefont{Croton}},
  \bibinfo{author}{\bibfnamefont{L.}~\bibnamefont{Hernquist}},
  \bibinfo{author}{\bibfnamefont{D.}~\bibnamefont{Keres}},
  \bibnamefont{et~al.}, \bibinfo{journal}{Astrophys. J.}
  \textbf{\bibinfo{volume}{715}}, \bibinfo{pages}{202} (\bibinfo{year}{2010}),
  \eprint{0906.5357}.

\bibitem[{\citenamefont{Ponce et~al.}(2012)\citenamefont{Ponce, Faber, and
  Lombardi}}]{Ponce:2011kv}
\bibinfo{author}{\bibfnamefont{M.}~\bibnamefont{Ponce}},
  \bibinfo{author}{\bibfnamefont{J.~A.} \bibnamefont{Faber}}, \bibnamefont{and}
  \bibinfo{author}{\bibfnamefont{J.}~\bibnamefont{Lombardi},
  \bibfnamefont{James~C.}}, \bibinfo{journal}{Astrophys. J.}
  \textbf{\bibinfo{volume}{745}}, \bibinfo{pages}{71} (\bibinfo{year}{2012}),
  \eprint{1107.1711}.

\bibitem[{\citenamefont{{Lippai} et~al.}(2008)\citenamefont{{Lippai}, {Frei},
  and {Haiman}}}]{Lippai:2008fx}
\bibinfo{author}{\bibfnamefont{Z.}~\bibnamefont{{Lippai}}},
  \bibinfo{author}{\bibfnamefont{Z.}~\bibnamefont{{Frei}}}, \bibnamefont{and}
  \bibinfo{author}{\bibfnamefont{Z.}~\bibnamefont{{Haiman}}},
  \bibinfo{journal}{Astrophys. J. Lett.} \textbf{\bibinfo{volume}{676}},
  \bibinfo{pages}{L5} (\bibinfo{year}{2008}), \eprint{0801.0739}.

\bibitem[{\citenamefont{Rossi et~al.}(2010)\citenamefont{Rossi, Lodato,
  Armitage, Pringle, and King}}]{Rossi:2009nk}
\bibinfo{author}{\bibfnamefont{E.~M.} \bibnamefont{Rossi}},
  \bibinfo{author}{\bibfnamefont{G.}~\bibnamefont{Lodato}},
  \bibinfo{author}{\bibfnamefont{P.}~\bibnamefont{Armitage}},
  \bibinfo{author}{\bibfnamefont{J.}~\bibnamefont{Pringle}}, \bibnamefont{and}
  \bibinfo{author}{\bibfnamefont{A.}~\bibnamefont{King}},
  \bibinfo{journal}{Mon. Not. Roy. Astron. Soc.}
  \textbf{\bibinfo{volume}{401}}, \bibinfo{pages}{2021} (\bibinfo{year}{2010}),
  \eprint{0910.0002}.

\bibitem[{\citenamefont{Corrales et~al.}(2009)\citenamefont{Corrales, Haiman,
  and MacFadyen}}]{Corrales:2009nv}
\bibinfo{author}{\bibfnamefont{L.~R.} \bibnamefont{Corrales}},
  \bibinfo{author}{\bibfnamefont{Z.}~\bibnamefont{Haiman}}, \bibnamefont{and}
  \bibinfo{author}{\bibfnamefont{A.}~\bibnamefont{MacFadyen}}
  (\bibinfo{year}{2009}), \eprint{0910.0014}.

\bibitem[{\citenamefont{Milosavljevic and
  Phinney}(2005)}]{Milosavljevic:2004cg}
\bibinfo{author}{\bibfnamefont{M.}~\bibnamefont{Milosavljevic}}
  \bibnamefont{and} \bibinfo{author}{\bibfnamefont{E.}~\bibnamefont{Phinney}},
  \bibinfo{journal}{Astrophys. J.} \textbf{\bibinfo{volume}{622}},
  \bibinfo{pages}{L93} (\bibinfo{year}{2005}), \eprint{astro-ph/0410343}.

\bibitem[{\citenamefont{Schnittman and Krolik}(2008)}]{Schnittman:2008ez}
\bibinfo{author}{\bibfnamefont{J.~D.} \bibnamefont{Schnittman}}
  \bibnamefont{and} \bibinfo{author}{\bibfnamefont{J.~H.}
  \bibnamefont{Krolik}}, \bibinfo{journal}{Astrophys. J.}
  \textbf{\bibinfo{volume}{684}}, \bibinfo{pages}{835} (\bibinfo{year}{2008}),
  \eprint{0802.3556}.

\bibitem[{\citenamefont{Campanelli
  et~al.}(2006{\natexlab{c}})\citenamefont{Campanelli, Lousto, and
  Zlochower}}]{Campanelli:2006uy}
\bibinfo{author}{\bibfnamefont{M.}~\bibnamefont{Campanelli}},
  \bibinfo{author}{\bibfnamefont{C.~O.} \bibnamefont{Lousto}},
  \bibnamefont{and}
  \bibinfo{author}{\bibfnamefont{Y.}~\bibnamefont{Zlochower}},
  \bibinfo{journal}{Phys. Rev.} \textbf{\bibinfo{volume}{D74}},
  \bibinfo{pages}{041501(R)} (\bibinfo{year}{2006}{\natexlab{c}}),
  \eprint{gr-qc/0604012}.

\bibitem[{\citenamefont{Schnittman}(2004)}]{Schnittman:2004vq}
\bibinfo{author}{\bibfnamefont{J.~D.} \bibnamefont{Schnittman}},
  \bibinfo{journal}{Phys. Rev.} \textbf{\bibinfo{volume}{D70}},
  \bibinfo{pages}{124020} (\bibinfo{year}{2004}), \eprint{astro-ph/0409174}.

\bibitem[{\citenamefont{Kesden et~al.}(2010)\citenamefont{Kesden, Sperhake, and
  Berti}}]{Kesden:2010ji}
\bibinfo{author}{\bibfnamefont{M.}~\bibnamefont{Kesden}},
  \bibinfo{author}{\bibfnamefont{U.}~\bibnamefont{Sperhake}}, \bibnamefont{and}
  \bibinfo{author}{\bibfnamefont{E.}~\bibnamefont{Berti}},
  \bibinfo{journal}{Astrophys. J.} \textbf{\bibinfo{volume}{715}},
  \bibinfo{pages}{1006} (\bibinfo{year}{2010}), \eprint{1003.4993}.

\end{thebibliography}

\end{document}